\documentclass{aa}         
\usepackage{txfonts}
\usepackage{graphicx}
\bibpunct{(}{)}{;}{a}{}{,}      

\begin{document}

\title{Wave heating in gravitationally stratified coronal loops in the presence of resistivity and viscosity.}

\author{K. Karampelas\inst{1} \and T. Van Doorsselaere\inst{1} \and M. Guo\inst{2,1} } 

\institute{Centre for mathematical Plasma Astrophysics, Department of Mathematics, KU Leuven, Celestijnenlaan 200B bus 2400, 3001 Leuven, Belgium \\ \email{kostas.karampelas@kuleuven.be} \and Institute of space sciences, Shandong University, Weihai 264209, China} 

\date{Received <date> /
Accepted <date>}

\abstract{In recent years, coronal loops have been the focus of studies related to the damping of different magnetohydrodynamic (MHD) surface waves and their connection with coronal seismology and wave heating. For a better understanding of wave heating, we need to take into account the effects of different dissipation coefficients such as resistivity and viscosity, the importance of the loop physical characteristics, and the ways gravity can factor into the evolution of these phenomena.}
{We aim to map the sites of energy dissipation from transverse waves in coronal loops in the presence and absence of gravitational stratification and to compare ideal, resistive, and viscous MHD.}
{Using the PLUTO code, we performed 3D MHD simulations of kink waves in single, straight, density-enhanced coronal flux tubes of multiple temperatures.}
{We see the creation of spatially expanded Kelvin-Helmholtz eddies along the loop, which deform the initial monolithic loop profile. For the case of driven oscillations, the Kelvin-Helmholtz instability develops despite physical dissipation, unless very high values of shear viscosity are used. Energy dissipation gets its highest values near the apex, but is present all along the loop. We observe an increased efficiency of wave heating once the kinetic energy saturates at the later stages of the simulation and a turbulent density profile has developed.}
{The inclusion of gravity greatly alters the dynamic evolution of our systems and should not be ignored in future studies. Stronger physical dissipation leads to stronger wave heating in our set-ups. Finally, once the kinetic energy of the oscillating loop starts saturating, all the excess input energy turns into internal energy, resulting in more efficient wave heating.}


\titlerunning{Wave heating in gravitationally stratified coronal loops.}
\authorrunning{Karampelas et al.}

\maketitle                      

\section{Introduction}
One of the open questions regarding the nature of the solar atmosphere is explaining its radial temperature profile. The extreme ultraviolet (EUV) and thermal X-ray emission from the solar corona reveal plasma temperatures above $1$ MK, while observations of active regions reveal temperatures of $\log T \geqslant 6.5$ for plasma confined into compact loops \citep{testareale2012}. These findings guide the study of solar atmospheric heating towards areas of stronger, structured magnetic fields in the context of both the active and the quiet Sun.

Coronal heating models are usually classified into direct current (DC) and alternating current (AC) models. In direct current models like Ohmic dissipation of current sheets and nanoflares, heating is induced by magnetic field braiding in timescales much larger than the Alfv\'{e}n crossing time along a coronal loop \citep{cargillklimchuk2004,klimchuk2006,chittahardi2018}. On the other hand, alternating current models focus on mechanisms with dynamic timescales that are shorter than the Alfv\'{e}n crossing time along a coronal loop and mainly consist of wave energy dissipation models \citep{Hollweg1981,ofman1994heat,pagano2017,pagano2018} and Alfv\'{e}n wave induced turbulence \citep{Ballegooijen2014,Ballegooijen2017,magyar2017}.

The increased interest in loop oscillations is also justified by the discovery of transverse magnetohydrodynamic (MHD) oscillations of loops \citep{aschwanden1999, nakariakov1999}. The physical characteristics of the loops allow them to dynamically connect different layers of the solar atmosphere, by acting as waveguides and transferring energy across those layers. The most well-studied loop model is the simple structure of a cylindrical flux tube; the theory of surface waves in \citet{zajtsev1975}, \citet{ryutov1976}, and \citet{edwin1983wave} described the different modes expected in such a structure. Observations by the Coronal Multi-channel Polarimeter (CoMP), the Solar Dynamics Observatory (SDO), and Hinode spacecraft have further sustained research interest in the environments where oscillating loops are found in abundance. A large number of studies have already proved the ubiquity of transverse perturbations along coronal loops, prominence threads and greater areas of the corona \citep{tomczyk2007, okamoto2007, tomczyk2009, mcintosh2011}, and the magnitude of the estimated energy carried by such waves is under strong debate \citep{depontieu2007,mcintosh2011,goossens2013energyApJ,tvd2014energyApJ,thurgood2014ApJ,morton2016ApJ}. 

The first step towards energy dissipation from waves in loop structures is the energy transfer to smaller scales, where it can be turned into internal energy of the plasma. The main mechanisms considered responsible for wave damping are resonant absorption for the case of standing modes \citep{Ionson1978ApJ,sakurai1991, goossens1992resonant, goossens2002coronal, ruderman2002damping, arregui2005resonantly, goossens2011resonant,yudaejung2017ApJ} and its analogous mechanism of mode coupling \citep{pascoe2010, demoortel2016} for propagating waves. Both mechanisms use a resonance to transfer the energy of the global mode to local azimuthal Alfv\'{e}n modes at the resonant layer, reducing the amplitude of the transverse oscillations. In the presence of a varying Alfvén speed profile transverse to the propagation direction, smaller scales are further created through phase mixing \citep{heyvaerts1983,soler2015}. Once the smaller scales have developed, dissipation mechanisms such as resistivity or viscosity can lead to heating \citep{ofman1998, pagano2017}. A disadvantage of this approach, however, is the spatial confinement of the resonant layer. \citet{cargill2016ApJ} showed that, unless broadband drivers \citep{ofman1998} or additional
heating mechanisms are taken into account, this localized heating would not be capable of sustaining a fixed density gradient between the loop and the environment once radiative cooling was considered. 

Another way to spread the resonant layer across a flux tube cross section, in the case of standing waves, is the development of the Kelvin-Helmholtz instability (KHI) from strong shear velocities generated by the azimuthal Alfv\'{e}n waves \citep{heyvaerts1983,zaqarashvili2015ApJ}. The KHI creates a turbulent layer at the loop edges, where resonant absorption and phase mixing can effectively transfer energy to smaller scales. Three-dimensional simulations of straight flux tubes confirmed the non-linear connection between resonant absorption, phase mixing, and KHI for driver generated azimuthal Alfv\'{e}n waves  \citep{uchimoto1991,ofman1994nonlinear,poedts1997a,poedts1997b}. More recent numerical studies \citep{terradas2008,magyar2016damping,antolin2017,howson2017twisted,terradas2018,antolin2018ApJ...856...44A} have confirmed the development of transverse wave induced Kelvin-Helmholtz (TWIKH) rolls for standing kink waves in flux tubes in different environments, which lead to mixing between the loop cross section and the surrounding plasma. 

Studies of continuous footpoint driven standing waves in flux tubes \citep{karampelas2017,karampelas2018fd}, inspired by the recently observed, decayless low-amplitude kink oscillations in coronal loops \citep{nistico2013,anfinogentov2015,nakariakov2016}, have focussed on the effects of KHI on coronal loop heating. The constant input of energy in these simulations causes the developed TWIKH rolls to expand across the loop cross section, leading its initial monolithic density profile into a turbulent state and fully deforming this profile in the process. This deformation spreads the resonant layer, where energy dissipation takes place across the loop in the presence of resistivity and viscosity. Its imprint on the magnetic field near the footpoint also causes the resitive heating rate there to spread gradually inside the loop, leading to energy dissipation and temperature increase. However, this heating is easily masked by the mixing between plasma of different temperatures.

In the current study, we expand upon our previous work, aiming to model low-amplitude, decayless kink waves in active region coronal loops that are driven by footpoint motions. We incorporate gravity into our models to study its effects on the loop dynamics alongside the potential effects on wave heating. Physical resistivity and shear viscosity have been introduced alongside gravity, allowing us to study their effects on the development of the TWIKH rolls for driven oscillations \citep[for impulsively oscillating loops without gravity]{howson2017} and on the wave heating process. Finally the energy evolution of different models is considered, giving us insight into the underlying mechanics of wave heating. 

\section{Numerical model}
\subsection{Equilibrium}
\begin{figure*}[t]
\centering
\resizebox{\hsize}{!}{\includegraphics[trim={0cm 0.4cm 0.4cm 0.4cm},clip,scale=0.3]{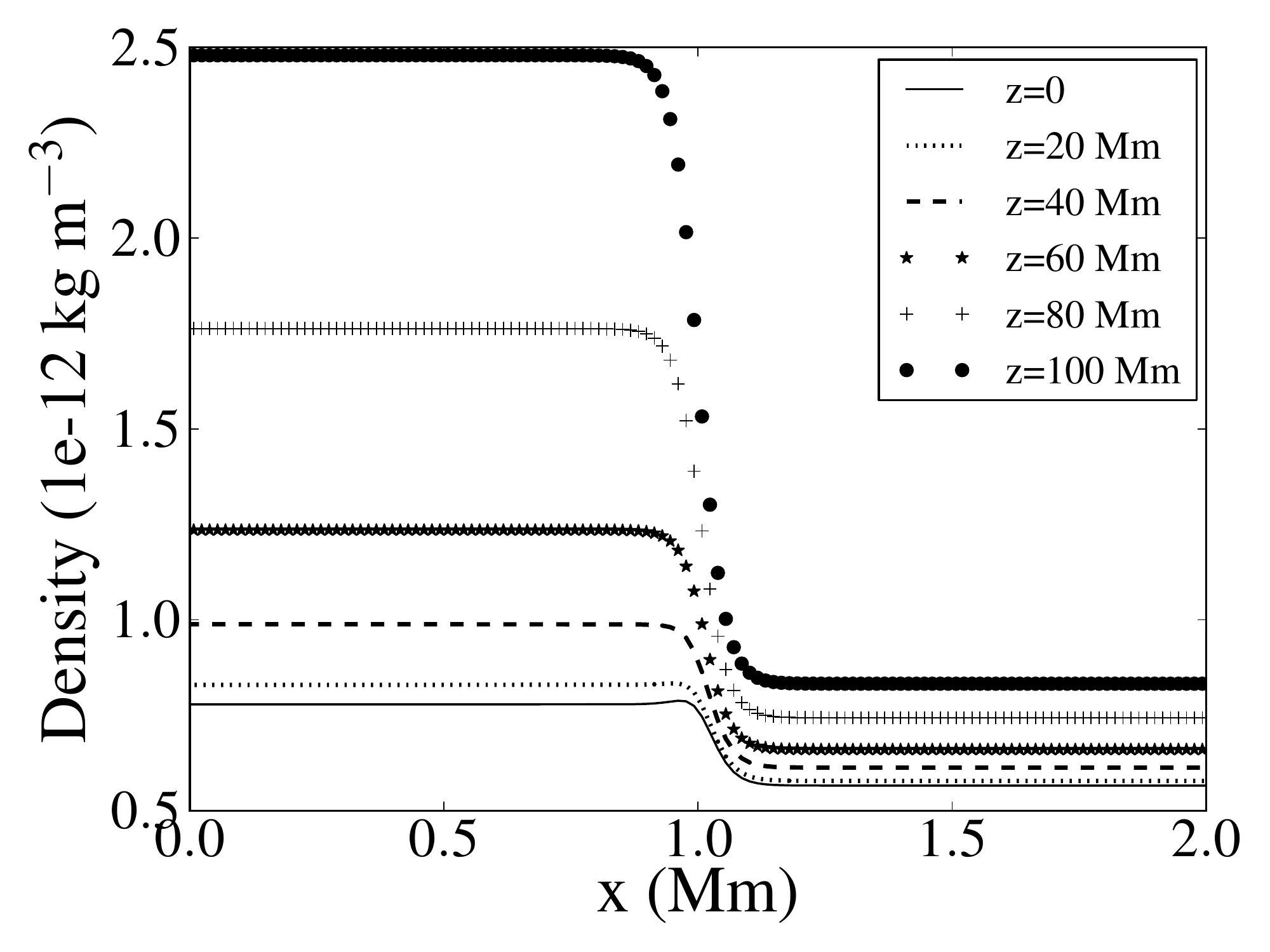}
\includegraphics[trim={0cm 0.4cm 0.4cm 0.4cm},clip,scale=0.3]{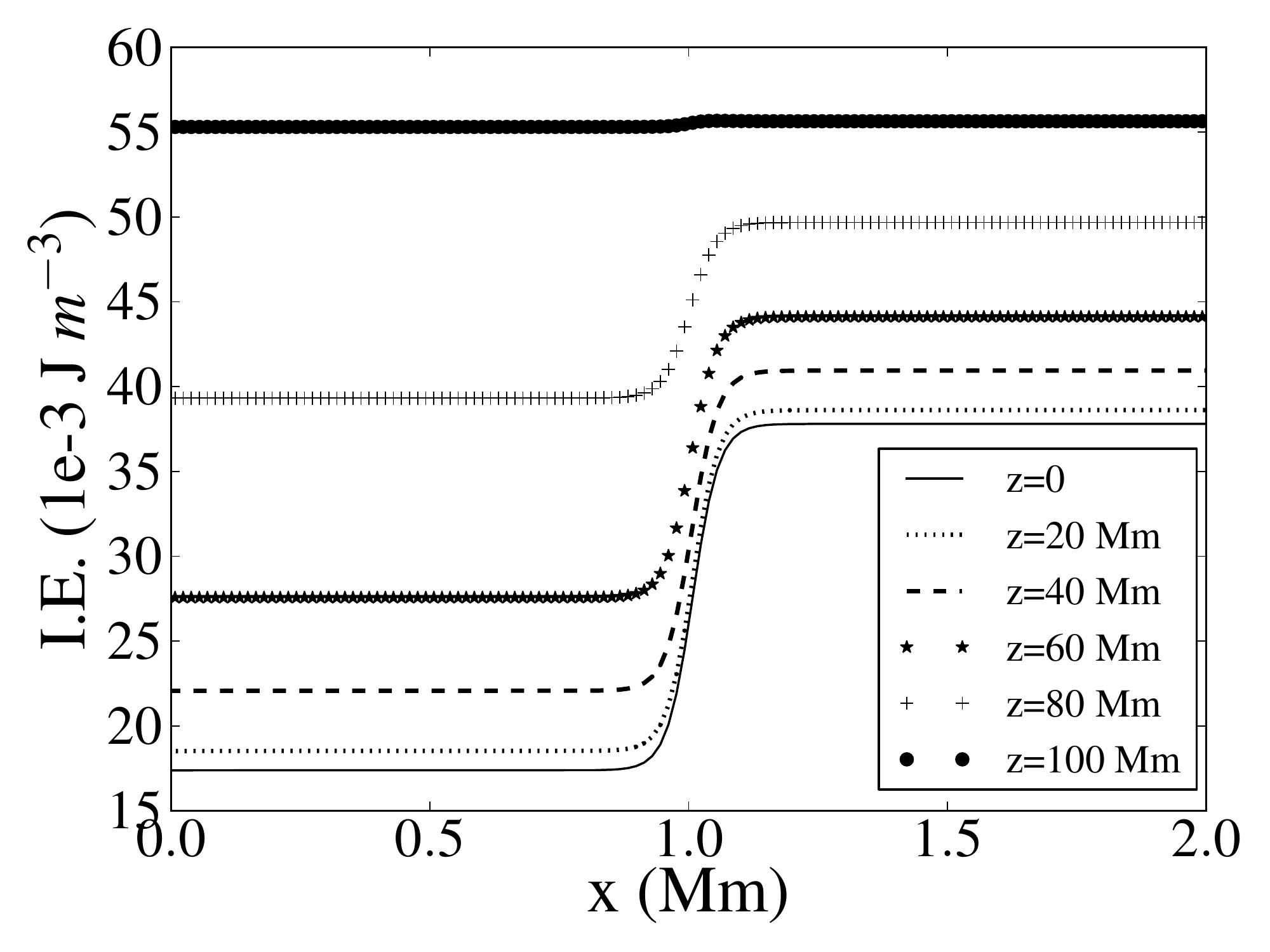}
\includegraphics[trim={0.4cm 0.4cm 0.4cm 0.4cm},clip,scale=0.3]{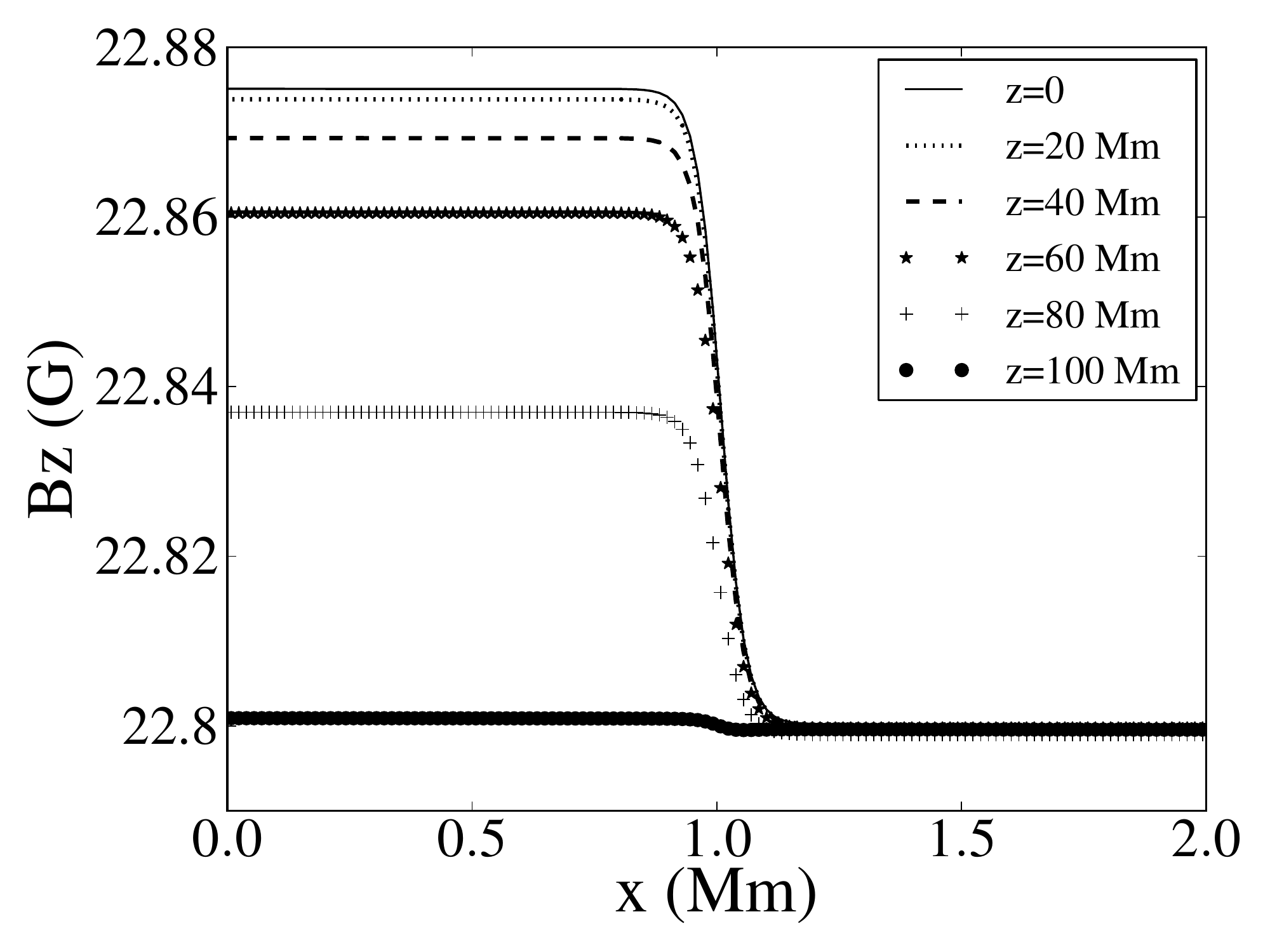}}
\caption{Radial profile of density (left), internal energy (middle), and $B_z$ magnetic field (right) for the gravitationally stratified models at different heights after the relaxation period. The profiles are considered before initiating the driver. The apex is located at $z=0$ and the footpoint at $z=100$ Mm. $x=0$ is the centre of the loop at $t=0$.}\label{fig:setup}
\end{figure*}

For our $3D$ simulations, we use straight, density-enhanced magnetic flux tubes in a low-$\beta$ coronal environment, similar to that in \citet{karampelas2017}. We mainly focus on gravitationally stratified, active region coronal loops in ideal, resistive, and viscous MHD, while also including two models of non-stratified loops in ideal MHD used as reference. Each loop has a full length (L) of $200$ Mm and an initial radius (R) of $1$ Mm, which is constant with height. In the following analysis, we denote the basic values of our physical parameters with the index i (e) for internal (external) values, with respect to our tube. 

The radial density profile, for all models is given by the relation
\begin{equation}
\rho(x,y) = \rho_e  + (\rho_i - \rho_e)\zeta(x,y), 
\end{equation}
\begin{equation}
\zeta(x,y) = \dfrac{1}{2}(1-\tanh((\sqrt{x^2+y^2}/R-1)\,b)),
\end{equation}
where $\rho_e$ is the external density and $\rho_i$ the internal or loop density. For the gravitationally stratified loops, we define the internal and external density at the footpoint as $\rho_i$ and $\rho_e$, with $\rho_e = 10^9 \mu \, m_p$ cm$^{-3} = 0.836 \times 10^{-12}$ kg m$^{-3}$ ($\mu = 0.5$ and $m_p$ is the proton mass). For the non-stratified models, the density radial profile is constant with height. By $x$ and $y$ we denote the coordinates in the plane perpendicular to the loop axis, $z$  along its axis and $b$ sets the width of the boundary layer. We consider $b=20$, which gives us an inhomogeneous layer of width $\ell \approx 0.3 R$. We choose a density ratio of $\rho_i/\rho_e = 3$ for all models at the footpoint, within the range of estimated ratios \citep{aschwanden2003VD}, which is suitable for fast transfer of energy from transverse to azimuthal motions, through resonant absorption. For all models studied, we set the temperature to be constant with height. Radial dependence of temperature depends on the different cases considered. Finally, we consider an initial uniform magnetic field and parallel to the flux tube axis, along the $z$-axis ($B_z=22.8$ G).

In the cases where gravity is included, it varies sinusoidally along the flux tube, taking a zero value at the loop apex ($z=0$) and maximum absolute value at the footpoints ($z=\pm 100$ Mm). We use this variation per height to model the effects of the curvature along the loop axis, while retaining a straight flux tube. Thus, we have stratification of pressure and temperature along the loop according to the hydrostatic equilibrium,\begin{equation}
\dfrac{\partial p_{i,e}}{\partial z}=-g\, \rho_{i,e}\, \sin(\dfrac{\pi z}{L})
.\end{equation}
As a consequence of gravitational stratification, there is a pressure imbalance at the loop boundary, leading to a jump in total pressure. This imbalance is countered by the restructuring of a stratified magnetic field inside the loop, which causes a weak standing oscillation, with velocities of only a small fraction of the amplitude of our driver. By letting our system relax for a period, it reaches a quasi-equilibrium state (Fig. \ref{fig:setup}) and the aforementioned perturbation does not affect the global dynamics of our system. After the relaxation, the magnetic field shows a slight increase towards the apex inside the loop due to the chosen density and temperature configuration. During the relaxation of the systems, temperature, pressure, and density do not deviate significantly from their initial state.

\begin{figure}
\centering
\resizebox{\hsize}{!}{\includegraphics[trim={0cm 9cm 12cm 5cm},clip,scale=0.16]{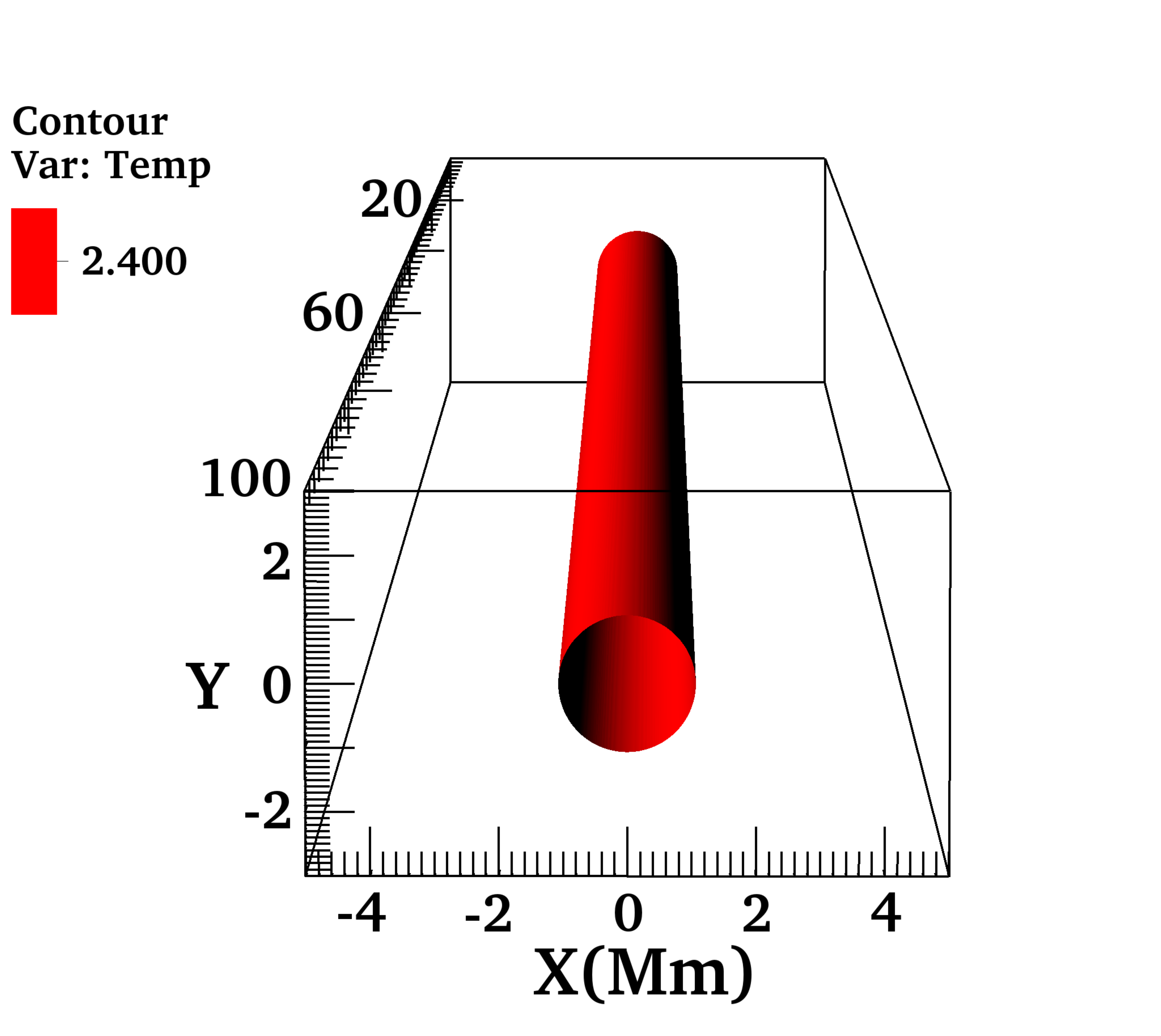}
\includegraphics[trim={18.5cm 9cm 12cm 5cm},clip,scale=0.16]{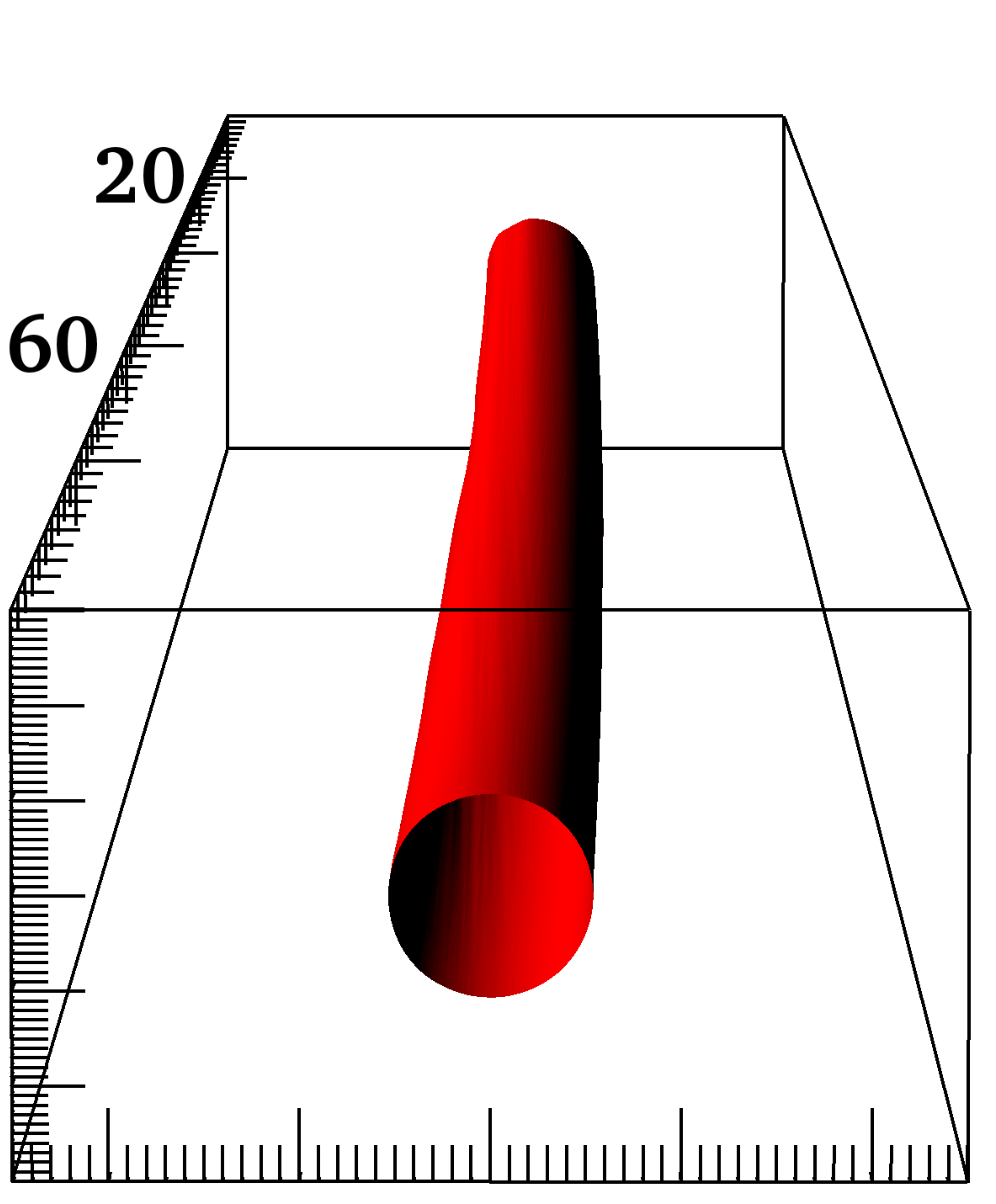}}
\resizebox{\hsize}{!}{\includegraphics[trim={0cm 0cm 12cm 5cm},clip,scale=0.16]{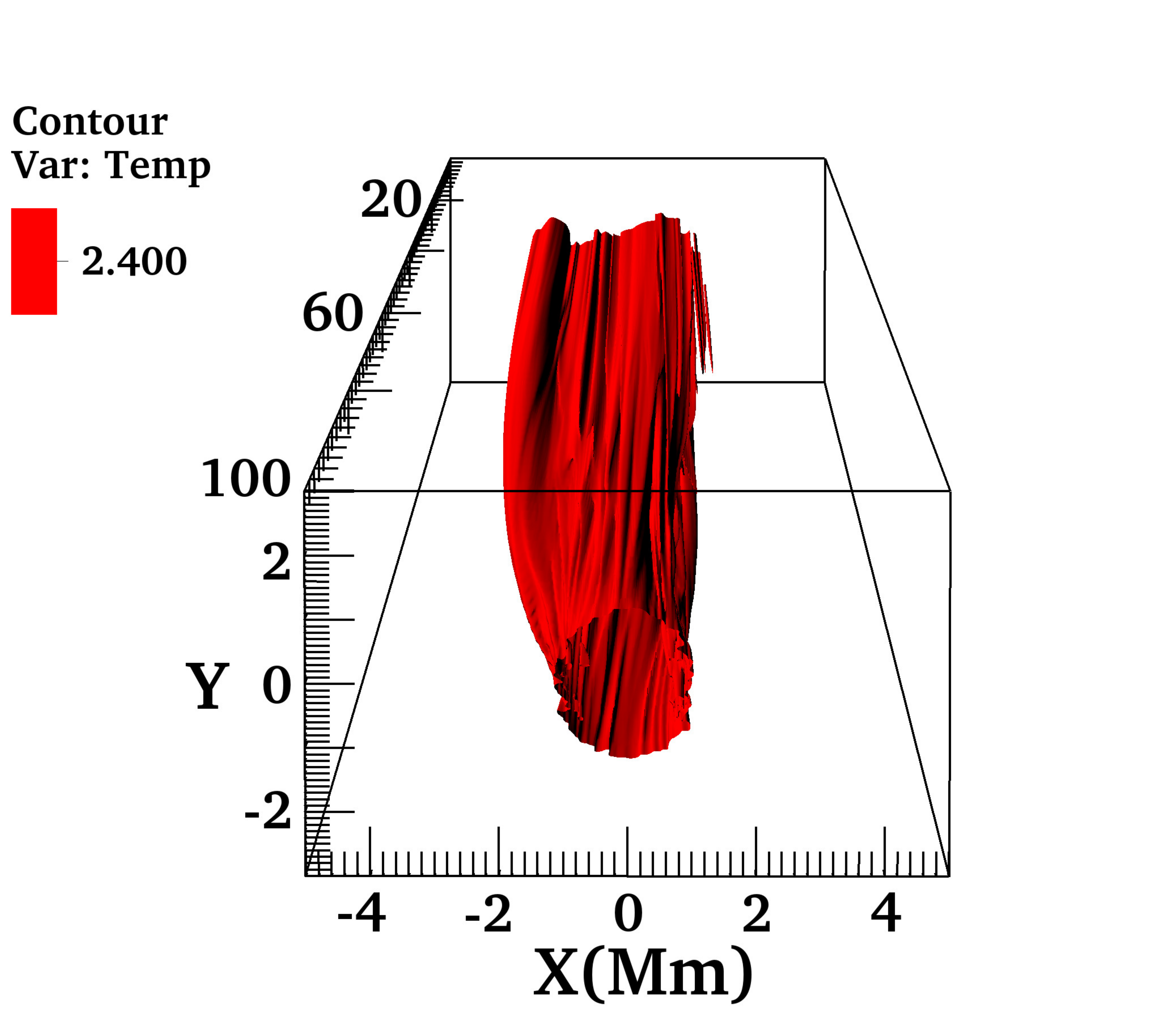}
\includegraphics[trim={18.5cm 0cm 12cm 5cm},clip,scale=0.16]{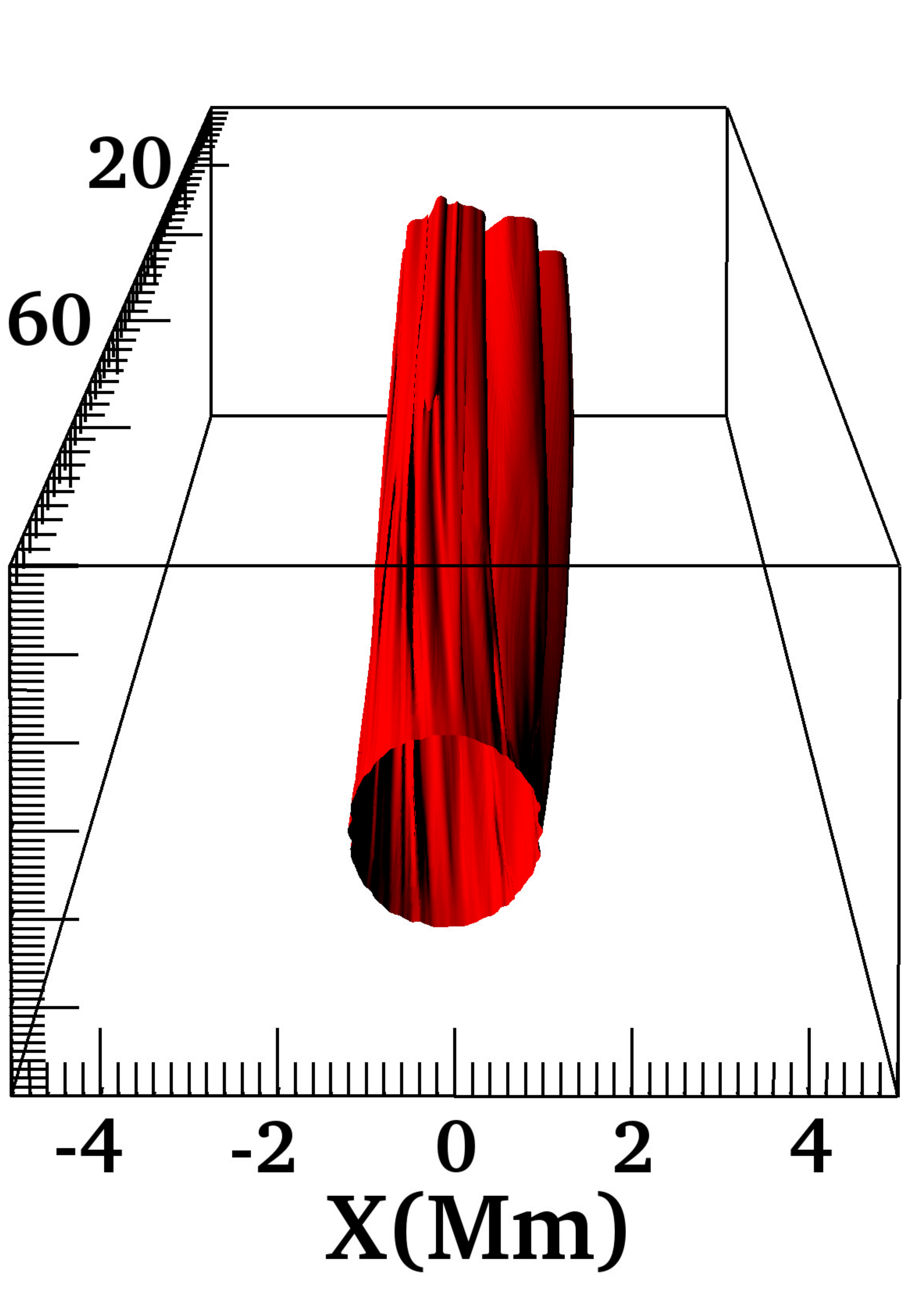}}
\caption{Three-dimensional temperature contour plot, measured in $10^6$ K, for a gravitationally stratified cold loop in a warm corona (model ColdI). Moving clockwise from the top left: t = $0$, $2.5\,P$, $4.75\,P,$ and $10\,P$, where $P=171$ s is the period of the driver. An animation of these figures, showing the oscillation for the model in ideal MHD, is available on-line (Movie 1).} \label{fig:kink}
\end{figure}

The different cases considered in the current work are as follows:
\begin{enumerate}
\item A model of a loop in hydrostatic equilibrium between itself and the background plasma. No gravity is included. We consider a uniform temperature $T_i = T_e =10^6$ K, and use ideal MHD with an estimated magnetic Reynolds number $R_m =10^6$ and estimated Reynolds number $R_e =10^6$. This model is called "UniT". Total pressure is kept constant along and across the flux tube by changing the magnetic field along the radial direction from $B_{zi}=22.8$ G for the internal to $B_{ze}=22.95$ G for the external magnetic field.  
\item A loop model without gravity ("ColdIngr") in hydrostatic equilibrium between itself and the background plasma. We consider a temperature ratio of $T_i /T_e =1/3$ with $T_i=9\times 10^5$ K, and use ideal MHD ($R_m =10^6$ and $R_e =10^6$).
\item A gravitationally stratified loop ("ColdI") in hydrostatic equilibrium between itself and the background plasma. We consider temperature ratio of $T_i /T_e =1/3$ with $T_i=9\times 10^5$ K, and use ideal MHD ($R_m =10^6$ and $R_e =10^6$).
\item Model "ColdR". Same as model ColdI but for resistive MHD ($R_m =10^4$ and $R_e =10^6$).
\item Model "ColdV". Same as model ColdI but for viscous MHD ($R_m =10^6$ and  $R_e =10^4$), where shear viscosity is considered.
\item Model "ColdV2". Same as model ColdV but for a lower Reynolds number ($R_m =10^6$ and  $R_e =10^2$).
\end{enumerate}
All of the gravitationally stratified models start from the same initial state (after the relaxation) shown in Fig \ref{fig:setup} before applying the driver and physical dissipation.

A detailed overview of the physical parameters for each model a presented in Table \ref{tab:paramb}. The different temperature profiles are useful in identifying and studying the underlying heating mechanisms in the solar corona. Model UniT is a very similar model to the \textit{Driven-equalT} model from \citet{karampelas2017}, but for a stronger driver. We simulated this system to see the effects of numerical diffusion on energy dissipation for our current code and resolution in the same way as in our previous study. Models ColdI, ColdR, ColdV, and ColdV2 are the extensions of the \textit{Driven-diffT} model from \citet{karampelas2017}, when gravity and physical dissipation are introduced. ColdIngr is based on the same model, but for different values of density, which is useful for directly comparing with the gravitationally stratified models, as is demonstrated later. Through these models we want to study the effects of gravity, resistivity, and viscosity on cold flux tubes, for example like the loops in thermal non-equilibrium \citep{froment2015tne, froment2017tne} considered later during their cooling phase.

\begin{table*}
\begin{center}
\caption{Overview of the physical parameters for the different models in our simulations. The index i (e) denote internal (external) values, while the index f represent the footpoint values. Density is normalized by $\rho_u = 10^{-12}$ kg m$^{-3}$.}\label{tab:paramb}
\begin{tabular}{c c c c c c c c c c c c c}
\hline 
\hline
Model & Name & Gravity & Period (s) & $\upsilon_0$ (km s$^{-1}$) & $T_i/T_e$ & $T_i$ (K) & $\rho_{if} / \rho_u$ & $B_z$ (G) & $b$ & $\beta_f$ & $R_e$ & $R_m$ \\ 
\hline 
$1$ & UniT     & no  & $256$ & $4$ & $1$   & $10^6$         & $2.509$ & $22.8$ & $20$ & $0.02$  & $10^6$ & $10^6$ \\ 
$2$ & ColdIngr & no  & $171$ & $4$ & $1/3$ & $9\times 10^5$ & $1.129$ & $22.8$ & $20$ & $0.08$  & $10^6$ & $10^6$ \\
$3$ & ColdI    & yes & $171$ & $4$ & $1/3$ & $9\times 10^5$ & $2.509$ & $22.8$ & $20$ & $0.018$ & $10^6$ & $10^6$ \\ 
$4$ & ColdR    & yes & $171$ & $4$ & $1/3$ & $9\times 10^5$ & $2.509$ & $22.8$ & $20$ & $0.018$ & $10^6$ & $10^4$ \\
$5$ & ColdV    & yes & $171$ & $4$ & $1/3$ & $9\times 10^5$ & $2.509$ & $22.8$ & $20$ & $0.018$ & $10^4$ & $10^6$ \\
$6$ & ColdV2   & yes & $171$ & $4$ & $1/3$ & $9\times 10^5$ & $2.509$ & $22.8$ & $20$ & $0.018$ & $10^2$ & $10^6$ \\
\hline 
\end{tabular}
\end{center}
\end{table*}

\begin{figure}[t]
\centering
\resizebox{\hsize}{!}{\includegraphics[trim={0.3cm 0.3cm 0.3cm 0.33cm},clip,scale=0.3]{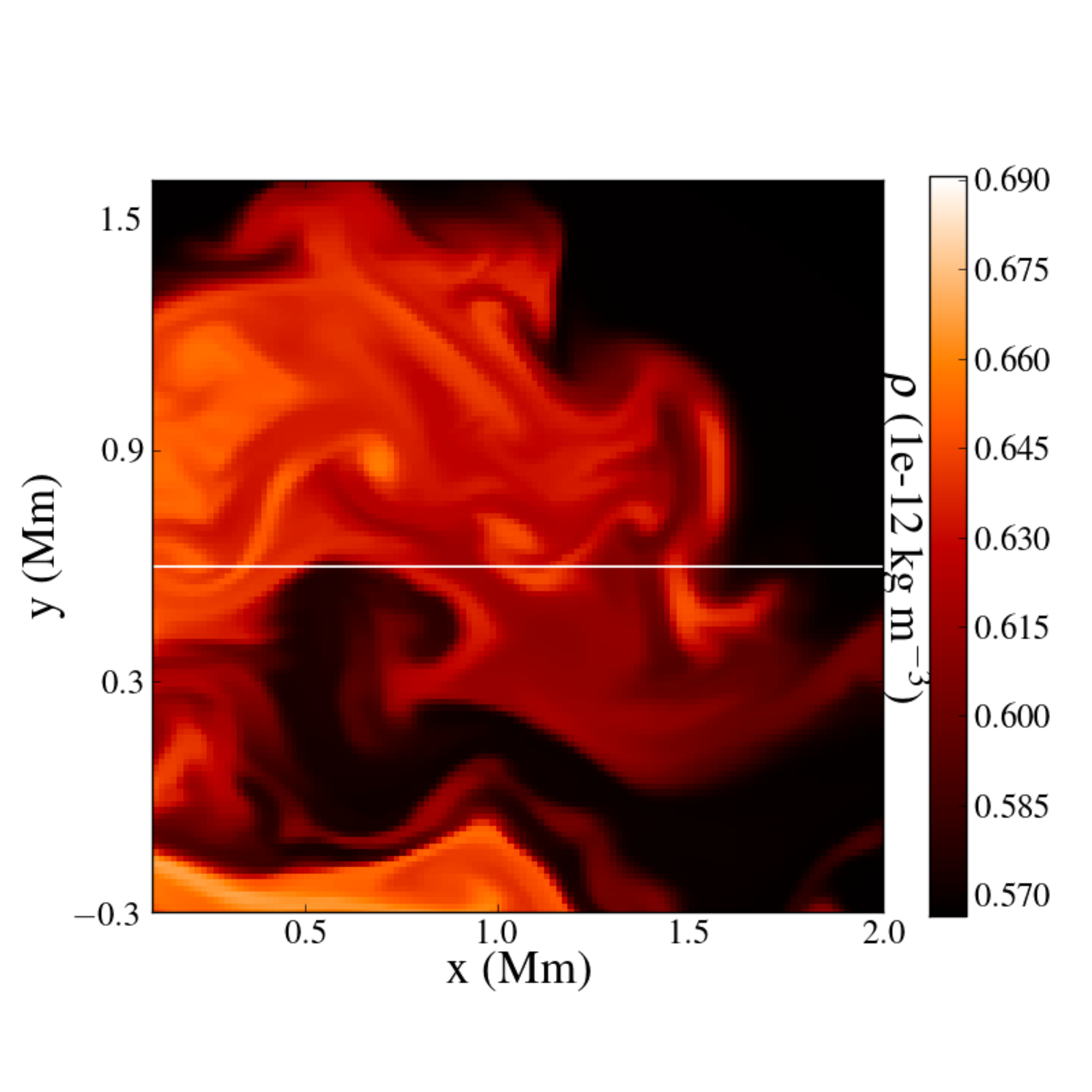}}
\resizebox{\hsize}{!}{\includegraphics[trim={0.3cm 0.3cm 0.3cm 0.33cm},clip,scale=0.4]{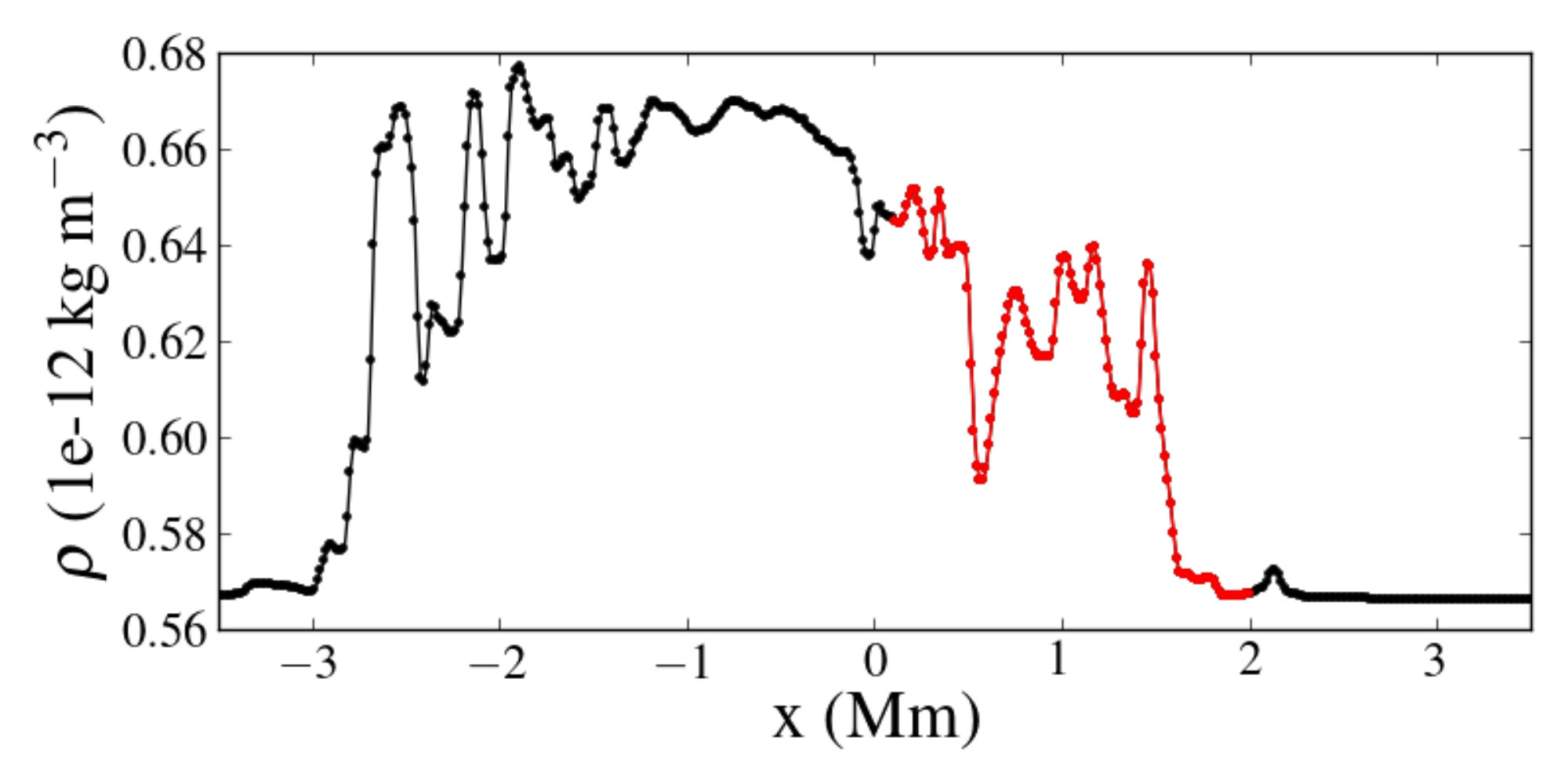}}
\caption{Top image: Part of the total density cross section for model ColdI, at the apex. We focus on the area with $-0.3\leq y$ (Mm) $\leq1.6$ and $0.1\leq x$ (Mm) $\leq2.0$ to highlight the resolution of smaller scale structures on the $x-y$ plane. Bottom image: The density structure at $y=0.6$ Mm, along the white line of the top image. The dots represent the grid points along the white line. The red line highlights the part visible in the image above. The plot shows time t = $10\,P$; $P=171$ s indicates the period of the driver.}\label{fig:smallscales}
\end{figure}

\begin{figure}
\centering
\includegraphics[trim={0.4cm 0cm 0.5cm 0.5cm},clip,scale=0.33]{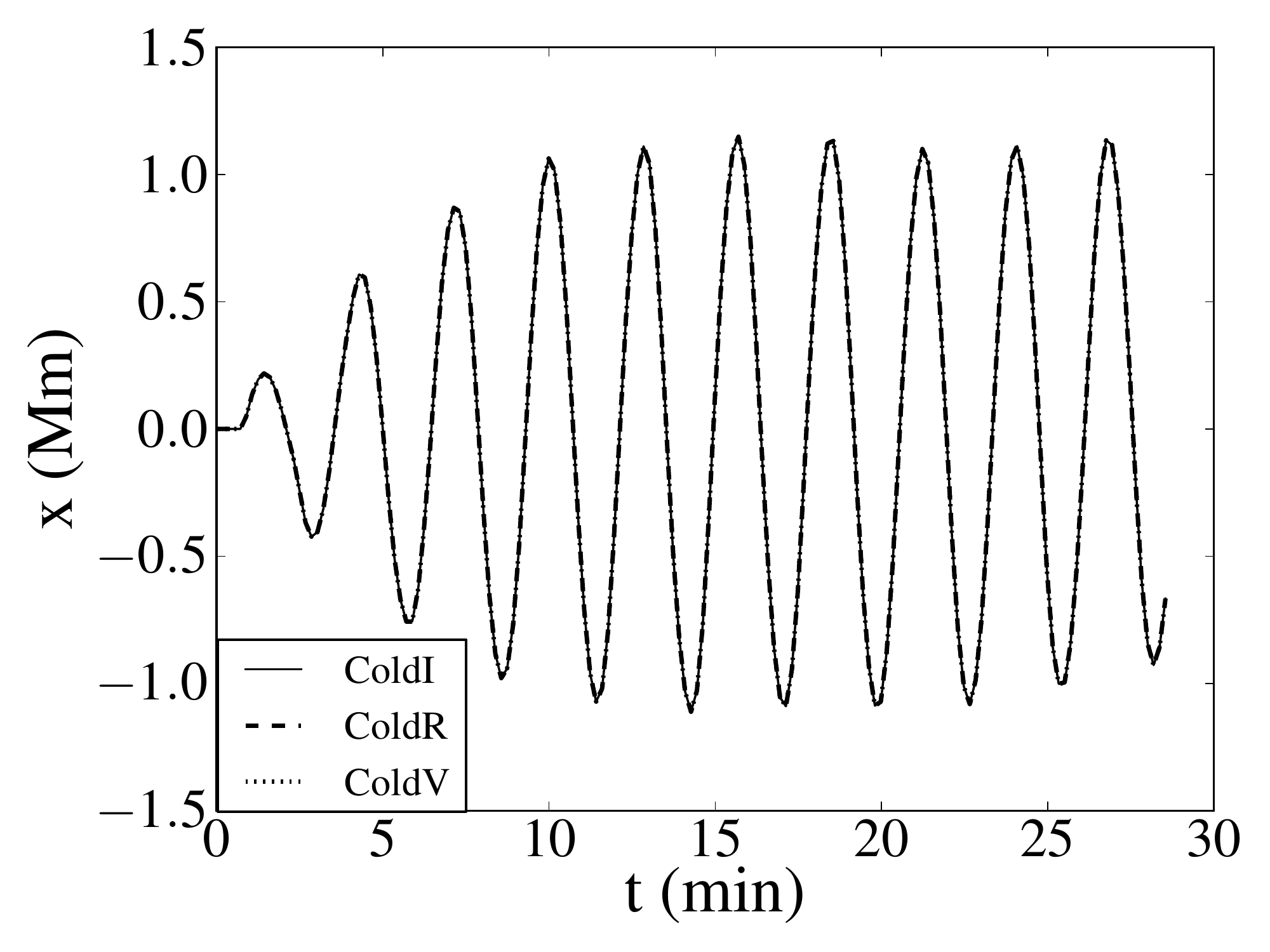}\\
\includegraphics[trim={0.4cm 0.5cm 0.5cm 0cm},clip,scale=0.33]{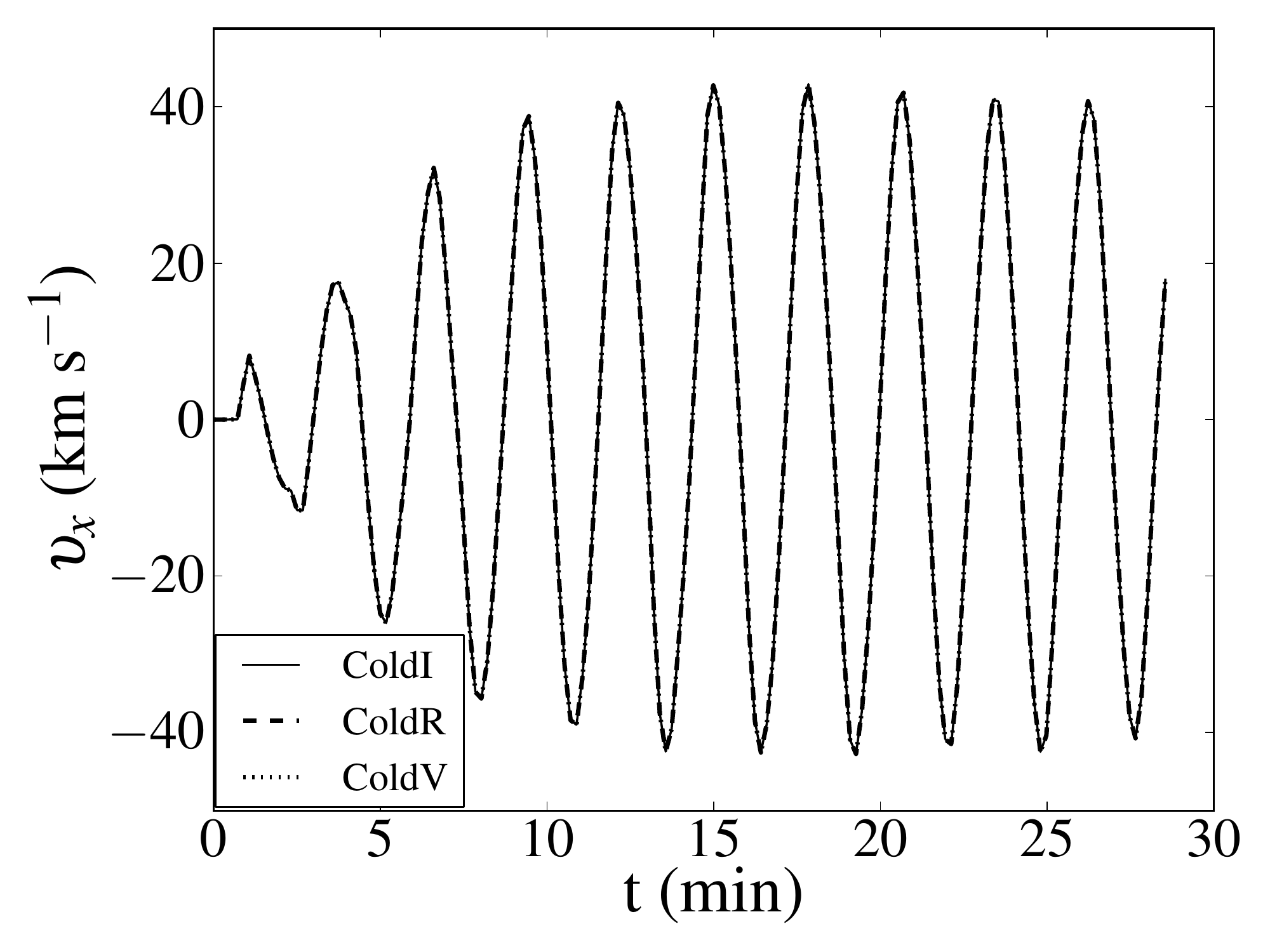}
\caption{Centre of mass displacement (top panel) and centre of mass $\upsilon_x$ velocity (bottom panel) at the apex for our different models.}\label{fig:cm}
\end{figure}

\subsection{Grid}
The 3D ideal MHD problem is solved using the PLUTO code \citep{mignonePLUTO2012, mignonePLUTO2018}, where the extended GLM method from \citet{dedner2002} is employed to keep the solenoidal constraint on the magnetic field. We use the finite volume piecewise parabolic method (PPM) with a second order spatial global accuracy, and the second order characteristic tracing method for calculating the timestep. For the resistivity and shear viscosity, an explicit method for recalculating the timestep is used.

The domain dimensions for models ColdIngr, ColdI, ColdR, ColdV, and ColdV2 are $(x,y,z) = (10,6,100)$ Mm. We use a uniform grid with a resolution of $640 \times 384 \times 64$ , which translates into cell dimensions of $15.625 \times 15.625 \times 1562.5$ km for all models. For the UniT model we use a domain of $(x,y,z) = (10,3,100)$ Mm, which have the same cell dimensions as in the rest of our models. The resolution is higher in the $x-y$ plane, to better resolve the small-scale structures that appear in the loop cross section, as we can see in Fig. \ref{fig:smallscales}. The resolution on the $z$-axis can sufficiently model the density stratification, since the lack of radiation or thermal conduction reduces the need for a finer grid along the tube axis. The footpoint of the loop is located at $z=100$ Mm and the apex at $z=0$. In all of our models, we have the inevitable numerical dissipation effects, which lead to an effective resistivity and viscosity many orders of magnitude larger than those expected in the solar corona. Through a parameter study of changing the values of physical resistivity and viscosity, we estimated the effective Reynolds and magnetic Reynolds number to be $R_e=10^6$ and $R_m=10^6$. These are the values for the ideal MHD cases in our simulations. 

\subsection{Driver}\label{driver}
Our tubes are driven from the footpoint ($z=100$ Mm), using a continuous, monoperiodic `dipole-like' driver \citep{karampelas2017}, inspired by that used by \citet{pascoe2010}. The period of the driver is $P\simeq 2L/c_k$, coinciding with the corresponding fundamental eigenfrequency for each model \citep{edwin1983wave, andries2005A&A430.1109A}. The values of the periods for each each model are listed in Table \ref{tab:paramb}.

The driver velocity is uniform inside the loop and time varying,
\begin{equation}
\lbrace \upsilon_x,\upsilon_y \rbrace=\lbrace \upsilon(t),0 \rbrace = \lbrace \upsilon_0 \cos(\dfrac{2\pi t}{P}),0 \rbrace ,
\end{equation}
where $\upsilon_0 = 4$ km s$^{-1}$ is the peak velocity amplitude, close to the observed photospheric motions. Outside the loop, the velocity follows the relation
\begin{equation}
\lbrace \upsilon_x,\upsilon_y \rbrace = \upsilon(t)R^2 \lbrace \frac{(x-\alpha(t))^2-y^2}{((x-\alpha(t))^2+y^2)^2},\frac{2(x-\alpha(t))y}{((x-\alpha(t))^2+y^2)^2} \rbrace,
\end{equation}
where $\alpha(t) = \upsilon_0 \, (0.5\,P/\pi)\, \sin(2\pi t/P)$ is a function that recentres the driver, following the movement of the footpoint. To avoid any numerical instabilities due to jumps in the velocity, a transition region following the density profile exists between the two areas. By moving the driver with the footpoint along the $x$ direction, we keep the base of the loop inside the central region of uniform velocity.

\begin{figure}
\centering
\includegraphics[trim={0.4cm 0.5cm 1cm 1cm},clip,scale=0.35]{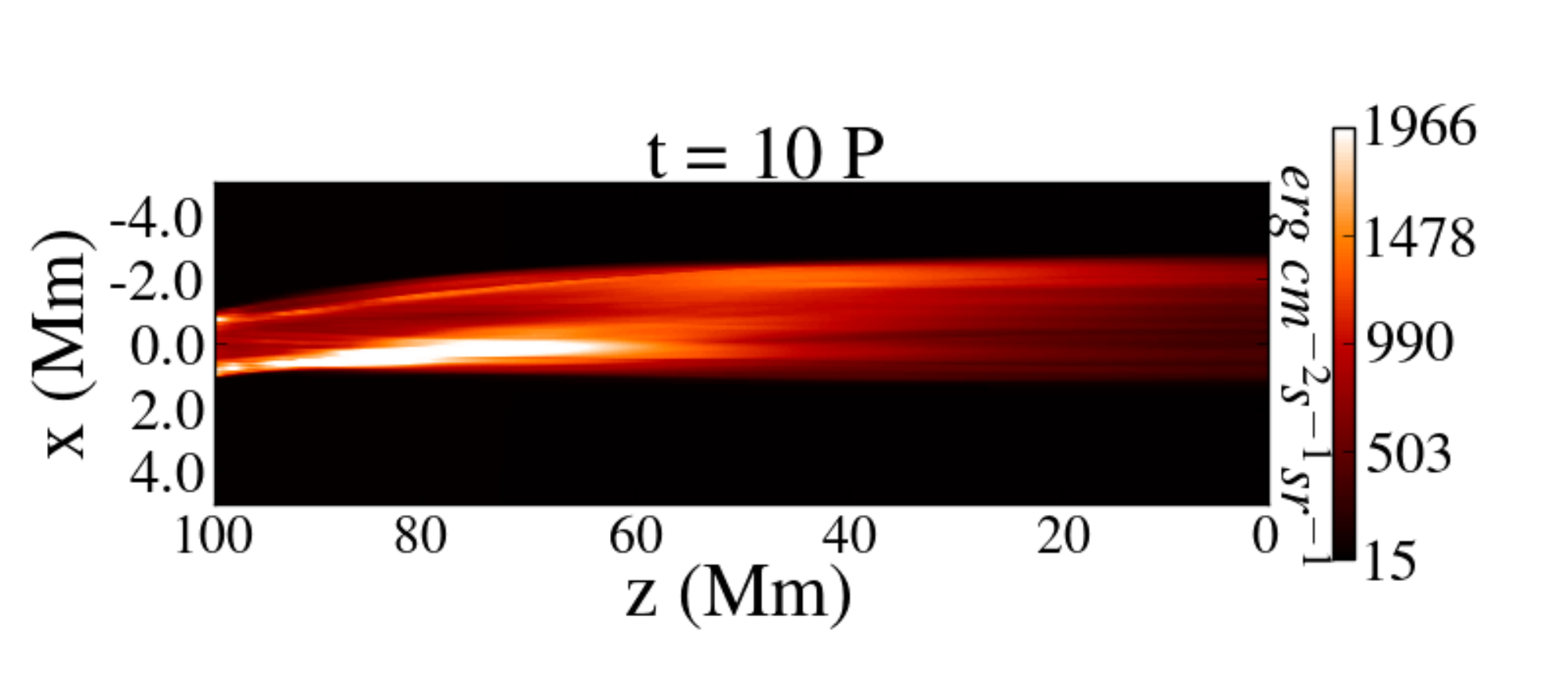}\\
\includegraphics[trim={0.4cm 0.5cm 1cm 1cm},clip,scale=0.35]{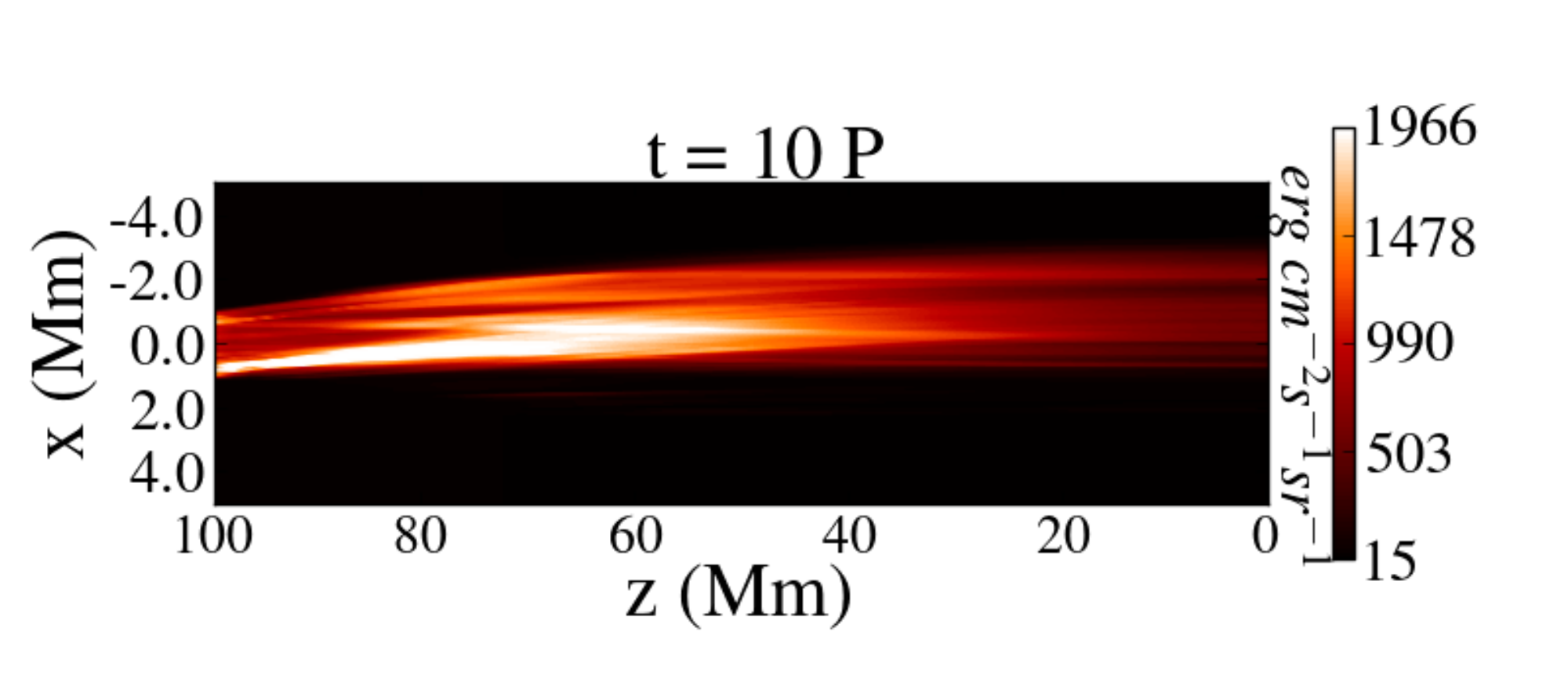}\\
\includegraphics[trim={0.4cm 0.5cm 1cm 1cm},clip,scale=0.35]{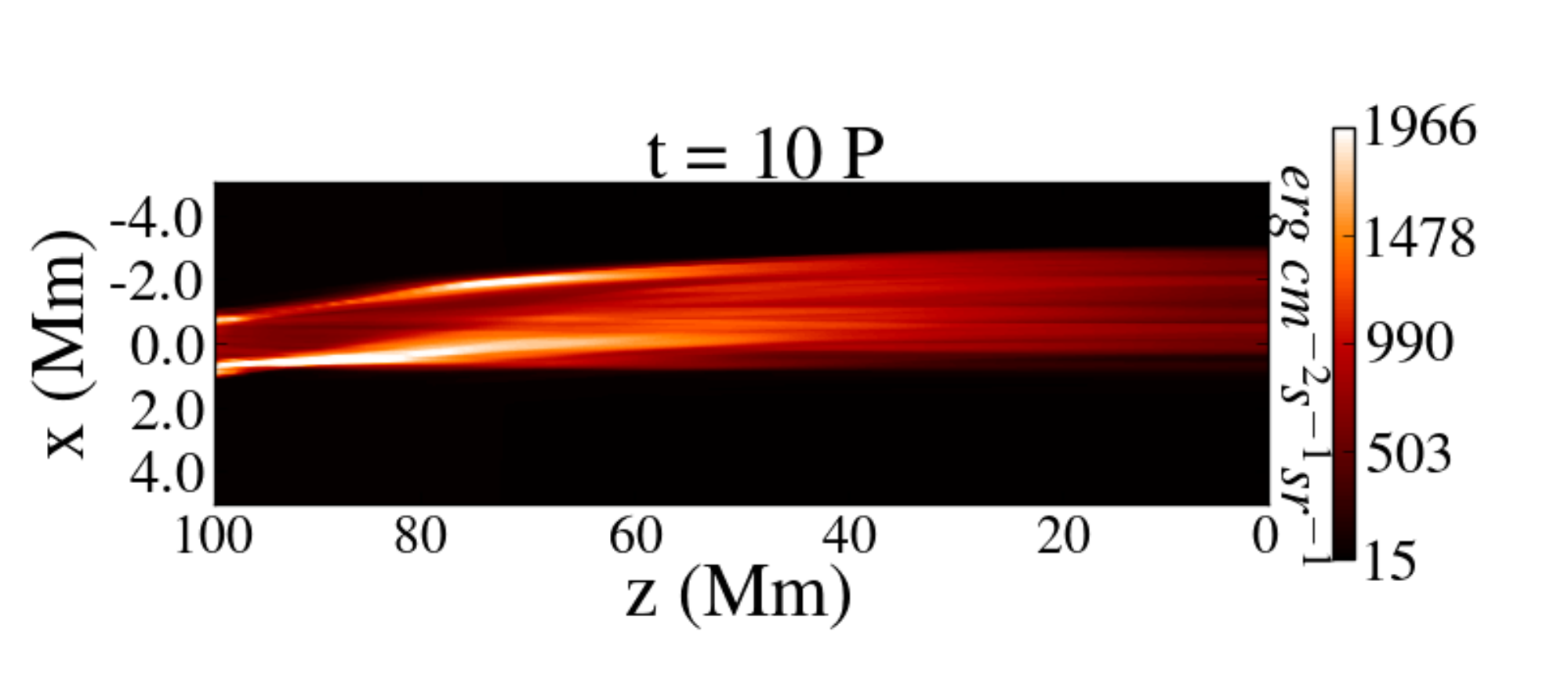}
\caption{Forward modelling images of the integrated emission intensity (in erg cm$^{-2}$s$^{-1}$sr$^{-1}$) of the cold tubes (the models ColdI, ColdR, and ColdV) for the $195.12 \, \AA$ line. The observer is at a $0^\circ$ LOS angle, perpendicular to the oscillatory motion. Half the loop length is modelled ($z=0-100$ Mm). From top to bottom the ideal case at t = $10\,P$, resisitive case at t = $10\,P$ , and viscous case at t = $10\,P$ are shown. The driver period is $P\simeq 171$ s. A movie with the forward modelling for model ColdI is available on-line (Movie 2).}\label{fig:fomo}
\end{figure}

\subsection{Boundary conditions}
We keep the velocity component parallel to the $z$-axis ($v_z$) antisymmetric at the bottom boundary ($z=100$ Mm) to prevent flows of mass through it. We also extrapolate the values for density and pressure, using the equations for hydrostatic equilibrium, while we use a zero normal gradient condition for the magnetic field,
\begin{equation}
B_i(z) = \dfrac{1}{11}\left( 2B_i(z-3) - 9B_i(z-2) +18B_i(z-1) \right) 
,\end{equation}
to extrapolate the values of each magnetic field component through the bottom boundary. Finally, the $v_x$ and $v_y$ velocities are defined by the driver. For the UniT and ColdIngr models, where no gravity is considered, we simply use Neumann-type, zero-gradient conditions for the density, pressure, and magnetic field. Studying the fundamental standing kink mode for an oscillating flux tube allows us to take advantage of the inherent symmetries of this mode, as well as the symmetric nature of our driver. In the top boundary ($z=0$), we kept $v_z$, $B_x$, and $B_y$ antisymmetric, in the $x-y$ plane at the apex, while all the other quantities are symmetric. Thus, only half the loop is simulated along the loop axis.

For the UniT model we also took into account the symmetric nature of the kink mode and our driver along the $y$-axis. The $v_y$ and $B_y$ are antisymmetric in the $x-z$ plane, while the other quantities are symmetric. Therefore, our computational time is reduced fourfold in total for this model, following our previous work \citep{karampelas2017,karampelas2018fd}. At the three lateral boundaries, we apply outflow (Neumann-type, zero-gradient condition) conditions, which allow waves to leave the domain.

For models ColdIngr, ColdI, ColdR, ColdV, and ColdV2, we do not employ the symmetry at the $x-y$ plane. This way, through the inevitable development of numerically induced asymmetries, we allow the loop to evolve in a non-symmetric environment, as we would generally expect in the solar corona. All the side boundaries in these models are set to outflow (Neumann-type, zero-gradient) conditions for all variables, which allow waves to leave the domain. 

To minimize their effect on the dynamics of our loops, we placed the $x$ side boundaries (along the direction of the oscillation) at a safe distance from the loop ($5$ R in $x$). On the $y$ direction (perpendicular to the oscillation), we placed the boundaries at $3$ R from the centre of the loop in order for them to not affect the development of our oscillations.

\subsection{Forward modelling}
We use the FoMo code \citep{fomo2016} to render spectroscopic images of our simulation data on the different models. We present snapshots of the emission intensities for different lines. In all images, we consider a line-of-sight plane perpendicular to the loop axis and we set the LOS angle perpendicular to the oscillation direction equal to $0^\circ$. By choosing to study the emission intensity for the Fe XII $195.12\, \AA$ line, we focus on the temperatures found predominately in the turbulent layer developing because of the KHI \citep{antolin2016,antolin2017}. 

\begin{figure*}
\centering
\resizebox{\hsize}{!}{\includegraphics[trim={1cm 0cm 0cm 0cm},clip,scale=0.16]{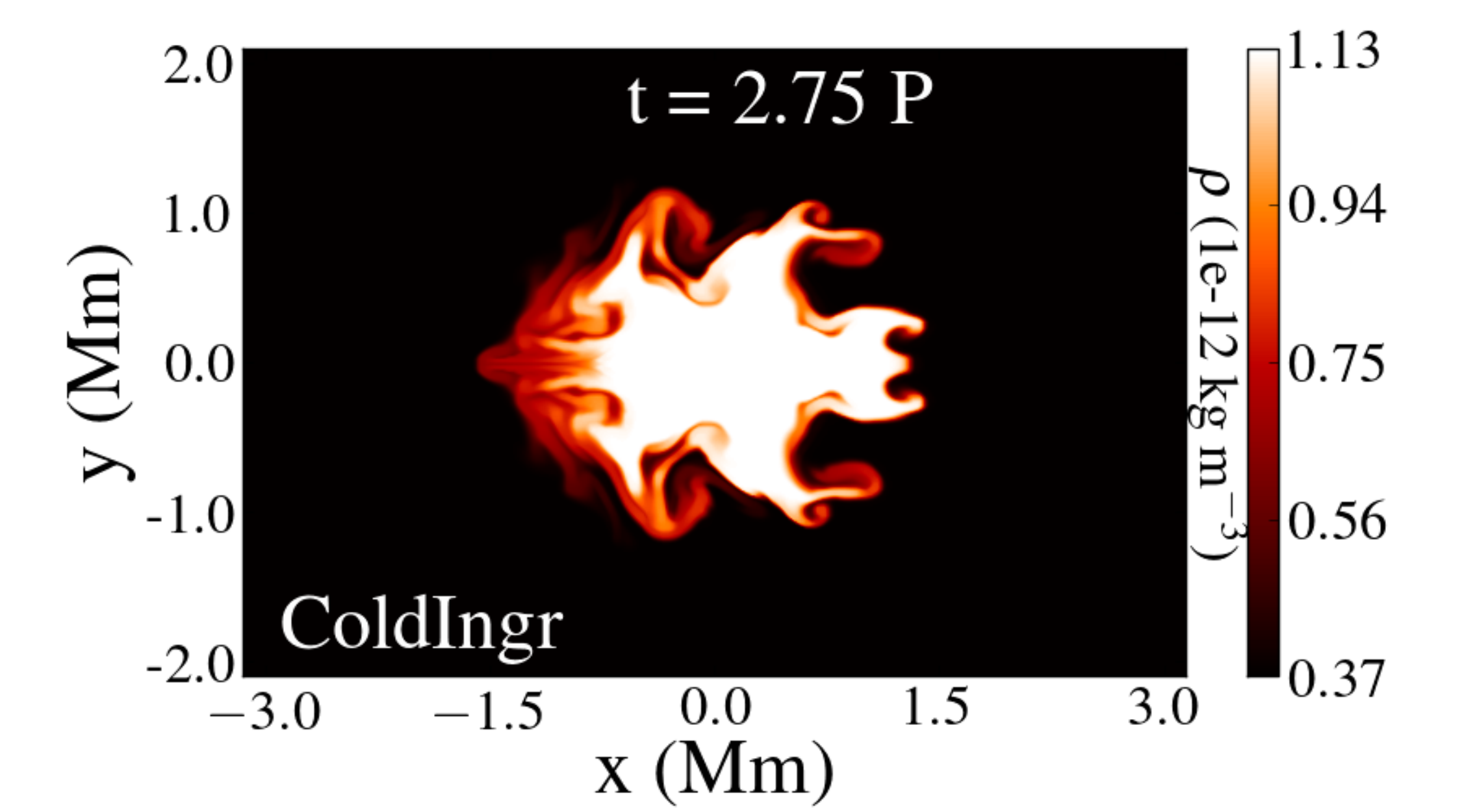}
\includegraphics[trim={1cm 0cm 0cm 0cm},clip,scale=0.16]{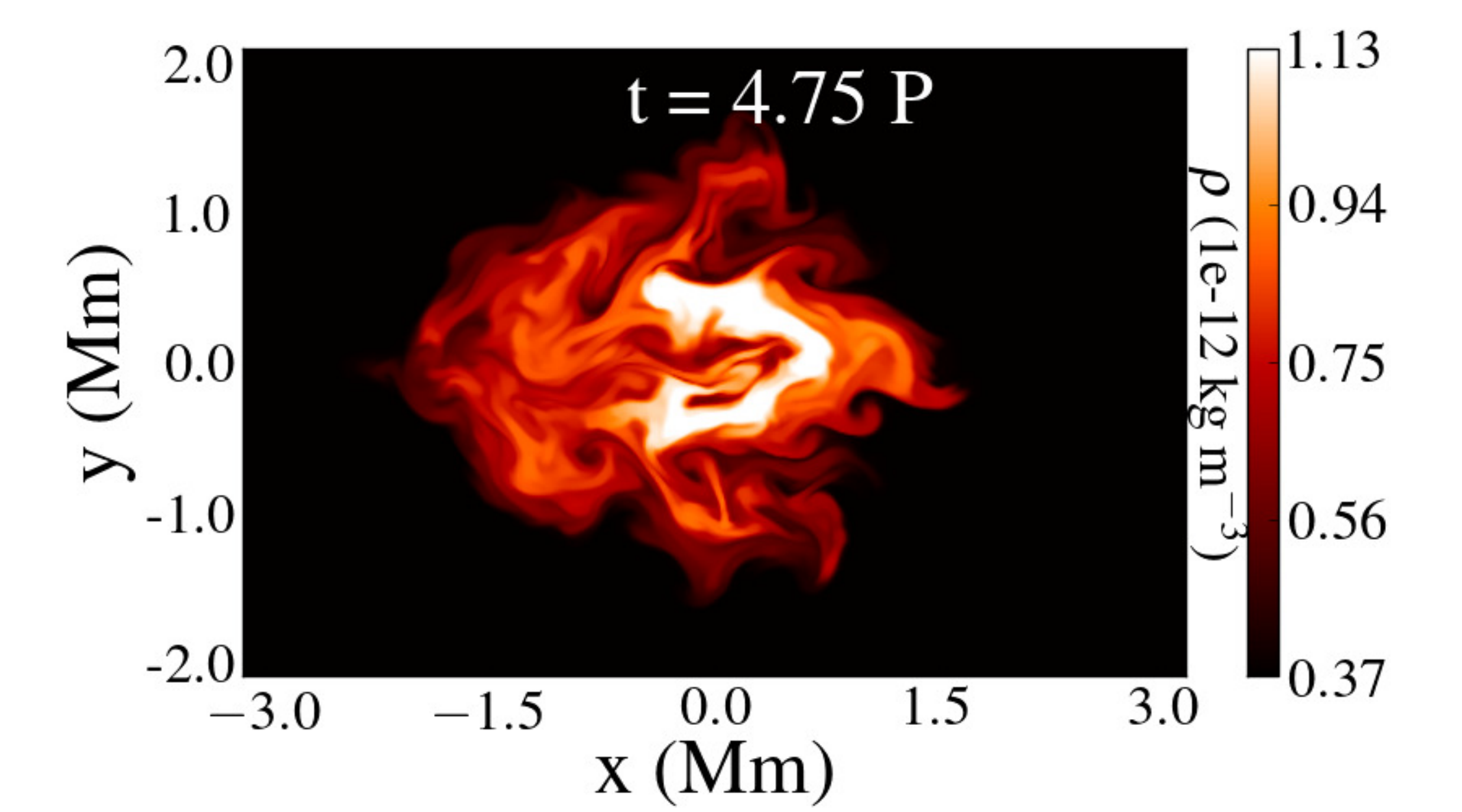}
\includegraphics[trim={1cm 0cm 0cm 0cm},clip,scale=0.16]{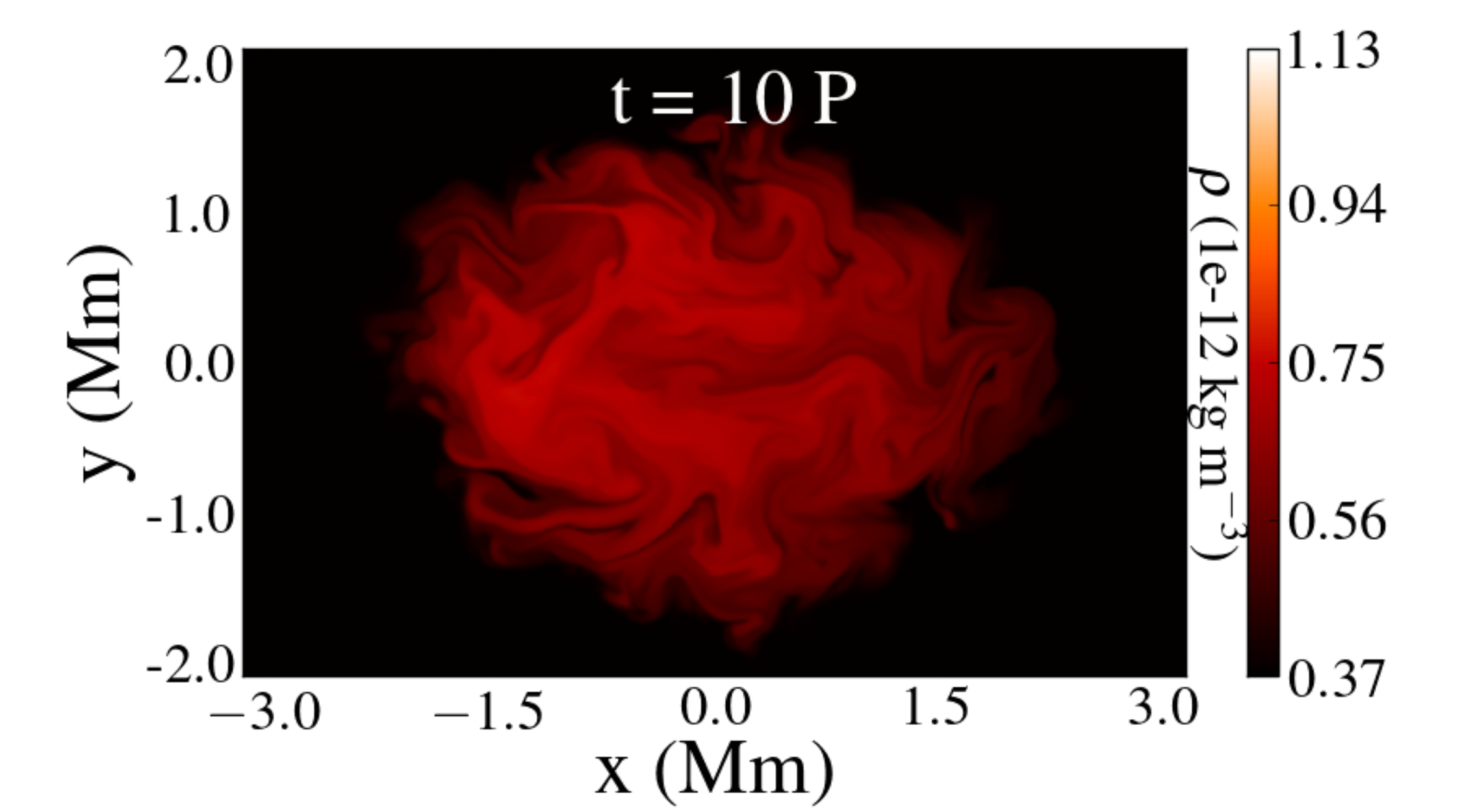}}
\resizebox{\hsize}{!}{\includegraphics[trim={1cm 0cm 0cm 0cm},clip,scale=0.16]{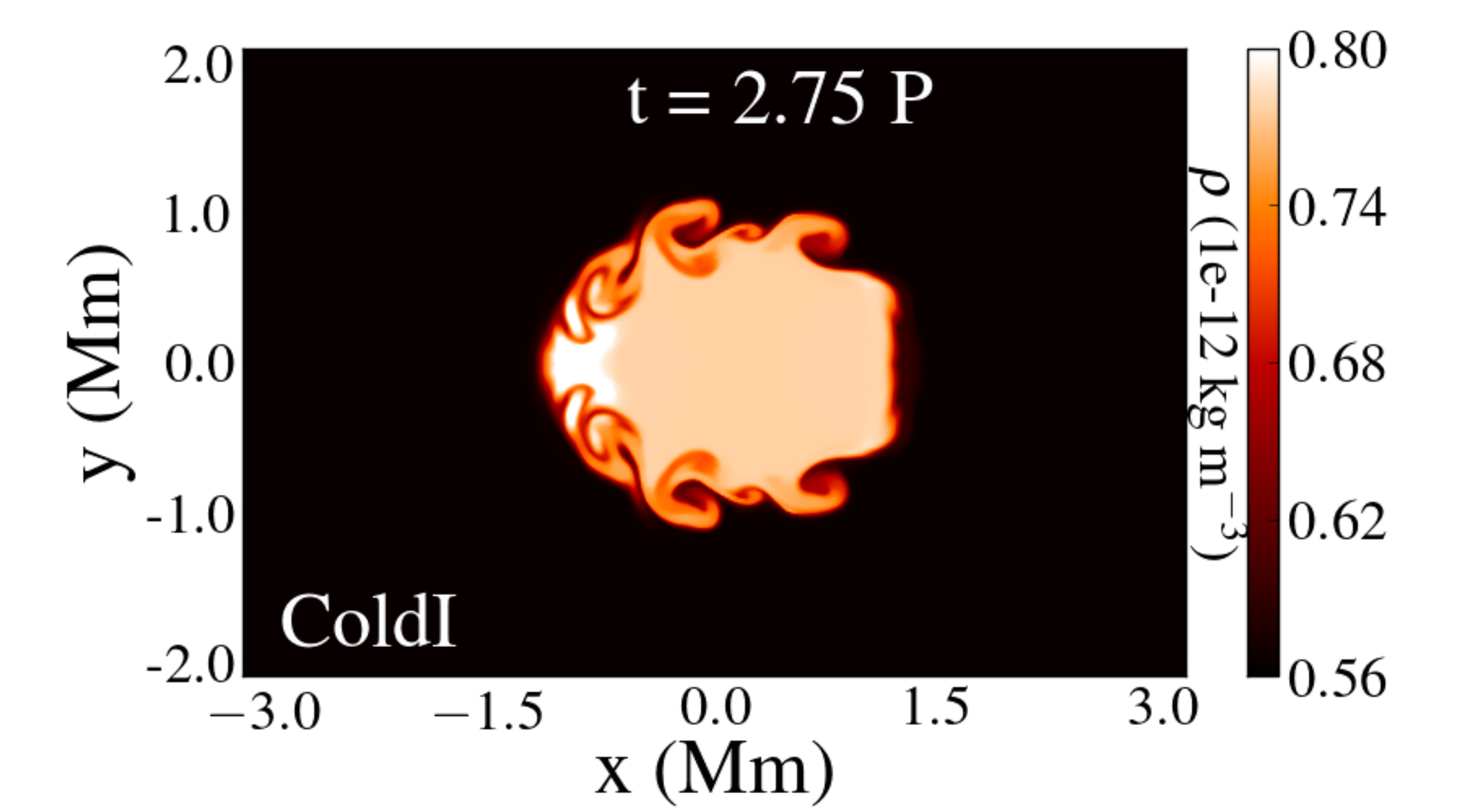}
\includegraphics[trim={1cm 0cm 0cm 0cm},clip,scale=0.16]{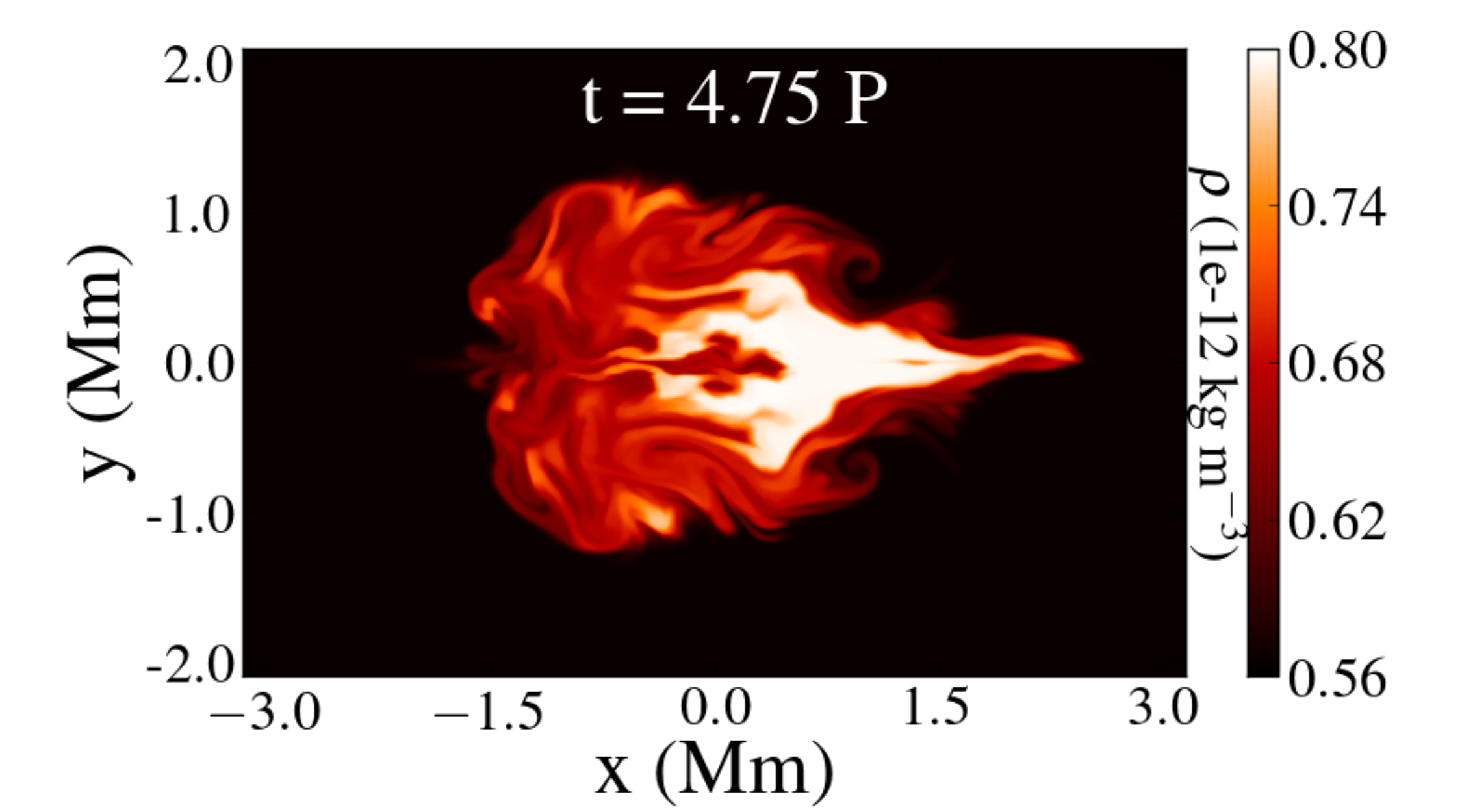}
\includegraphics[trim={1cm 0cm 0cm 0cm},clip,scale=0.16]{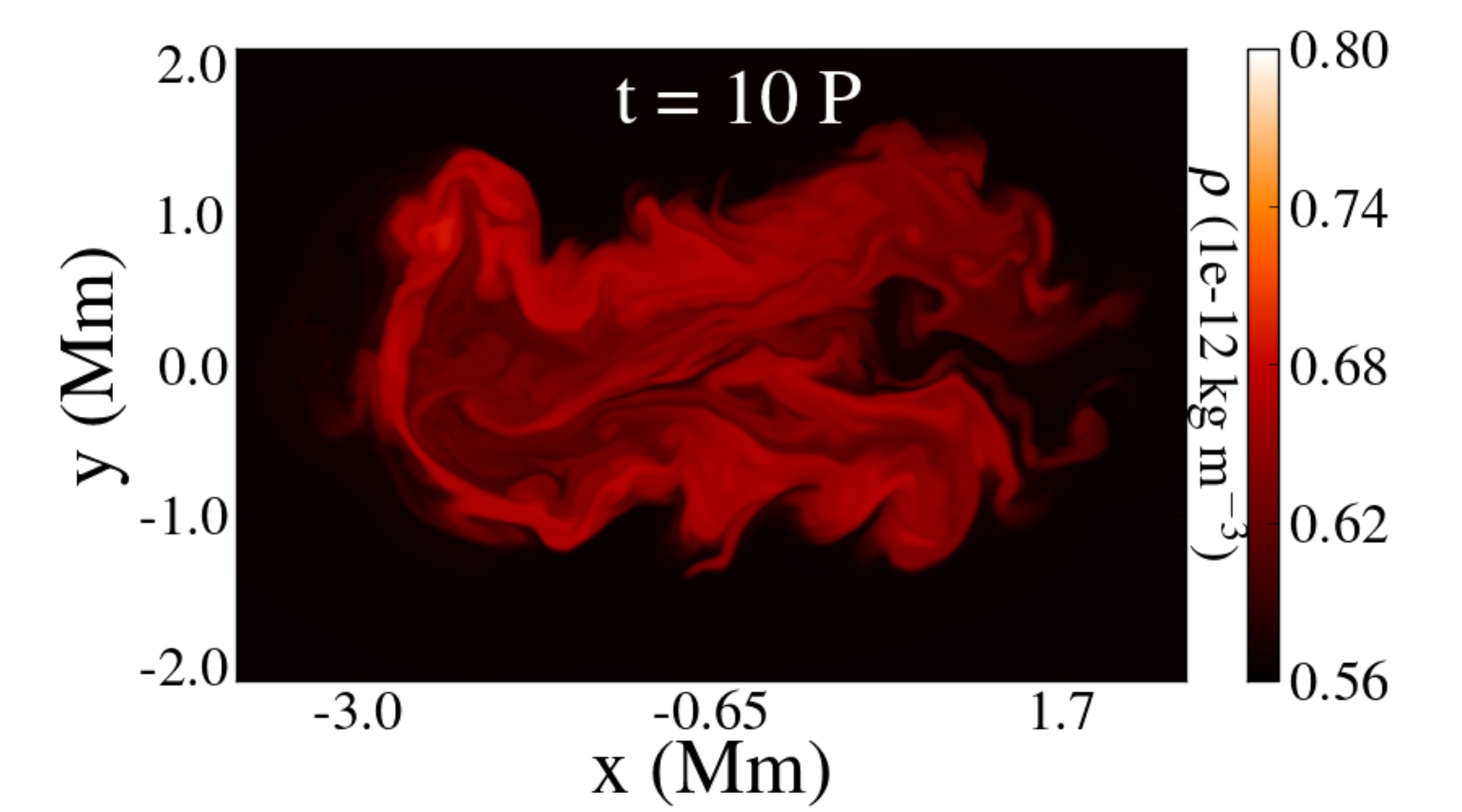}}
\resizebox{\hsize}{!}{\includegraphics[trim={1cm 0cm 0cm 0cm},clip,scale=0.16]{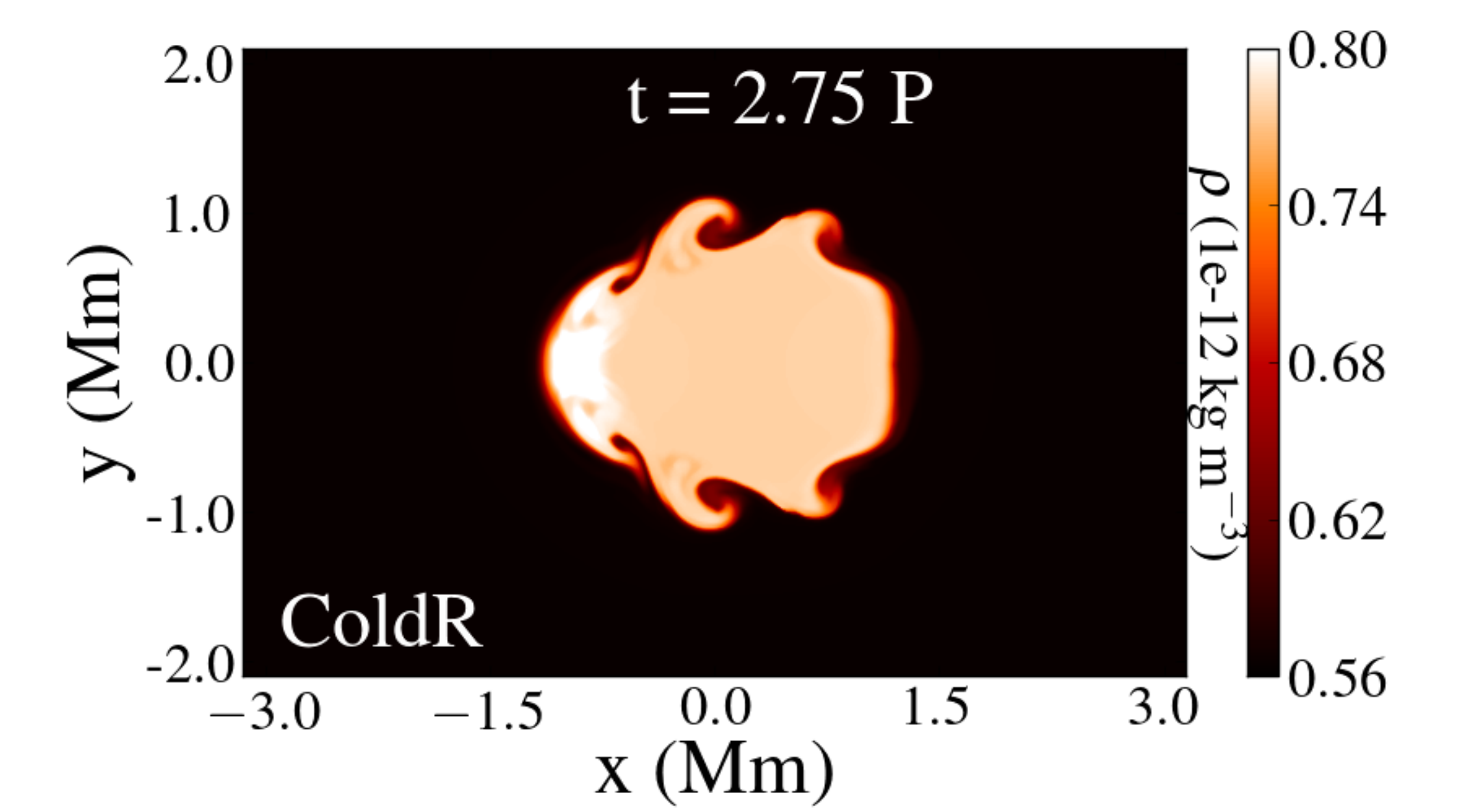}
\includegraphics[trim={1cm 0cm 0cm 0cm},clip,scale=0.16]{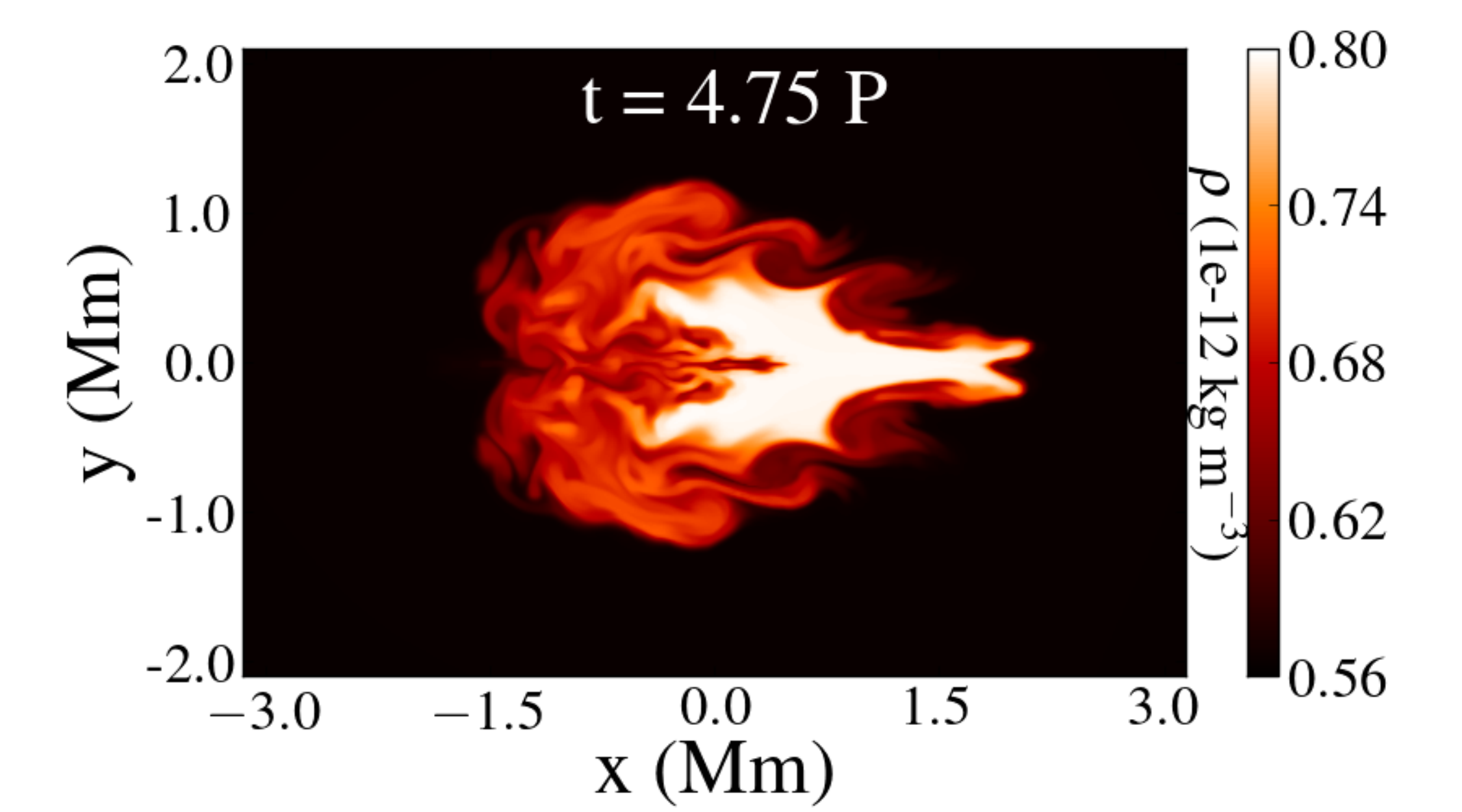}
\includegraphics[trim={1cm 0cm 0cm 0cm},clip,scale=0.16]{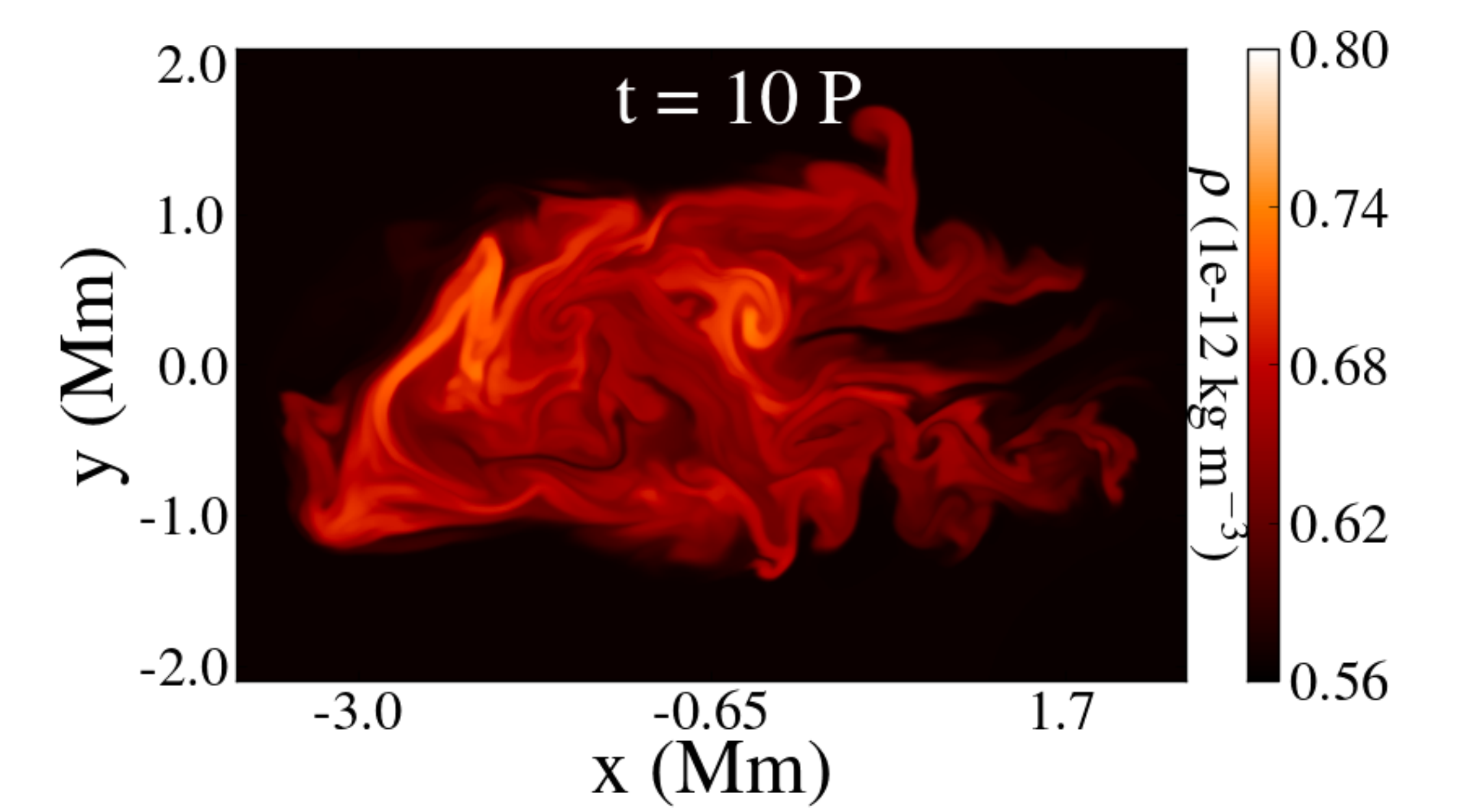}}
\resizebox{\hsize}{!}{\includegraphics[trim={1cm 0cm 0cm 0cm},clip,scale=0.16]{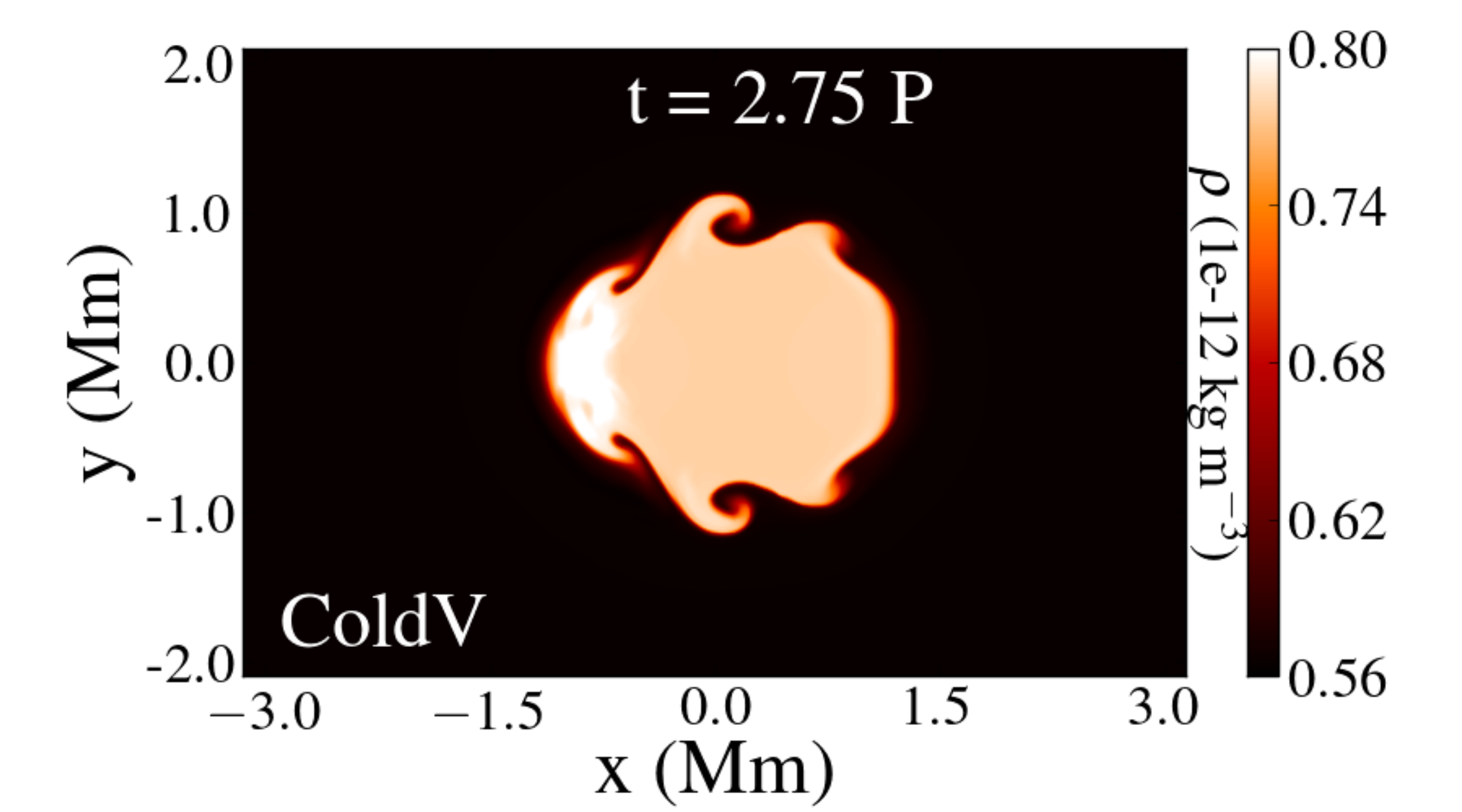}
\includegraphics[trim={1cm 0cm 0cm 0cm},clip,scale=0.16]{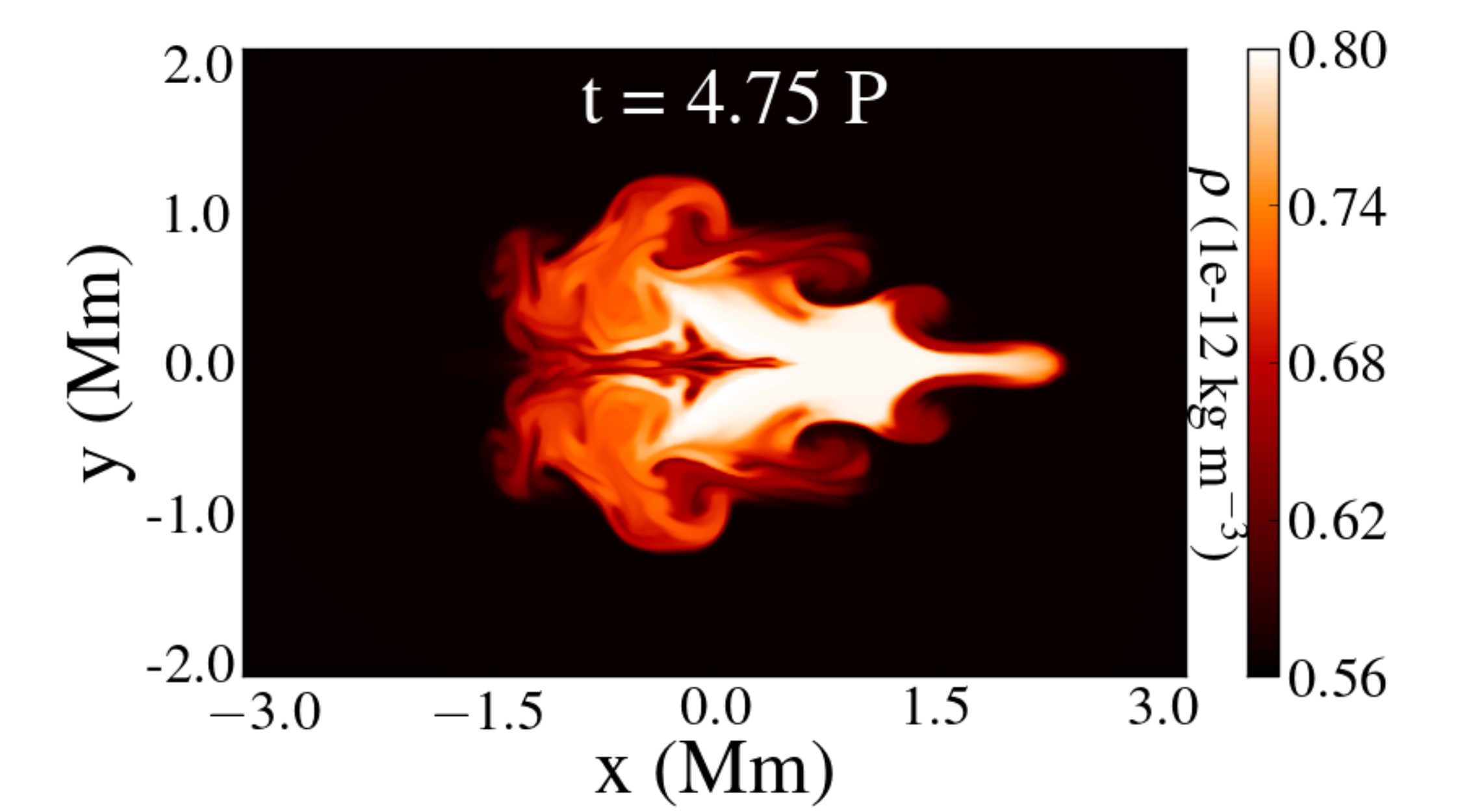}
\includegraphics[trim={1cm 0cm 0cm 0cm},clip,scale=0.16]{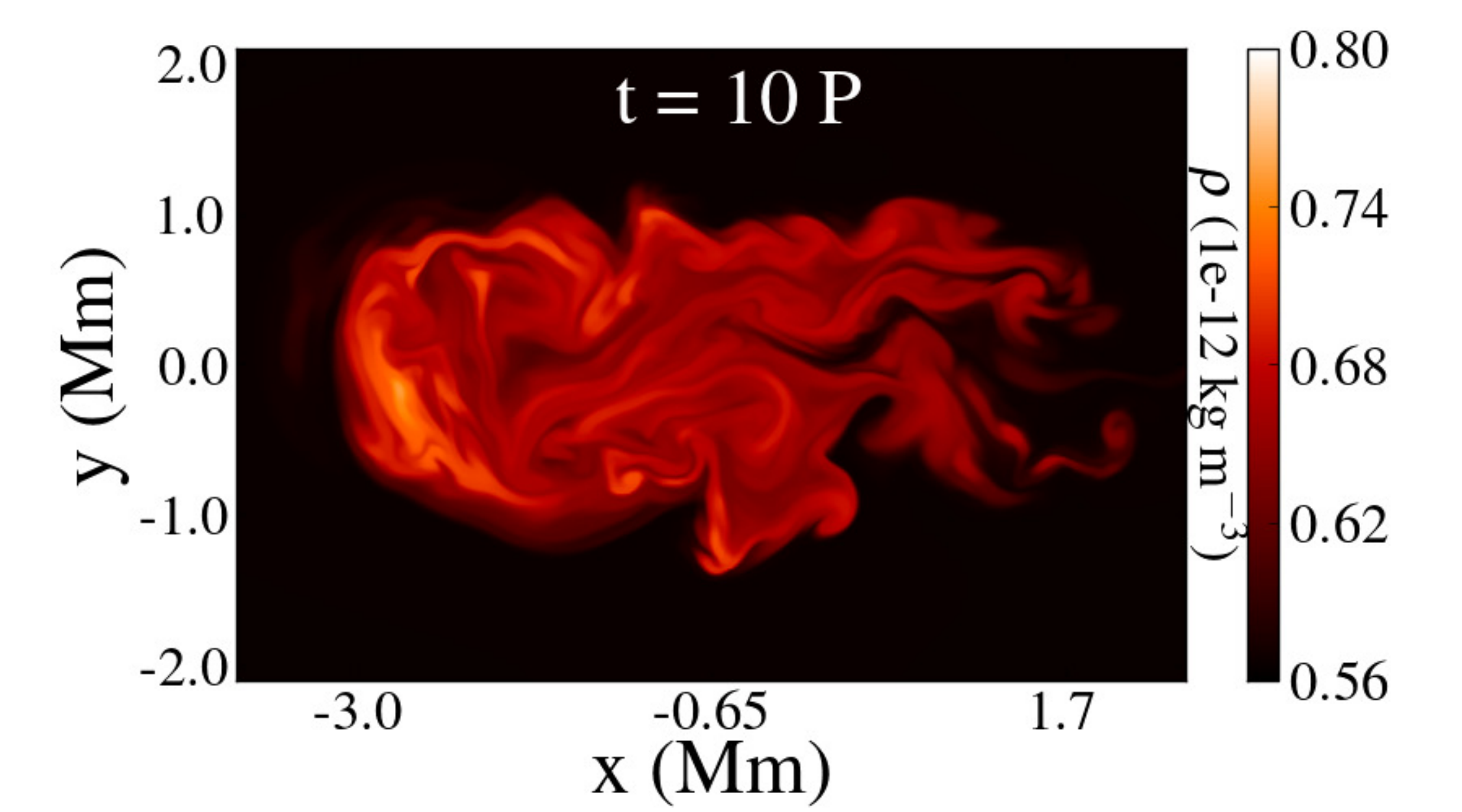}}
\resizebox{\hsize}{!}{\includegraphics[trim={1cm 0cm 0cm 0cm},clip,scale=0.16]{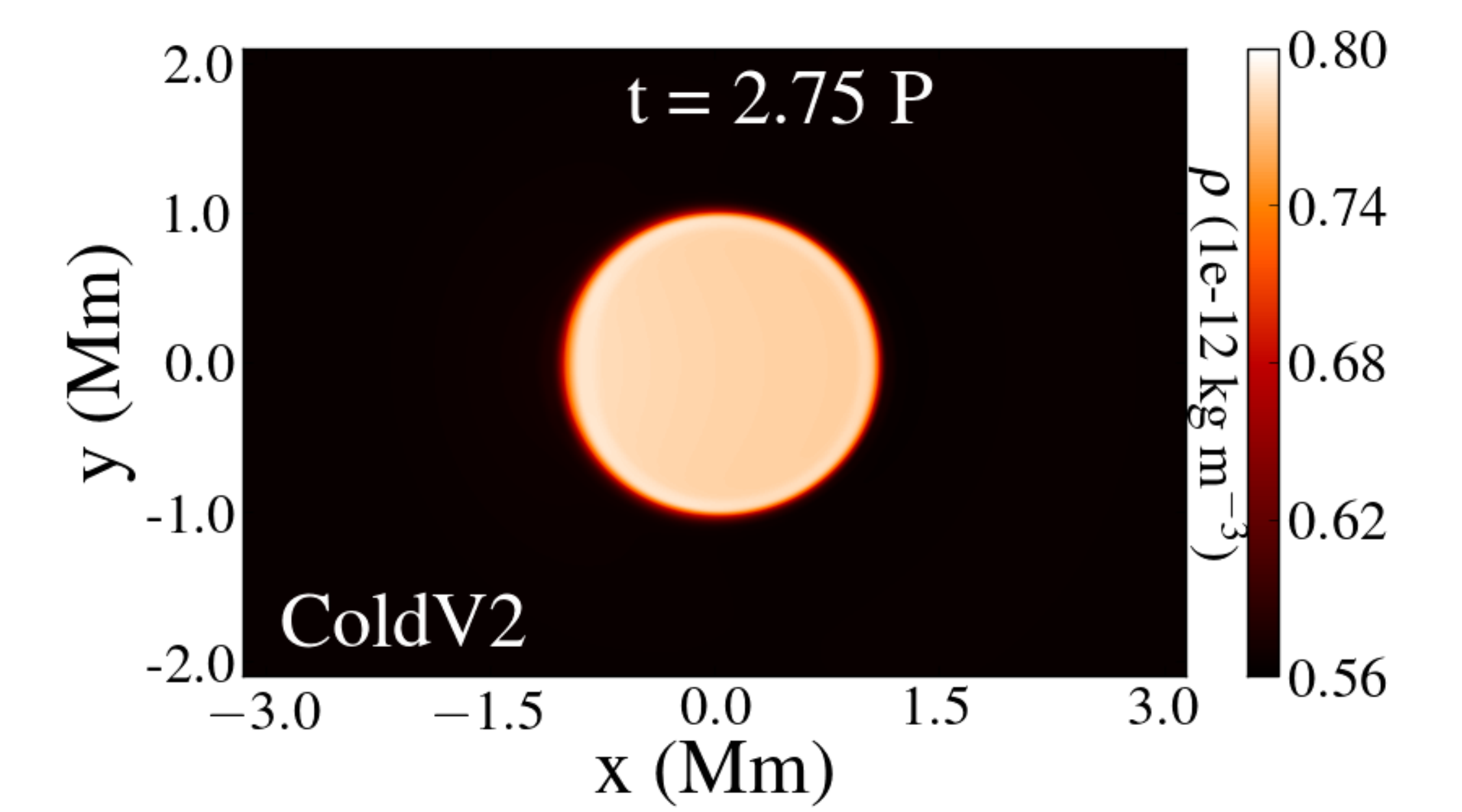}
\includegraphics[trim={1cm 0cm 0cm 0cm},clip,scale=0.16]{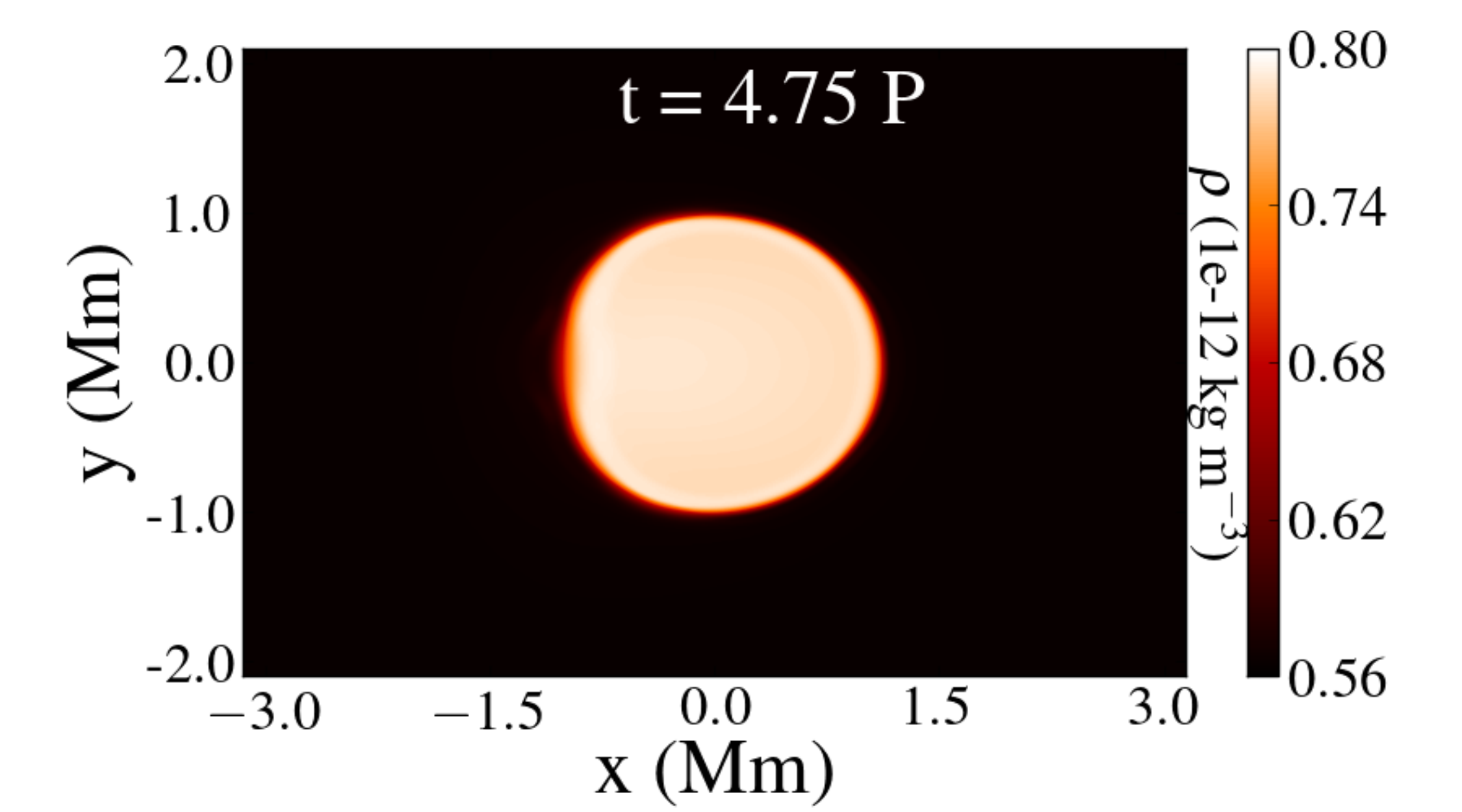}
\includegraphics[trim={1cm 0cm 0cm 0cm},clip,scale=0.16]{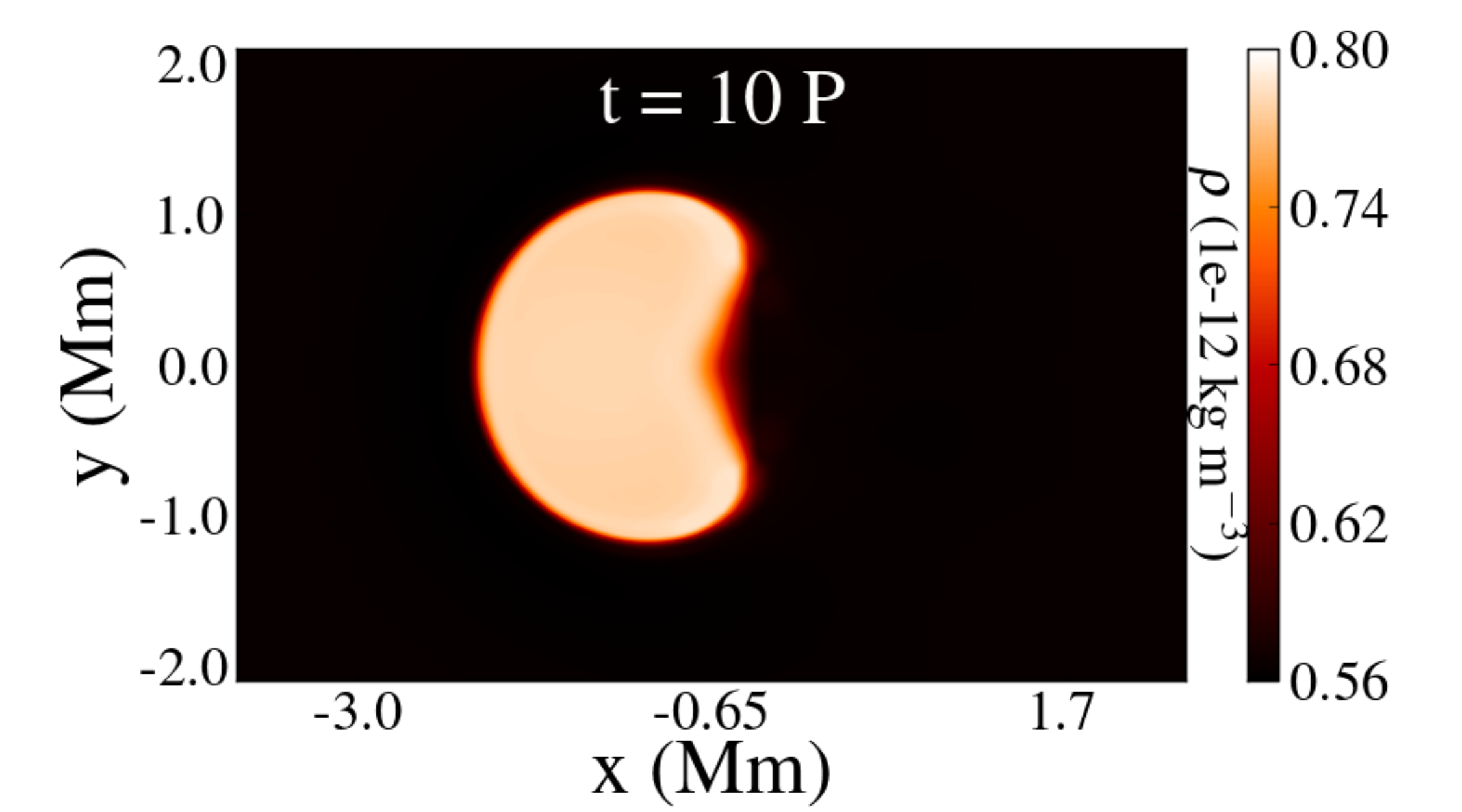}}
\caption{Contour plots of the density in the cross section at the apex for five different set-ups of cold loops at three different times. From top to bottom, models ColdIngr, ColdI, ColdR, ColdV and ColdV2. Panels are recentred to keep a clear view of the entire cross section. A different colourscale is chosen for the ColdIngr case, better adjusted to the density profile. All panels have the same dimensions of $6.3$ Mm in the $x$ direction and $4.2$ Mm in the $y$ direction. The period of the driver is $P=171$ s. Animations of these plots for the ColdI, ColdR and ColdV are available on-line (Movies 3, 4 and 5).}\label{fig:TWIKH}
\end{figure*}

\section{Results}
In the following analysis, we define the flux tube through a density threshold (normalized by $\rho_u = 10^{-12}$ kg m$^{-3}$) as follows:
\begin{equation}\label{eq:thres}
\rho_{tube} \geq 0.359 \times \dfrac{\rho_i}{\rho_u} \, f(x,y,z) = 0.9 \, \exp\left(\dfrac{-g_{0}\,L}{R_\mathrm{specific}\, \pi} \dfrac{\cos\left( \pi z/L \right)}{ T(x,y) }  \right) 
,\end{equation}
where $g_0=274$ m s$^{-2}$ is the solar gravitation in the surface of the sun, $R_\mathrm{specific}$ is the specific gas constant, and $T(x,y)$ is the initial temperature profile for each model, which is independent of height in our set-ups. For models $1$ and $2$ we have $\rho_{tube}(x,y) \geq 0.359 \rho_i / \rho_u$.

Regarding the models of cold loops embedded in a hot environment, our choices of the density profiles and magnetic field ensure that these five models have the same frequency of the fundamental standing kink mode, as we see in Table \ref{tab:paramb}. These models all have the same optimal driver frequency, almost the same initial magnetic field, and the exact same driver. As a result, the only differences in the input energy from the driver are due to the different dynamical evolution of the systems. This difference of the input energy eventually affects the evolution of wave dissipation and the development of heating. Therefore, most of the differences are attributed to the presence or absence of gravity for the ideal MHD, and in the different values of $R_m$ and $R_e$ for the gravitationally stratified cases.

\subsection{Loop dynamics and evolution}
We drive our loops for a total of  ten periods. As in \citet{karampelas2017}, the first waves to reach the apex ($z=0$) are the azimuthal Alfv\'{e}n waves at the boundary layer of our tube, thanks to their higher propagation speed, followed by the propagating kink waves. The propagating waves superpose with the counter propagating waves from the other footpoint (due to the symmetry at the apex), forming a standing wave. By choosing driving frequencies equal to the analytically predicted frequencies for the fundamental kink mode \citep{edwin1983wave, andries2005A&A430.1109A}, we forced our loops to perform an oscillation resembling the fundamental standing mode for the kink wave. An animation of the ColdI model is available in the electronic version of this paper, for Fig. ~\ref{fig:kink}, showing the evolution of that oscillation. We note that due to the finite speed of the waves originated at the footpoint, the apex starts oscillating later than the footpoint, where the driver is located. This is shown in Fig. \ref{fig:cm}, where the oscillation at the apex starts later than the start of the driver (at $t=0$). This leads to a phase difference between oscillation at the footpoint and the apex, similar to that observed for the driven models of our previous studies \citep{karampelas2017,karampelas2018fd}.

As we see in Fig. \ref{fig:cm}, the centre of mass at the apex shows a maximum displacement of $\approx 1$ Mm from the equilibrium position at $t=0$. This is larger than the $\approx 0.1-0.3$ Mm oscillation amplitudes observed for decayless transverse oscillations in coronal loops \citep{nistico2013,anfinogentov2015,nakariakov2016}, and is caused by the strength of the driver used in the current set-up. We note that the $\upsilon_x$ velocity at the location of the centre of mass are $\approx 3.3 \%$ of the initial internal Alfv\'{e}n velocity.

As is expected from theory \citep{heyvaerts1983,zaqarashvili2015ApJ} and simulations \citep{terradas2008,antolin2017}, the location of the antinode of the $x$-velocity (here the apex) is Kelvin-Helmholtz unstable. We already know from \citet{karampelas2018fd} that the  KHI for driven oscillations leads to the development of spatially extended eddies, the TWIKH rolls. Because of the frozen-in condition, the out of phase movement of the TWIKH rolls create elongated strand-like features along the flux tube, which we see in Fig. ~\ref{fig:kink}. The same structures are visible in Fig. \ref{fig:fomo}, where we present snapshots of the emission intensity for the Fe XII $195.12\, \AA$ line at the end of our simulation, for models ColdI, ColdR, and ColdV. This spectral line was chosen because it is better suited to detect the hotter plasma at the loop edges in our set-up \citep{antolin2017}. The resulting images are very similar to the non-stratified case \citep{karampelas2018fd}, showing that the introduction of gravity does not affect previous results on forward modelling of oscillating loops \citep{antolin2016,antolin2017}. Furthermore, the similarity of the results between the ideal, resistive, and viscous MHD models can potentially hinder the observational distinction between the dissipative effects in coronal loops should one focus only on studying the dynamical evolution of these systems.

In Fig. \ref{fig:TWIKH} we show cross sections at the apex, at multiple oscillation times for models ColdIngr, ColdI, ColdR, ColdV, and ColdV2. For $R_e \geqslant 10^4$ and $R_m \geqslant 10^4$ the presence of higher dissipation such as resistivity and viscosity delays the emergence of the KHI compared to the ColdI model, in agreement with \citet{howson2017}. However, the instability is fully developed for models ColdI, ColdR and ColdV within the first three driving periods. After almost five periods, the TWIKH rolls have already expanded across the loop cross section, deforming the initial density profile. By the end of the simulation, the loop surface area basically doubles for all three models at the apex, and the TWIKH rolls turn the initial monolithic density profile into a turbulent density profile. This evolution of the cross section is also observed in most of the other models considered for this study, and is responsible for some of the results regarding the temperature evolution in our models. By simulating the entire loop cross section, we observe numerically induced asymmetries in the development of KHI. These become more prominent in the second half of the simulations, creating an non-symmetric turbulent density profile, closer to what would be generally expected in the solar corona. Movies $3$, $4,$ and $5$ of Fig. \ref{fig:TWIKH} show the evolution of the loop cross section for the entirety of the simulation. 

The only exception to the aforementioned cases is  model ColdV2 ($R_e = 10^2$), where no TWIKH rolls are observed. The high value of the shear viscosity in that model leads to the complete suppression of the KHI for the duration of our simulation. A similar effect was observed before in \citet{howson2017} in impulsively oscillating coronal loops for combined high values of resistivity and viscosity ($R_e=10^4$ and $R_m=10^4$). The higher values of dissipation required to suppress the KHI in our work are due to the continuous driving of our loops. 

\subsection{Temperature evolution in cold loops}

\begin{figure}
\centering
\includegraphics[trim={0.5cm 0.5cm 0.5cm 0.4cm},clip,scale=0.3]{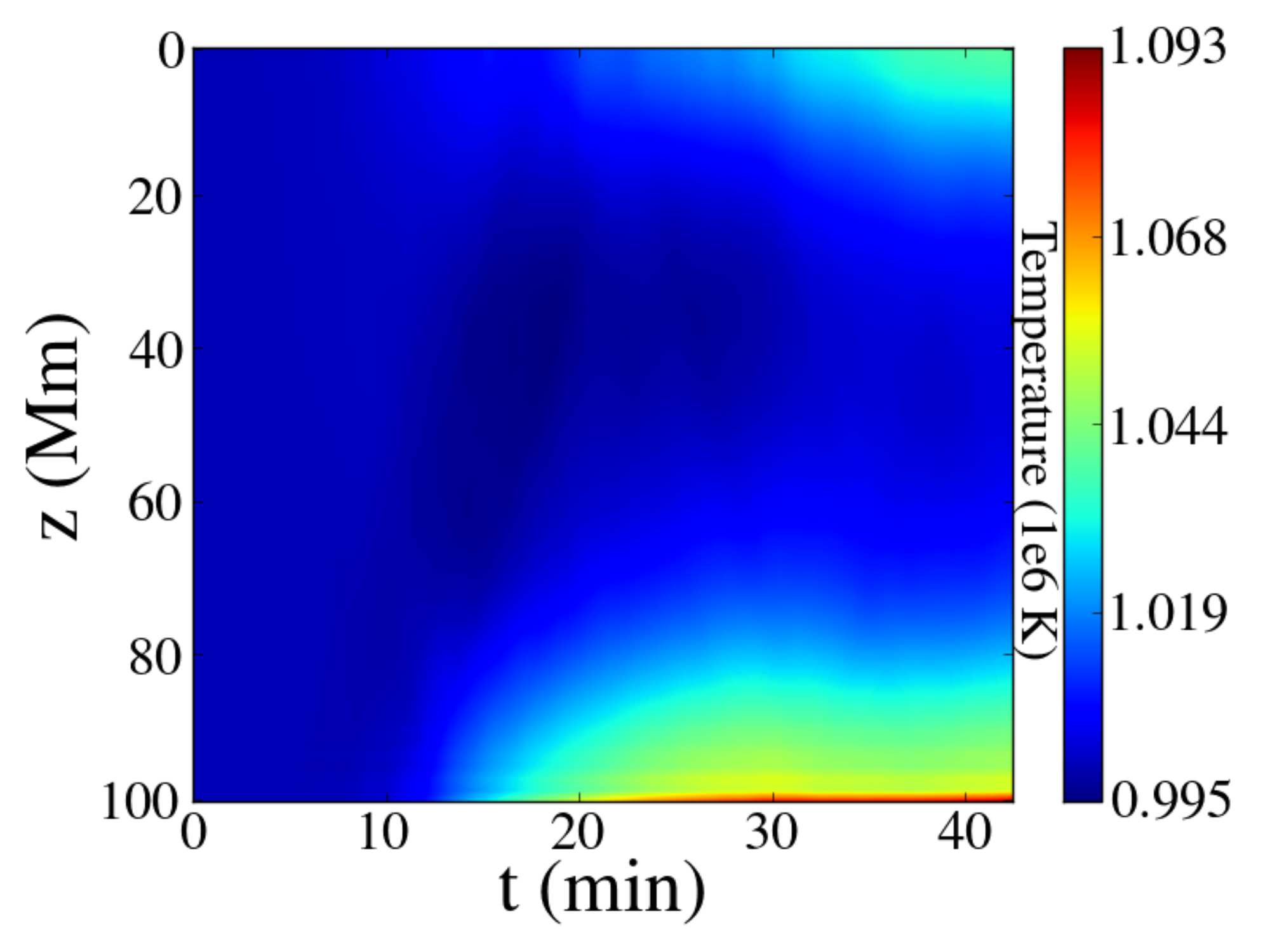}
\includegraphics[trim={0.5cm 0.5cm 0.5cm 0.4cm},clip,scale=0.3]{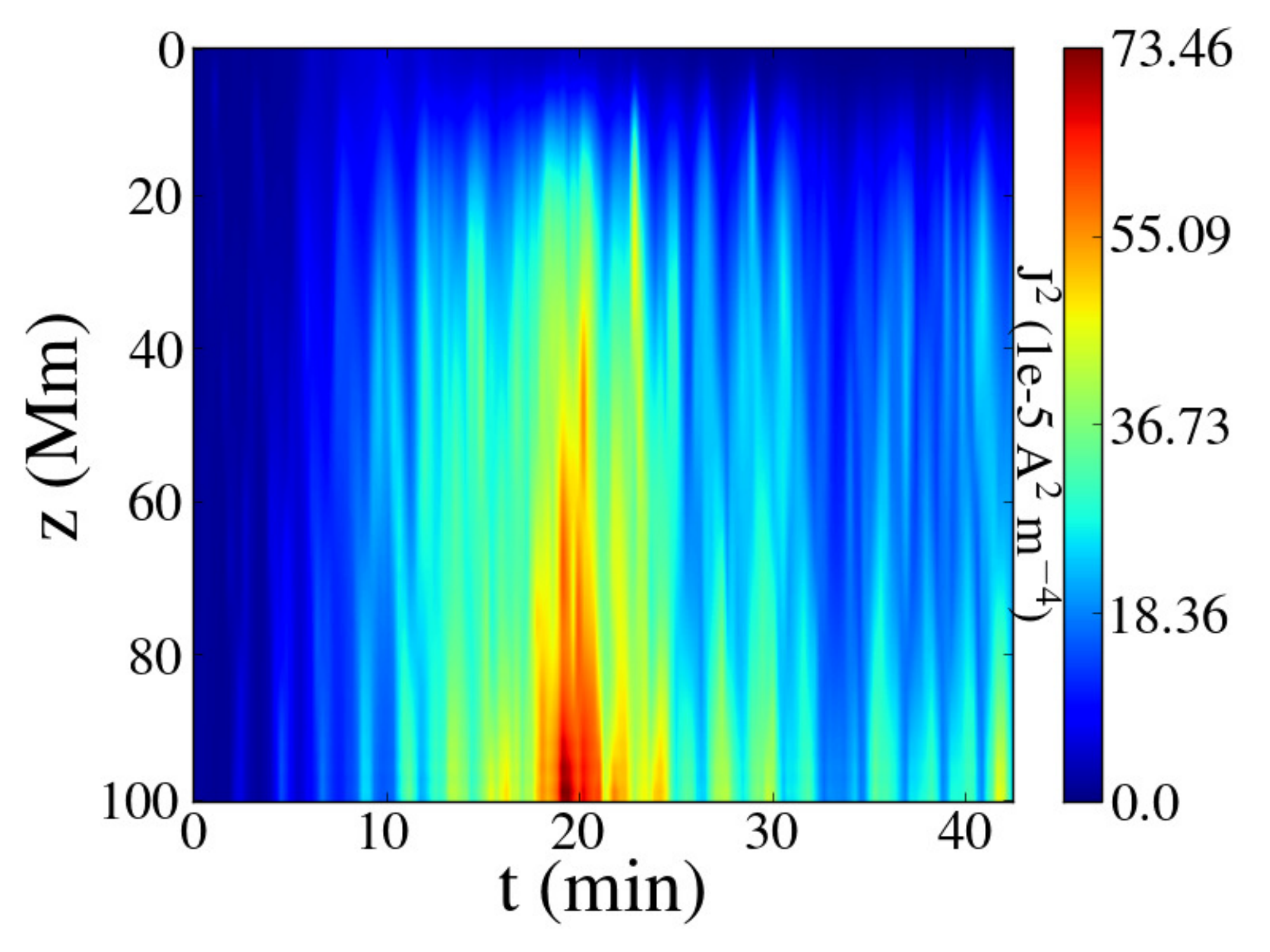}
\caption{Top panel: Average temperature of the flux tube for $\rho \geq 0.9 \times 10^{-12}$ kg m$^{-3}$ along the z-axis, for a non-stratified loop with uniform temperature (model $1$). Bottom panel: Average square current densities ($J^2$) of the flux tube for $\rho \geq 0.9 \, f(x,y,z)$ for the same model. The apex is located at $z=0$.}\label{fig:isot}
\end{figure}

In \citet{karampelas2017}, we proved that (numerical) resistivity increases the temperature of a non-stratified loop, with uniform initial temperature (Driven-equalT model), near the footpoint. In order to validate our previous results and study the effects of numerical dissipation in the current code to our results, we simulated a similar set-up for an increased resolution (UniT model). In Fig. \ref{fig:isot}, we examine the temperature profiles along the $z$-axis over time for this model and we plot the average temperature for the flux tube cross section (for $\rho \geq 0.9 \times 10^{-12}$ kg m$^{-3}$). We observe a gradual increase of the average temperature over time the closer we get to the footpoint and apex. The temperature increase is comparable in both regions and the highest values are observed near the footpoint, while the area at mid-length of the loop experiences a negligible temperature increase.

At the apex, the higher velocity and the developed KHI leads to stronger viscous heating. Resistive heating is not expected to be as pronounced there because of the lack of strong currents caused by the nature of the fundamental standing kink mode. Ohmic dissipation, however, is stronger near the footpoint, where the average square current densities (dominated by the $J_z^2$) are at their strongest \citep{tvd2007resist}. In \citet{karampelas2017} we observed similar temperature profiles, but the temperature increase at the apex was not as pronounced. The observed differences between the present and past results are caused partly by the stronger driver employed and partly by the different numerical dissipation in each code. However, the temperature increases at the apex is expected from our previous analysis, despite the apparent contradiction with the older results. The temperature increase at the footpoint is explained through the higher values of the resistive heating rate there \citep{karampelas2018fd}.

\begin{figure*}
\centering
\resizebox{\hsize}{!}{\includegraphics[trim={0.5cm 0cm 0cm 0cm},clip,scale=0.16]{a34309-18_pic20}
\includegraphics[trim={0.5cm 0cm 0cm 0cm},clip,scale=0.16]{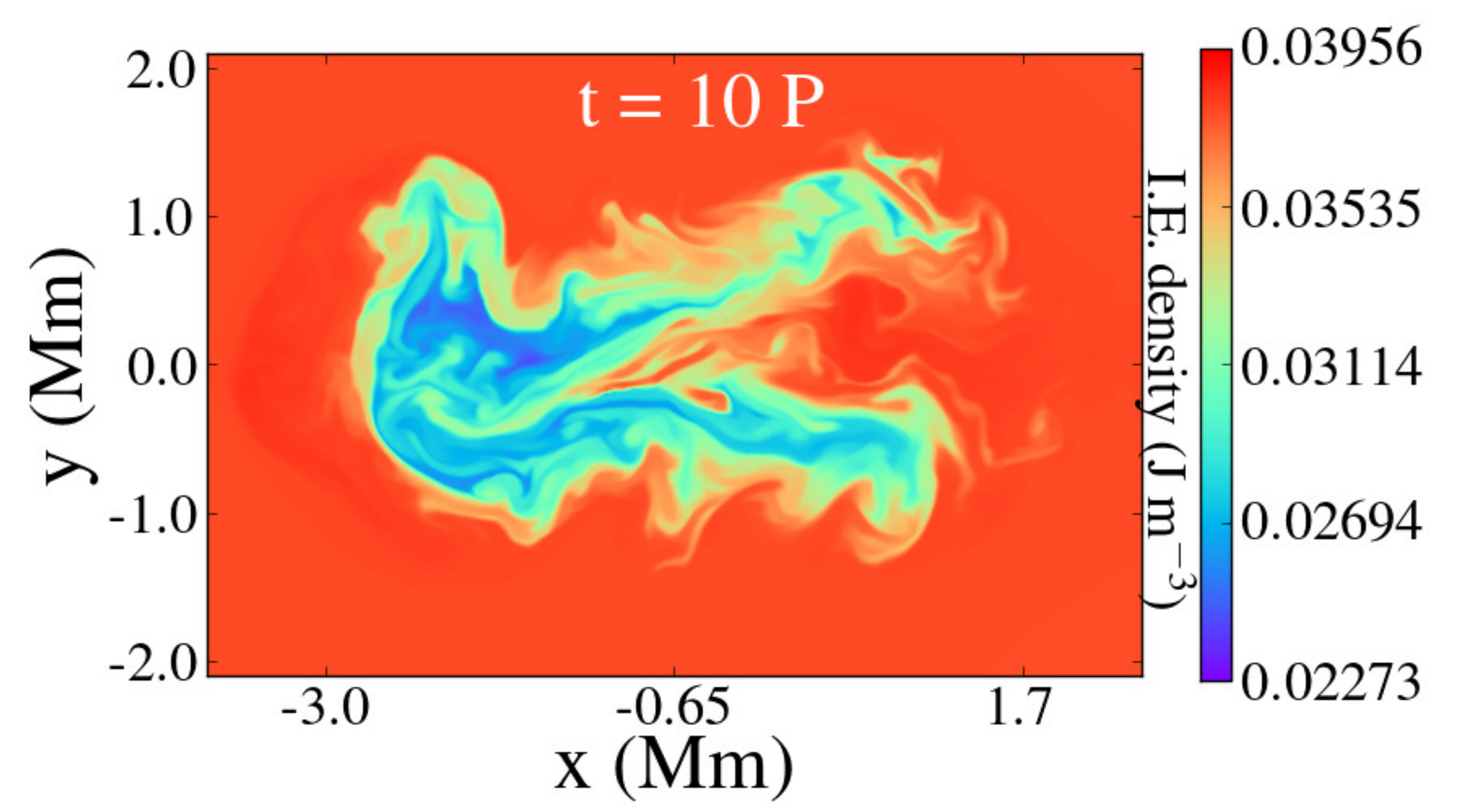}
\includegraphics[trim={0.5cm 0cm 0cm 0cm},clip,scale=0.16]{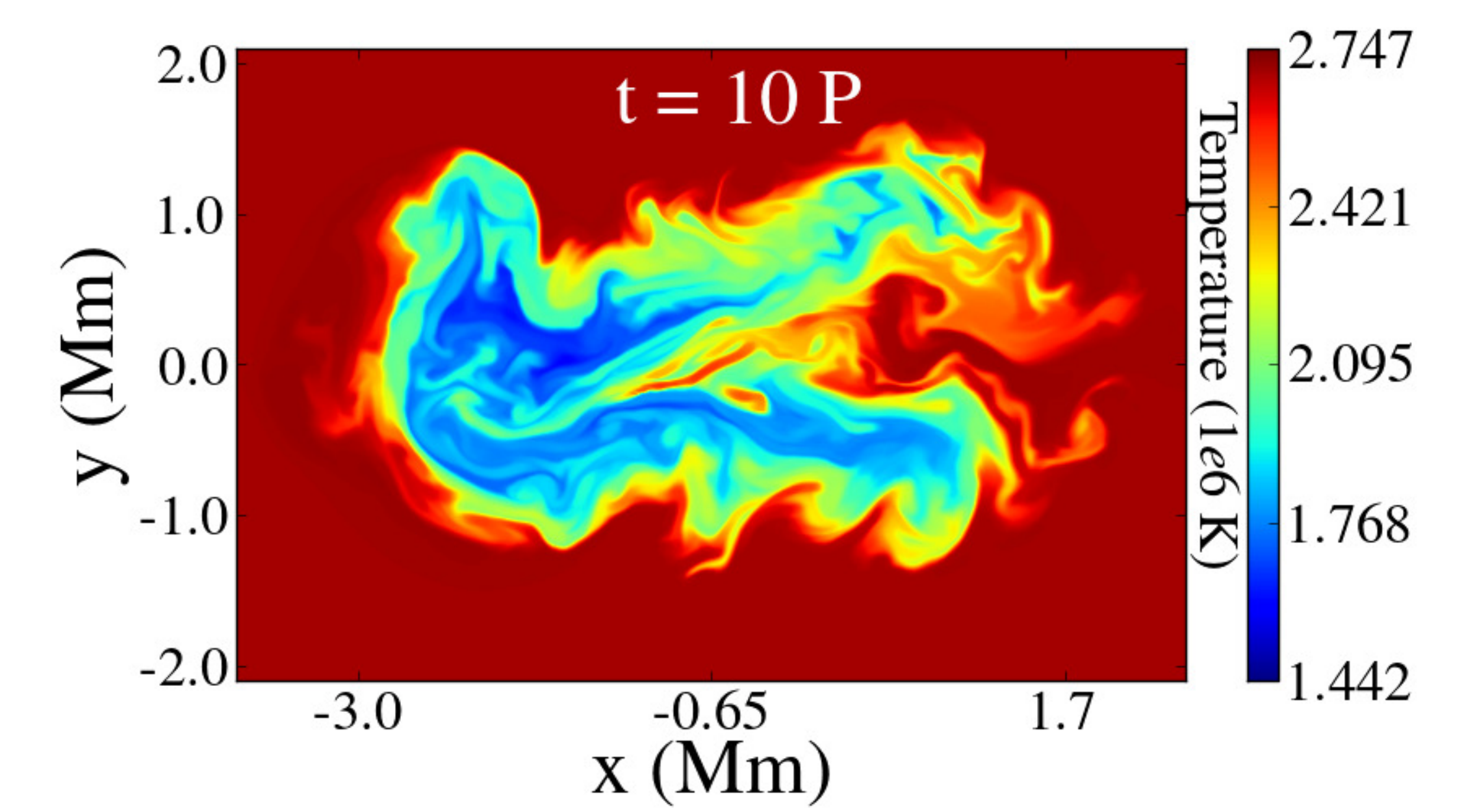}}
\resizebox{\hsize}{!}{\includegraphics[trim={0.5cm 0cm 0cm 0cm},clip,scale=0.16]{a34309-18_pic23}
\includegraphics[trim={0.5cm 0cm 0cm 0cm},clip,scale=0.16]{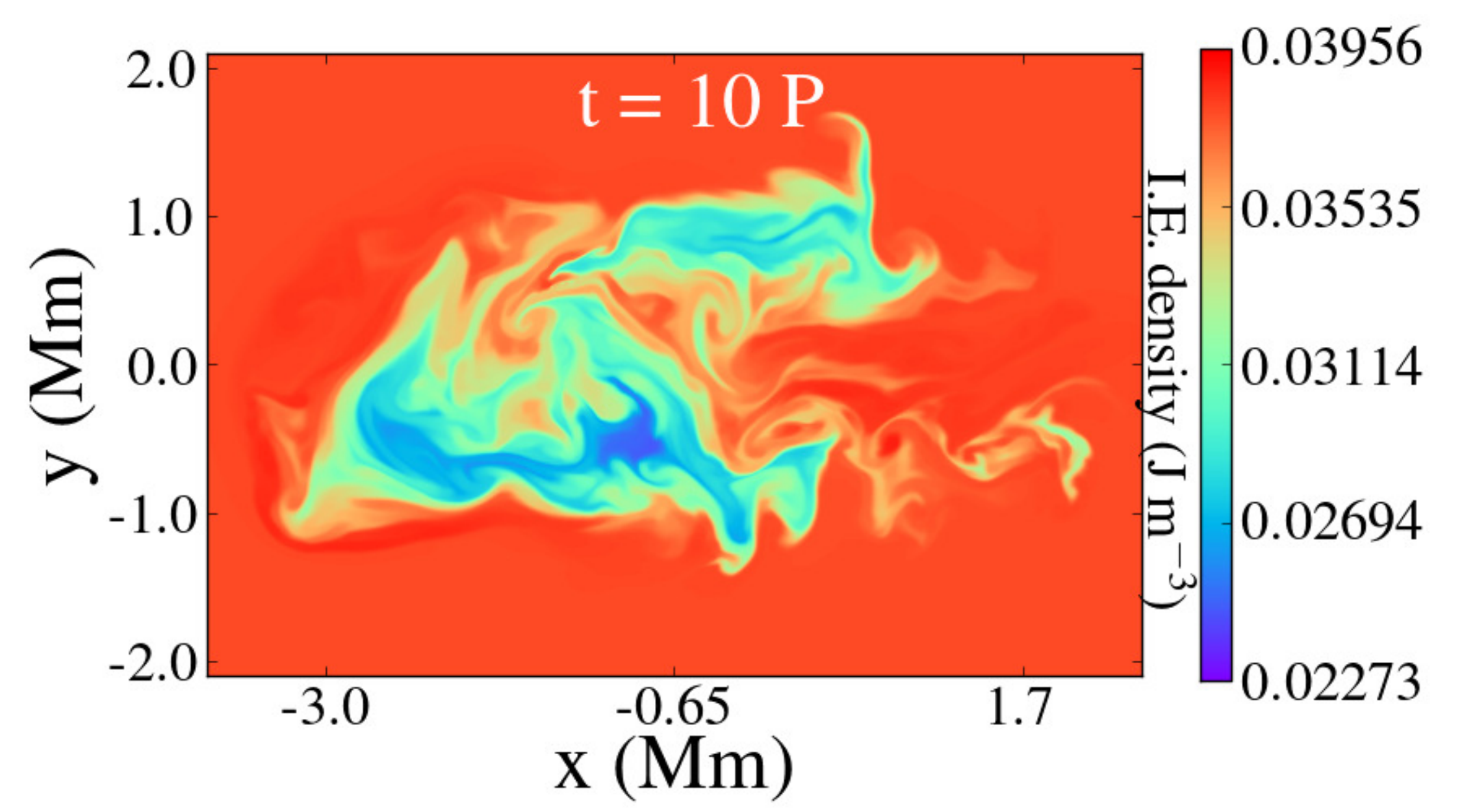}
\includegraphics[trim={0.5cm 0cm 0cm 0cm},clip,scale=0.16]{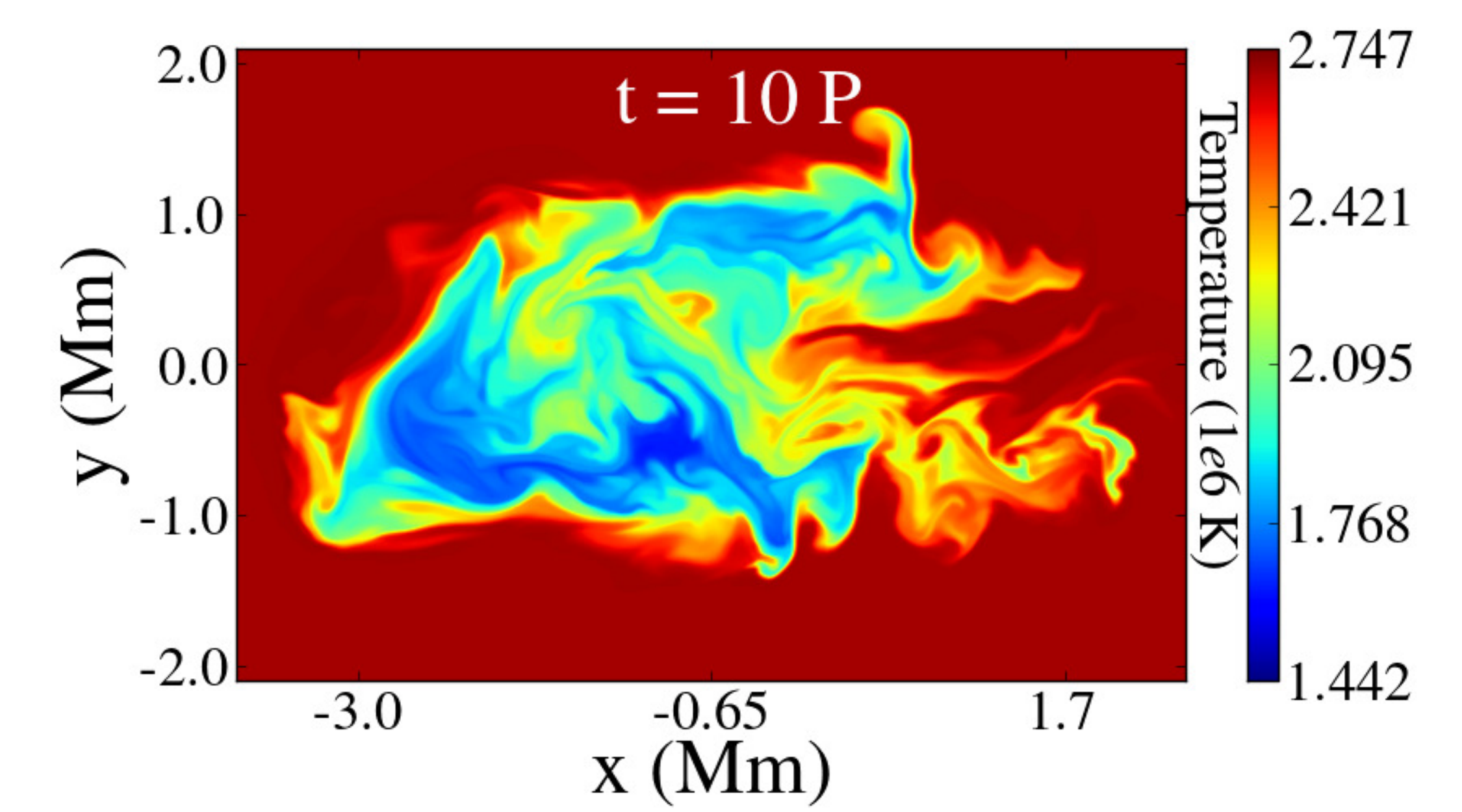}}
\resizebox{\hsize}{!}{\includegraphics[trim={0.5cm 0cm 0cm 0cm},clip,scale=0.16]{a34309-18_pic26}
\includegraphics[trim={0.5cm 0cm 0cm 0cm},clip,scale=0.16]{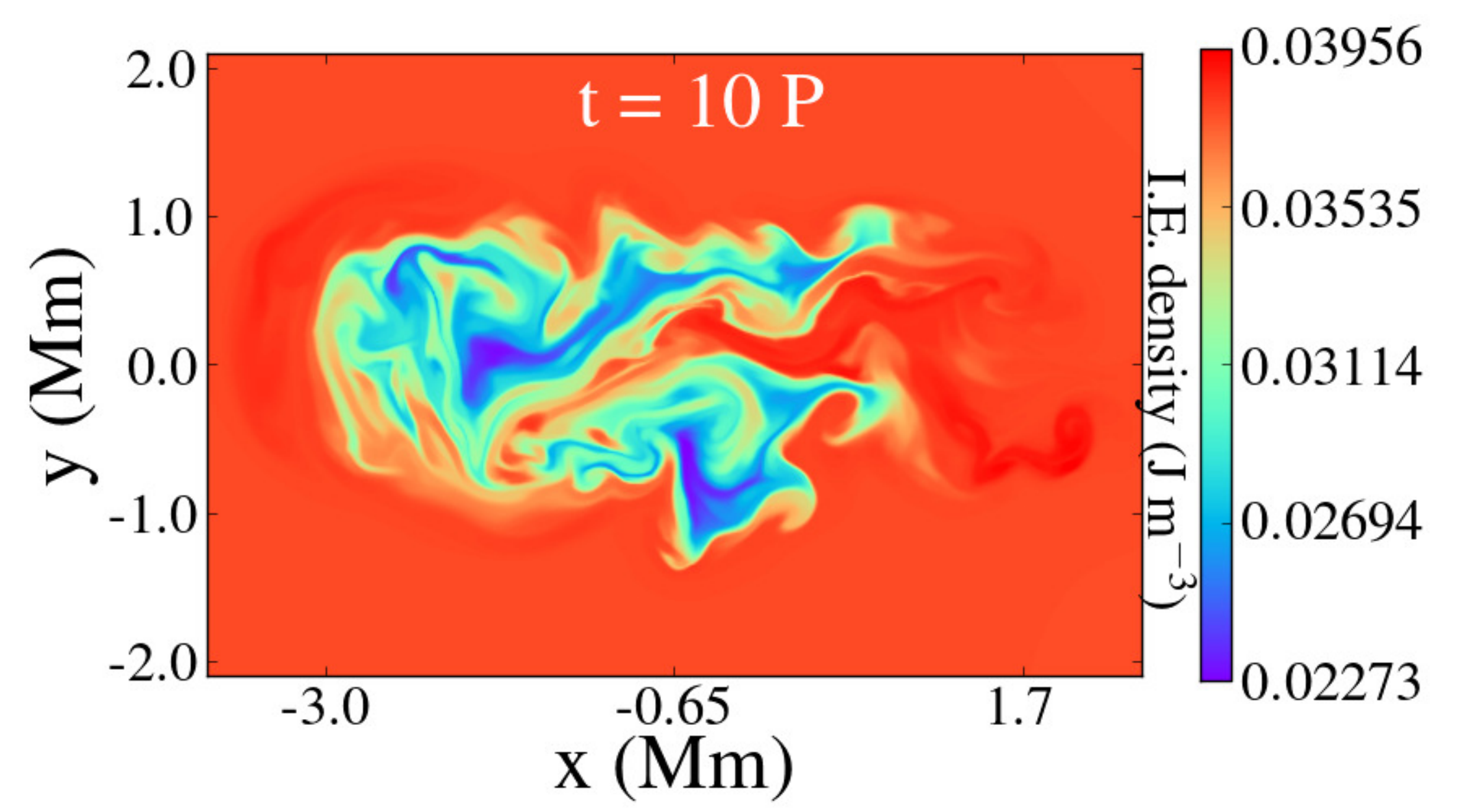}
\includegraphics[trim={0.5cm 0cm 0cm 0cm},clip,scale=0.16]{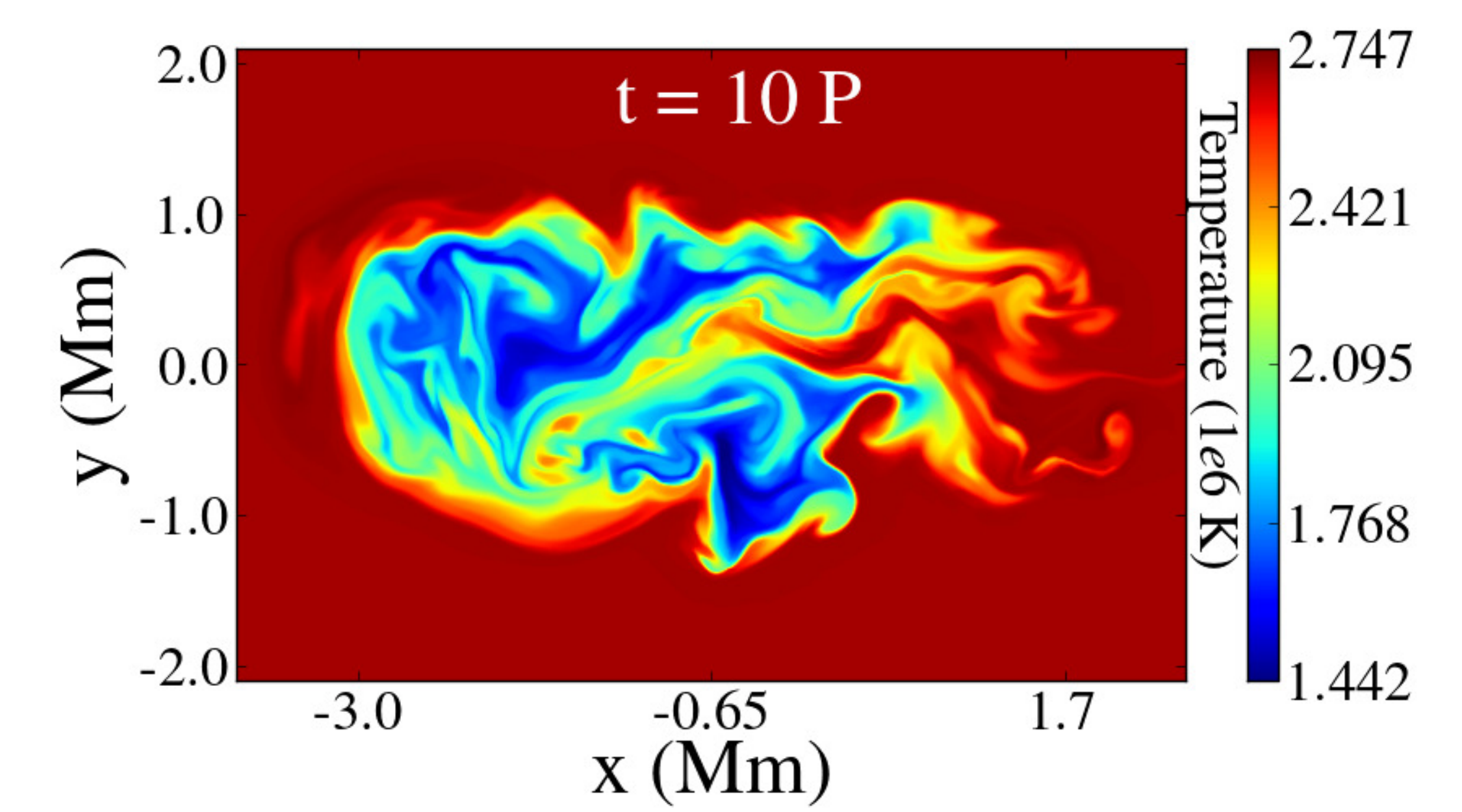}}
\caption{Contour plots of density (left), internal energy (middle), and temperature (right) at the apex for $-2.1\leq y$ (Mm) $\leq2.1,$ and $-3.8\leq x$ (Mm) $\leq2.5$. From top to bottom, the cases for ideal MHD, resistive MHD, and viscous MHD (models ColdI, ColdR and ColdV) are shown.   The driver period is $P\simeq 172$ s.}\label{fig:456c}
\end{figure*}

Considering again our models of gravitationally stratified cold loops embedded in a hot corona, we try to see where the energy dissipation takes place and how this affects the temperature distribution. In Fig. \ref{fig:456c} the cross sections at the apex of models ColdI, ColdR, and ColdV are shown for density, internal energy density, and temperature. As we see in the contours for internal energy density, the highest values lay on the interface between the expanded TWIKH rolls and the environment, while the internal part of the loop also shows increased values from those derived from the initial conditions (see Fig. \ref{fig:setup}). The highest values are found in the ColdV model, followed by the ColdR model, and finally by ColdI. The same can also be seen in the contours for the temperature and the ColdV case shows the highest values of temperature at the locations of the highest internal energy increase. Because of delayed mixing the viscous set-up also has some of the lowest temperatures in internal areas of the loop when compared to the other two cases. These profiles guide us into treating the observed temperature increase of $\sim 4.7\times 10^4$ K mainly as the result of dissipation, rather than the adiabatic temperature fluctuation that was observed in loops of uniform temperature \citep{antolin2017,karampelas2017}.

\begin{figure*}
\centering
\resizebox{\hsize}{!}{\includegraphics[trim={0cm 0cm 0cm 0cm},clip,scale=0.16]{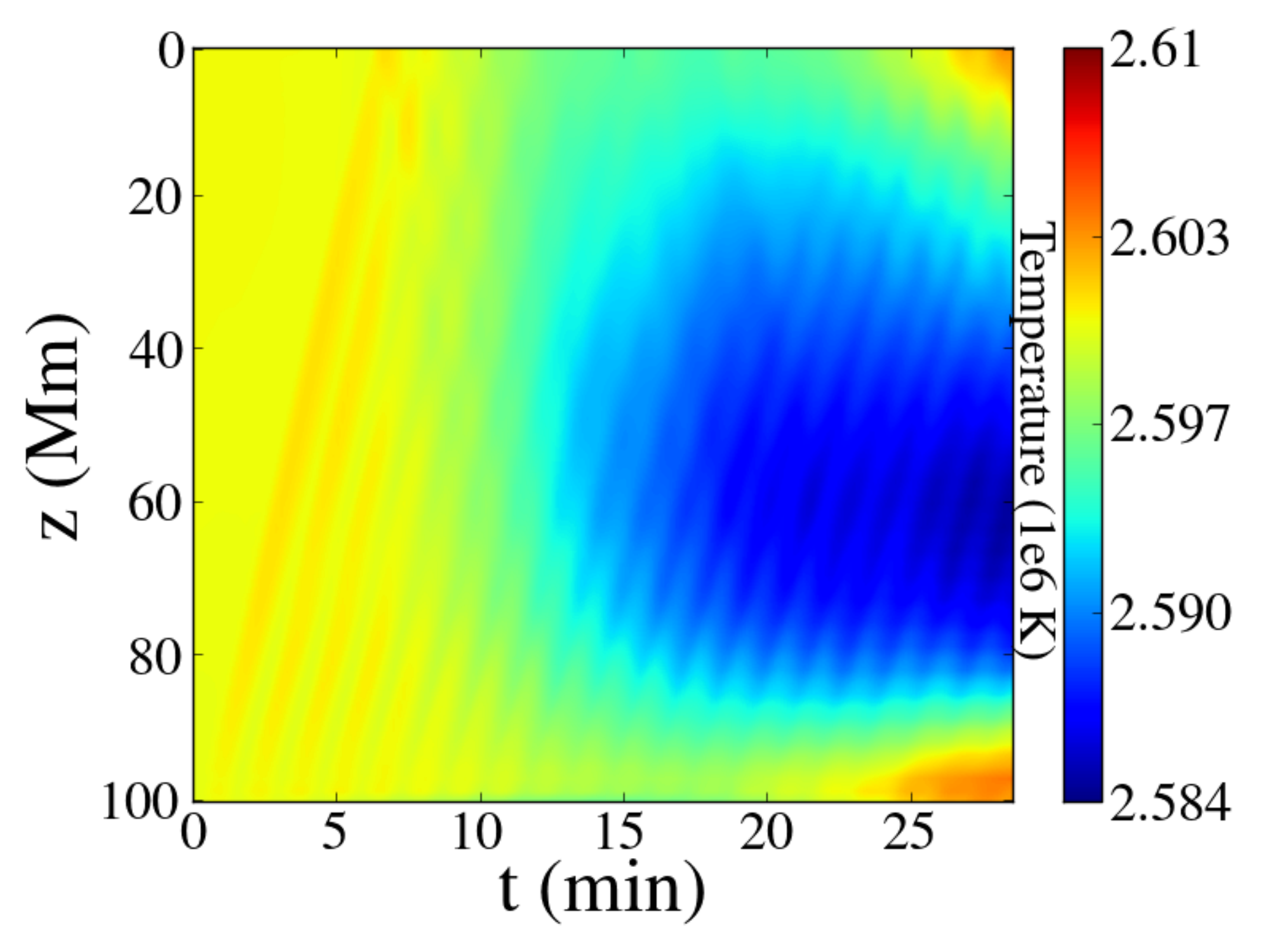}
\includegraphics[trim={0cm 0cm 0cm 0cm},clip,scale=0.16]{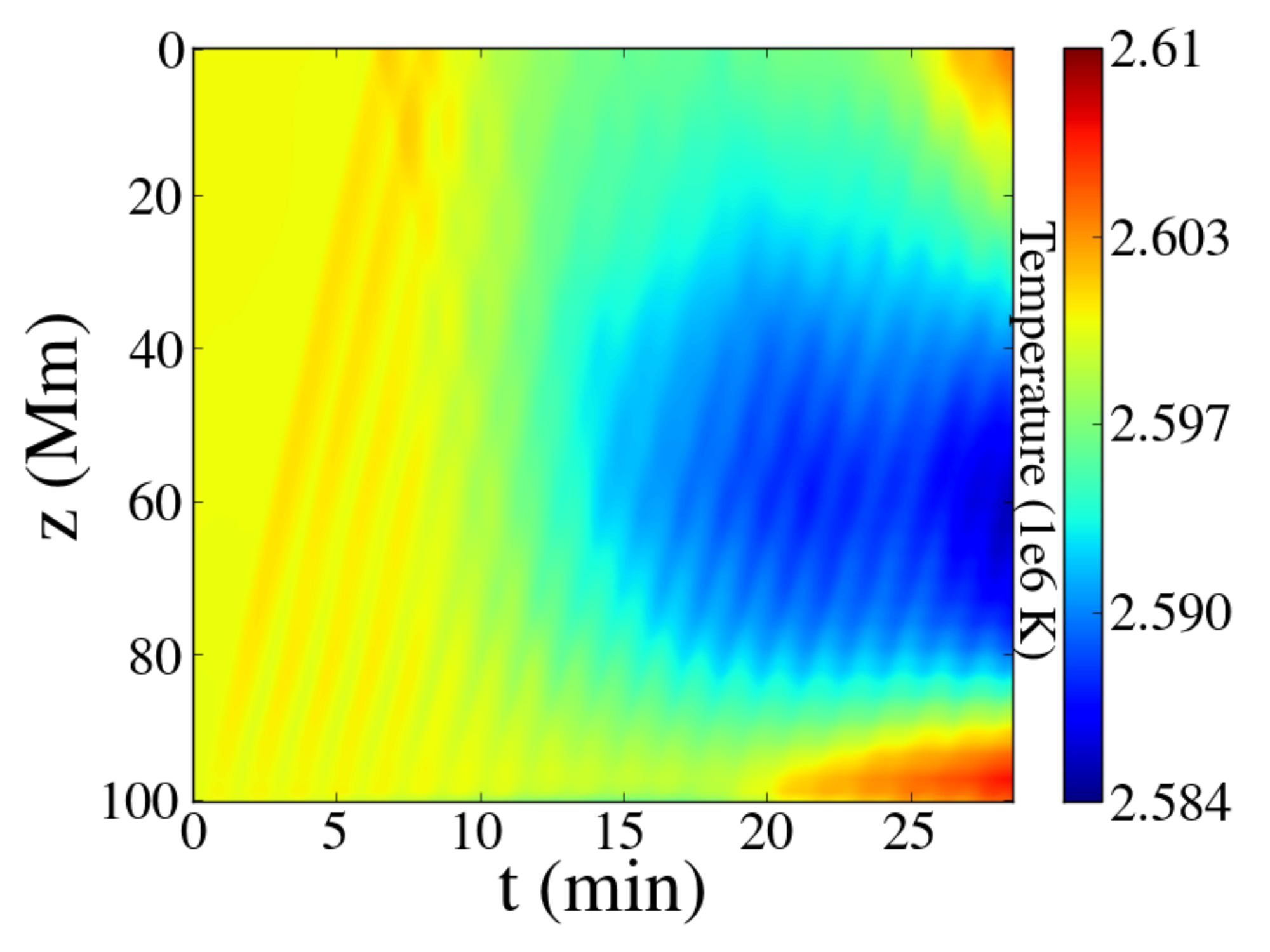}
\includegraphics[trim={0cm 0cm 0cm 0cm},clip,scale=0.16]{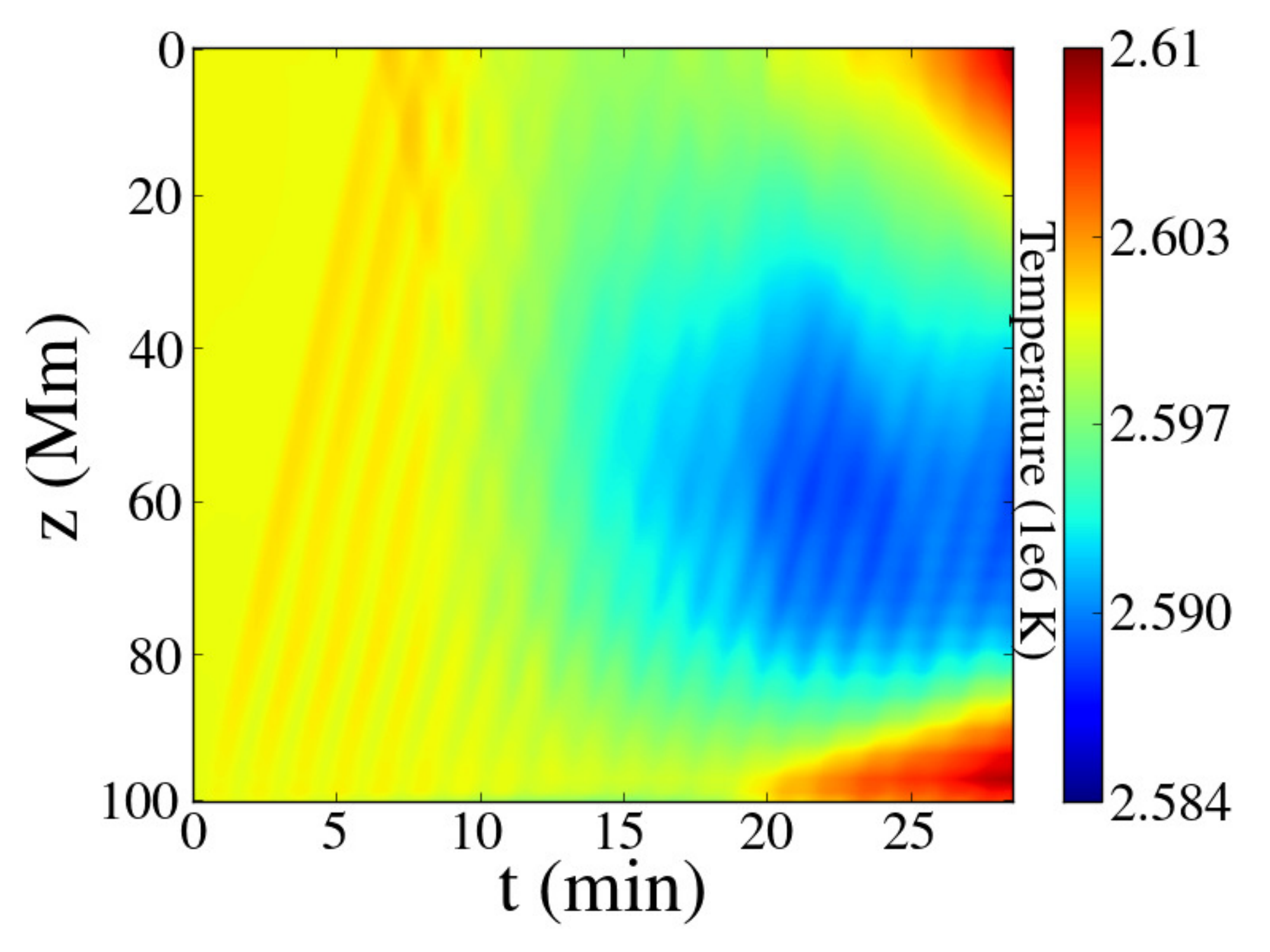}}
\resizebox{\hsize}{!}{\includegraphics[trim={0cm 0cm 0cm 0cm},clip,scale=0.16]{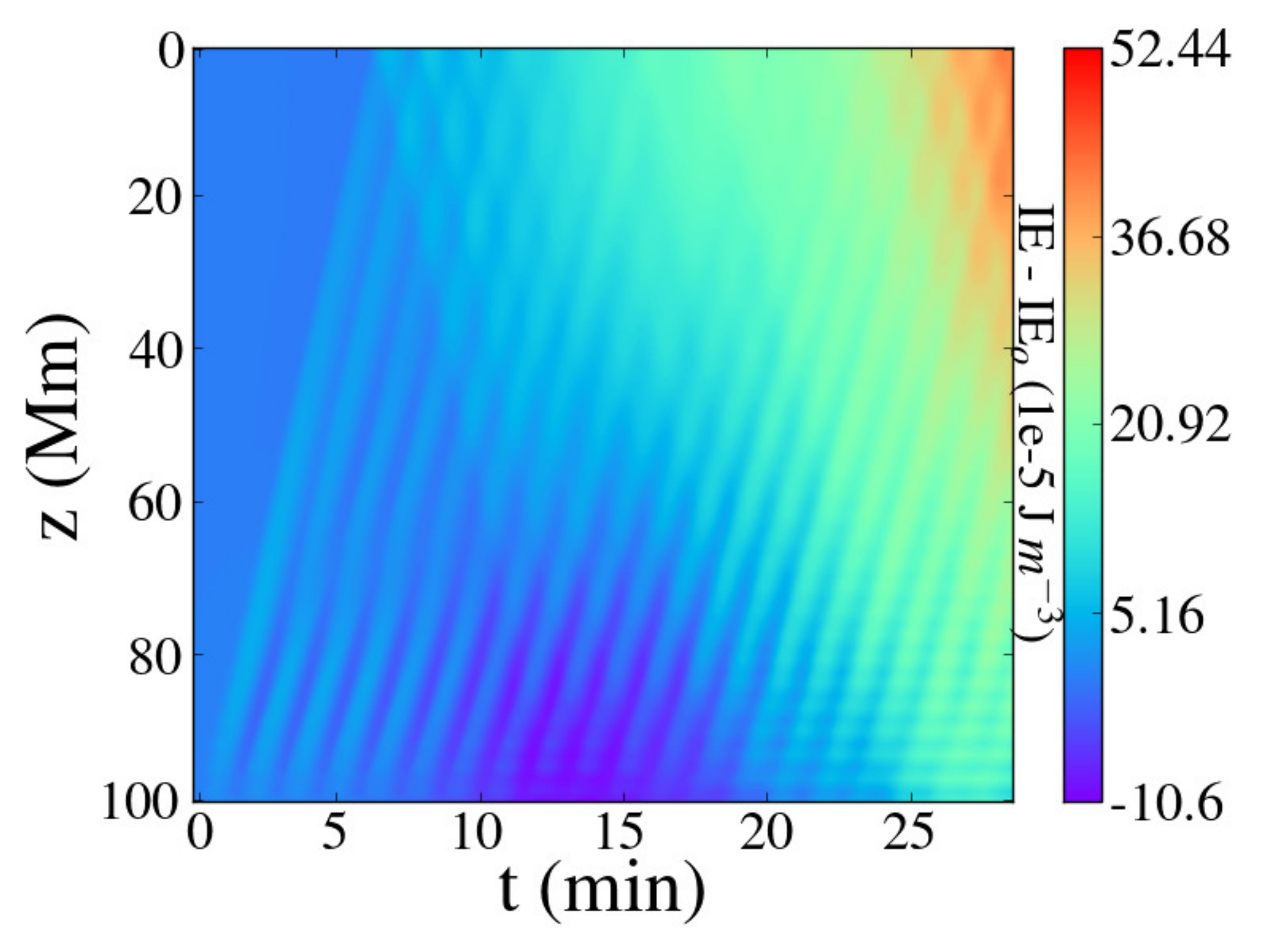}
\includegraphics[trim={0cm 0cm 0cm 0cm},clip,scale=0.16]{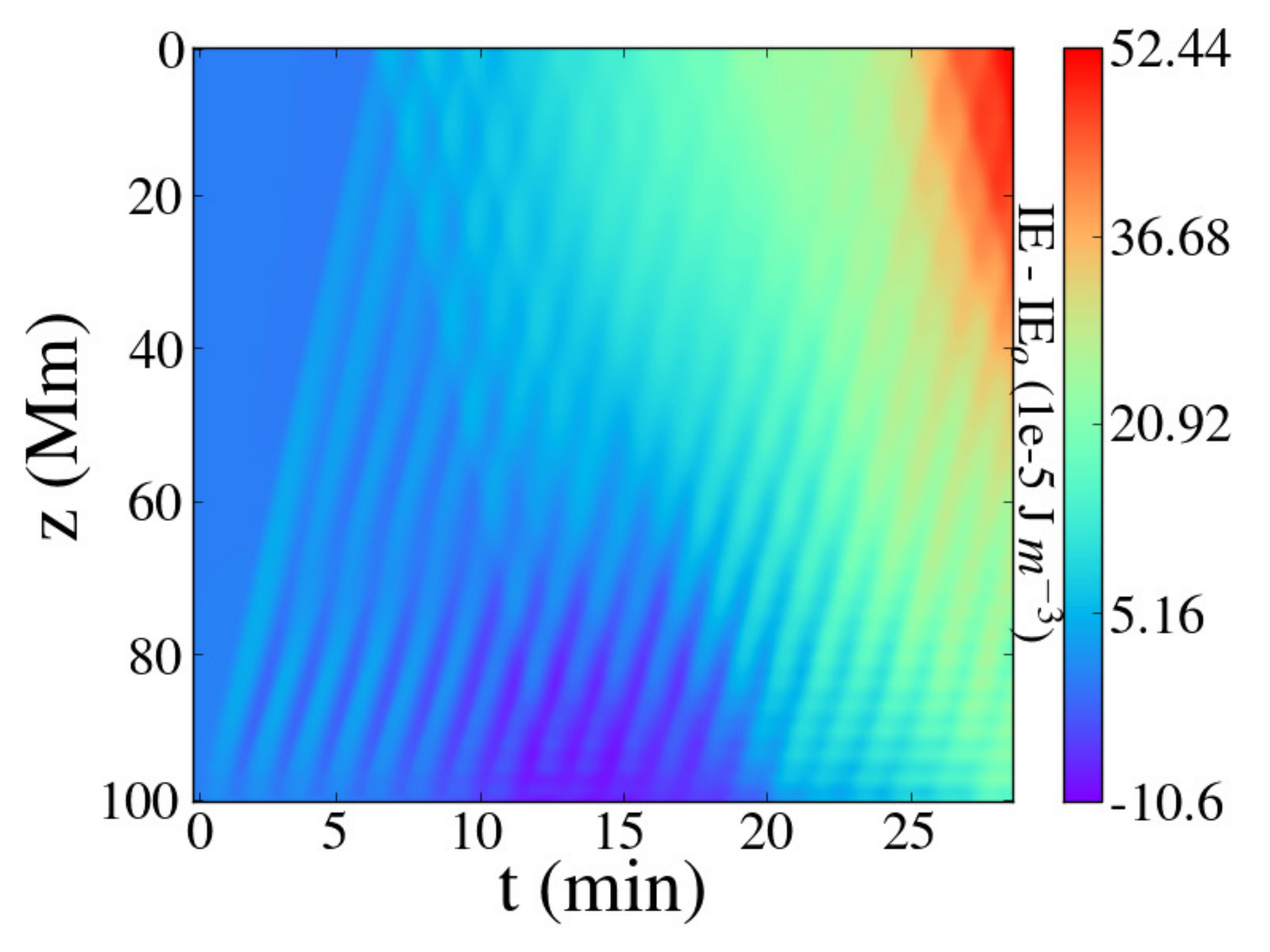}
\includegraphics[trim={0cm 0cm 0cm 0cm},clip,scale=0.16]{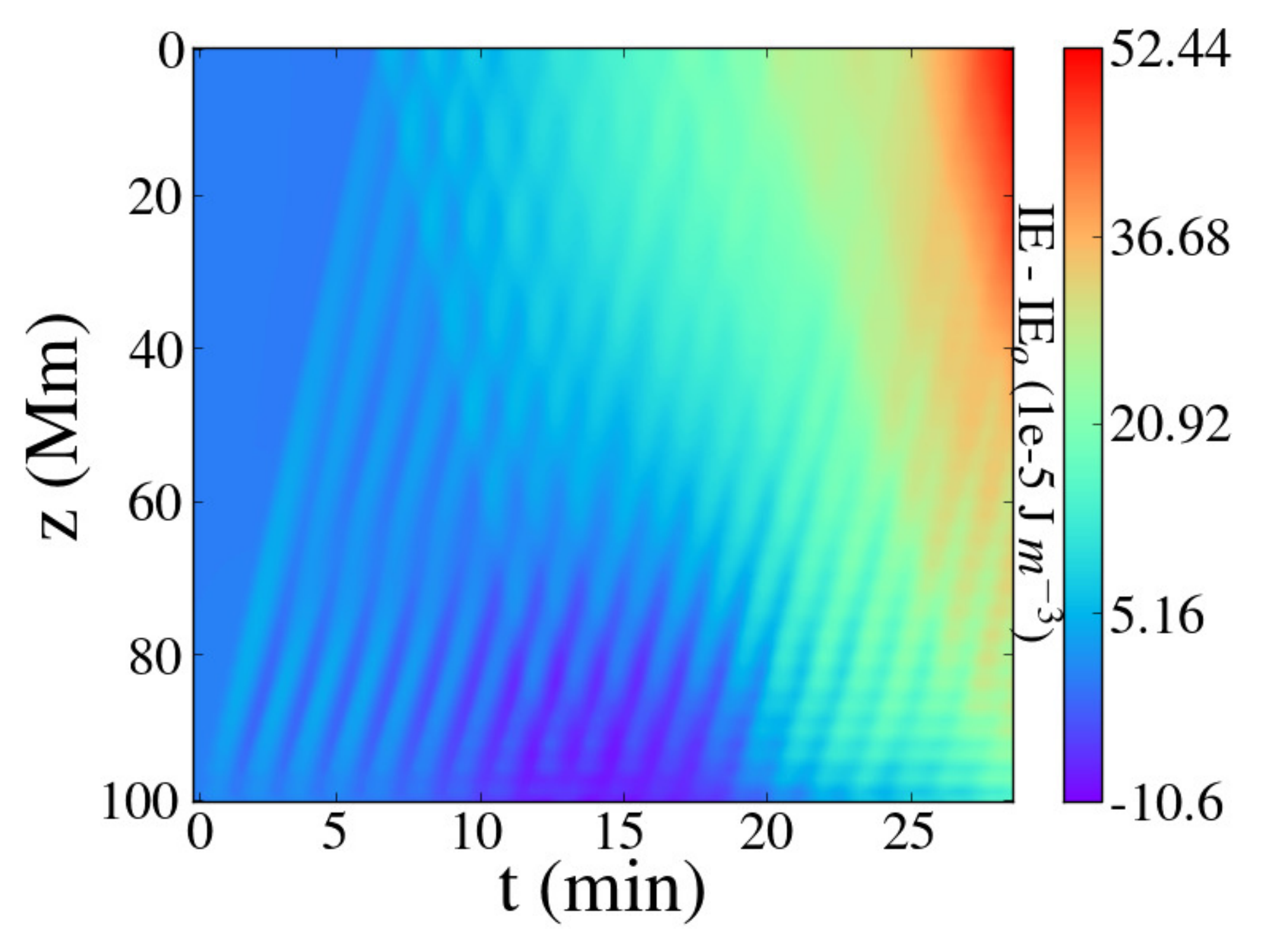}}
\resizebox{\hsize}{!}{\includegraphics[trim={0cm 0cm 0cm 0cm},clip,scale=0.16]{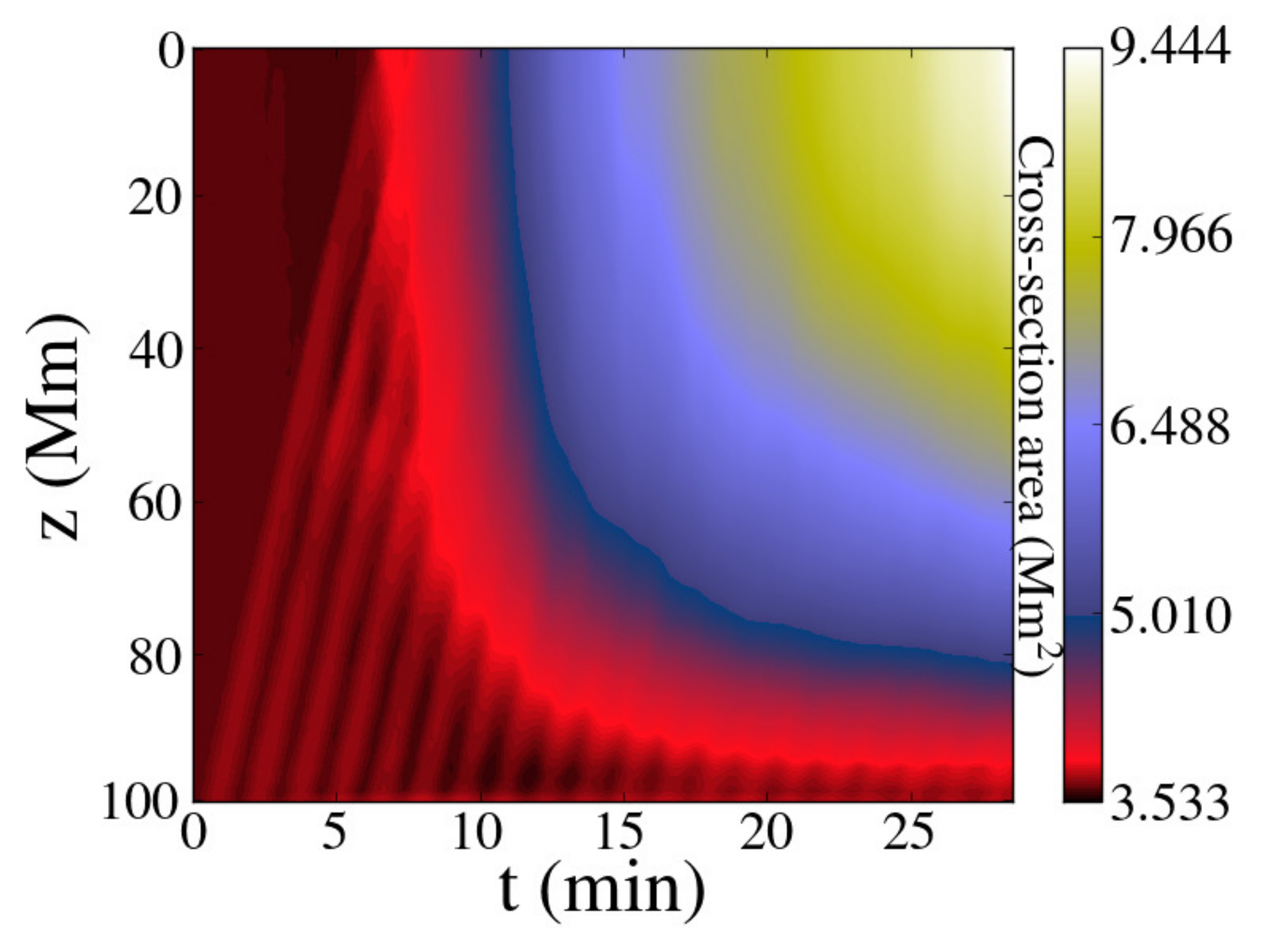}
\includegraphics[trim={0cm 0cm 0cm 0cm},clip,scale=0.16]{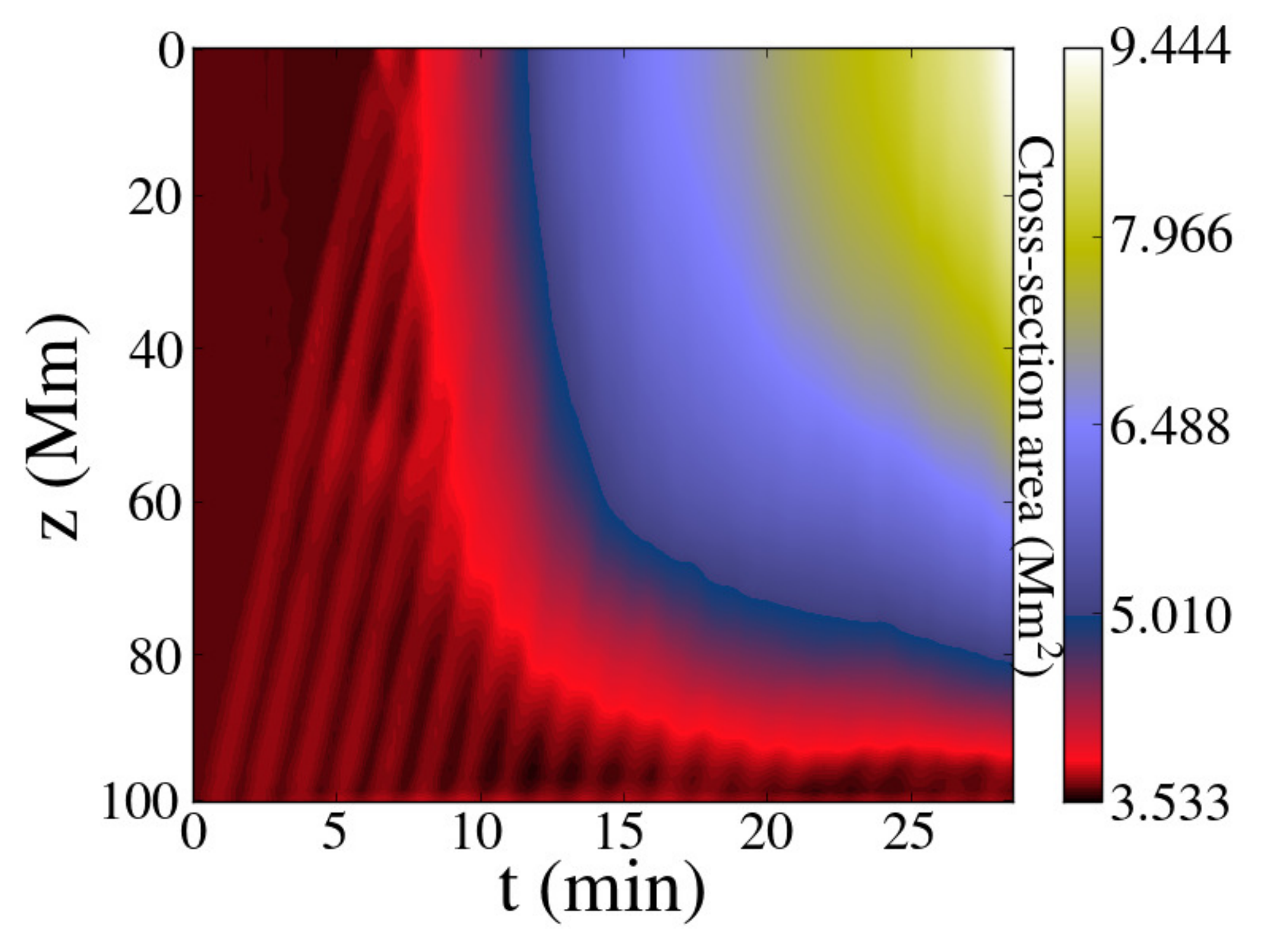}
\includegraphics[trim={0cm 0cm 0cm 0cm},clip,scale=0.16]{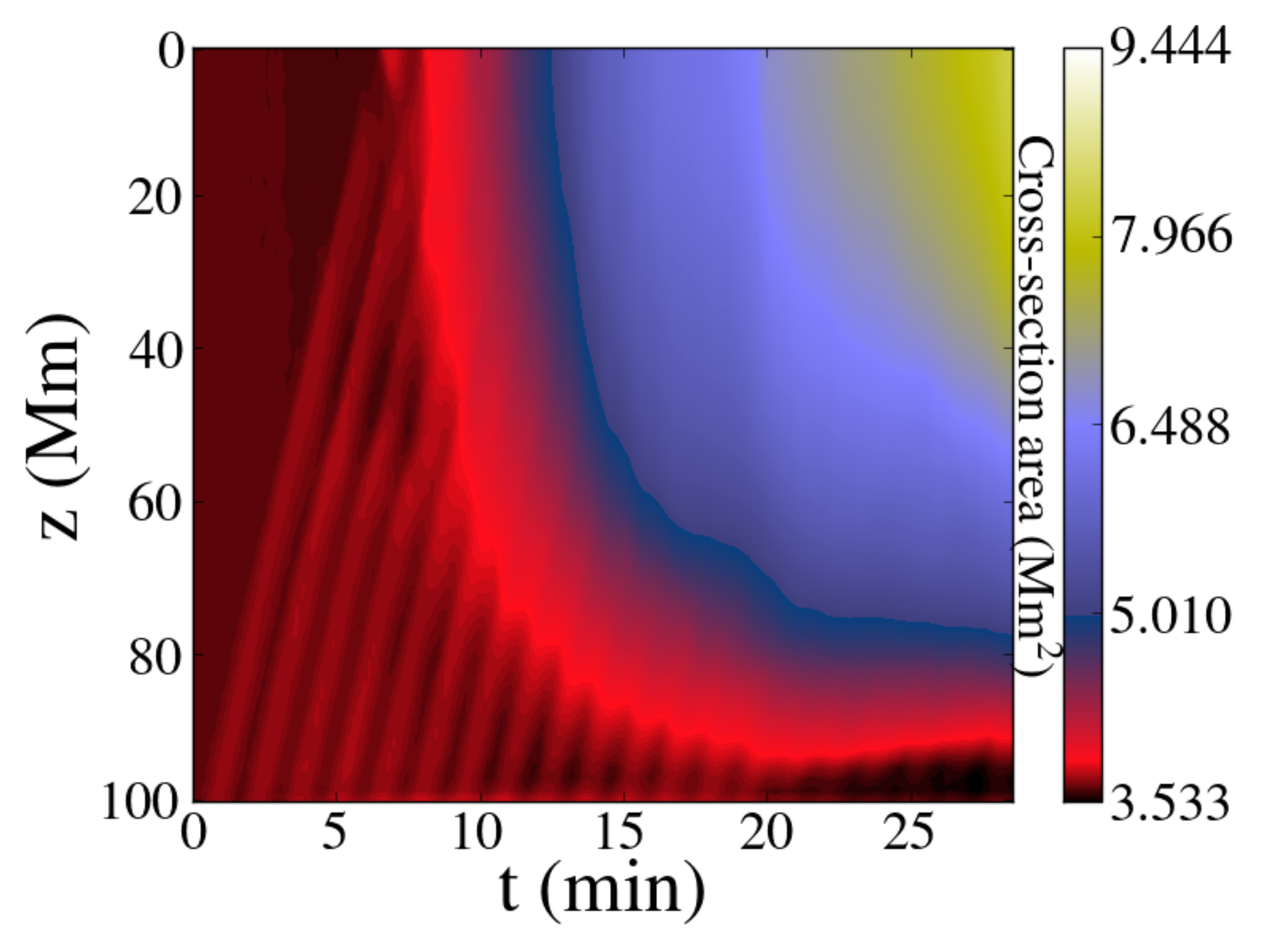}}
\resizebox{\hsize}{!}{\includegraphics[trim={0cm 0cm 0cm 0cm},clip,scale=0.16]{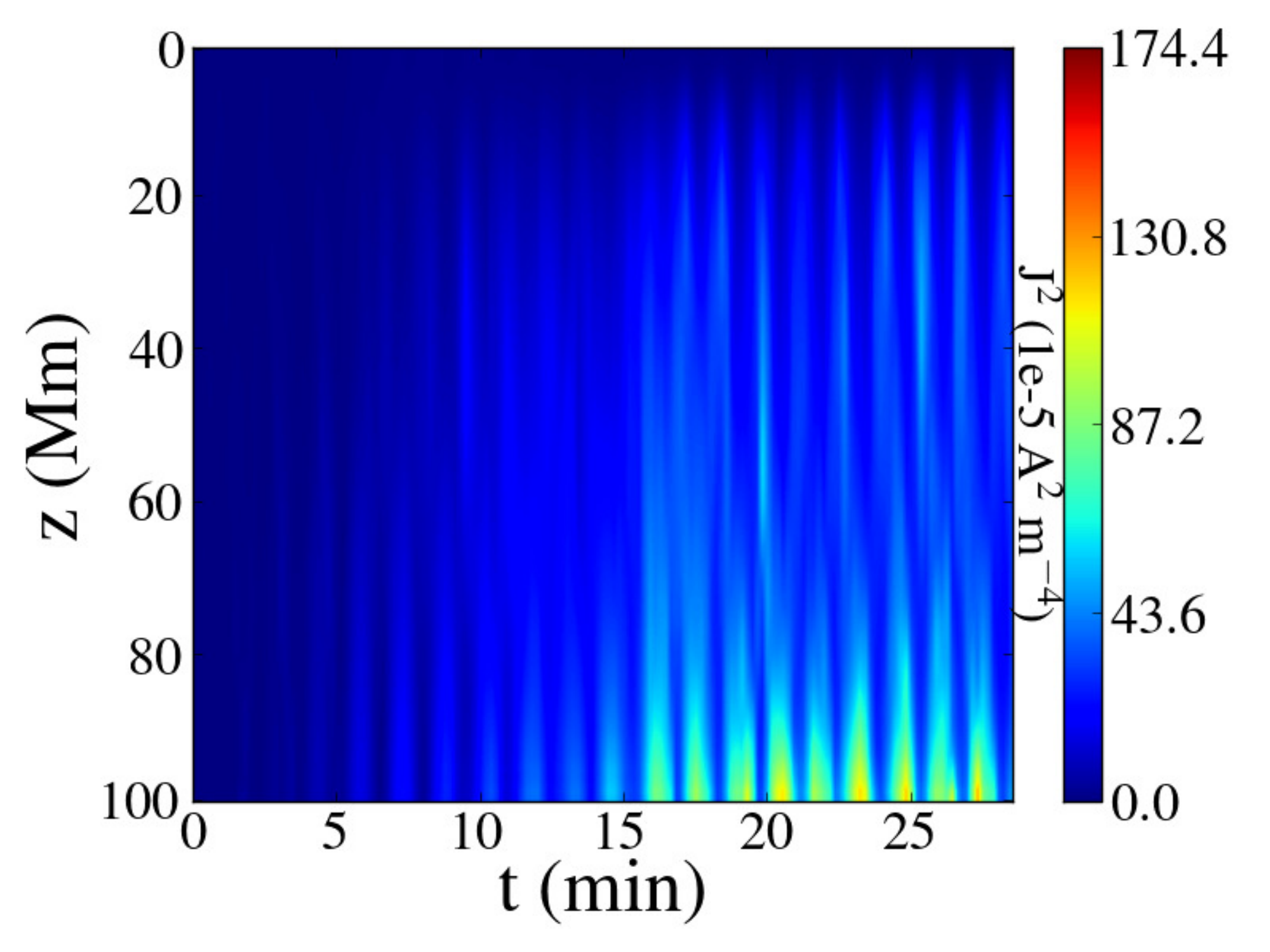}
\includegraphics[trim={0cm 0cm 0cm 0cm},clip,scale=0.16]{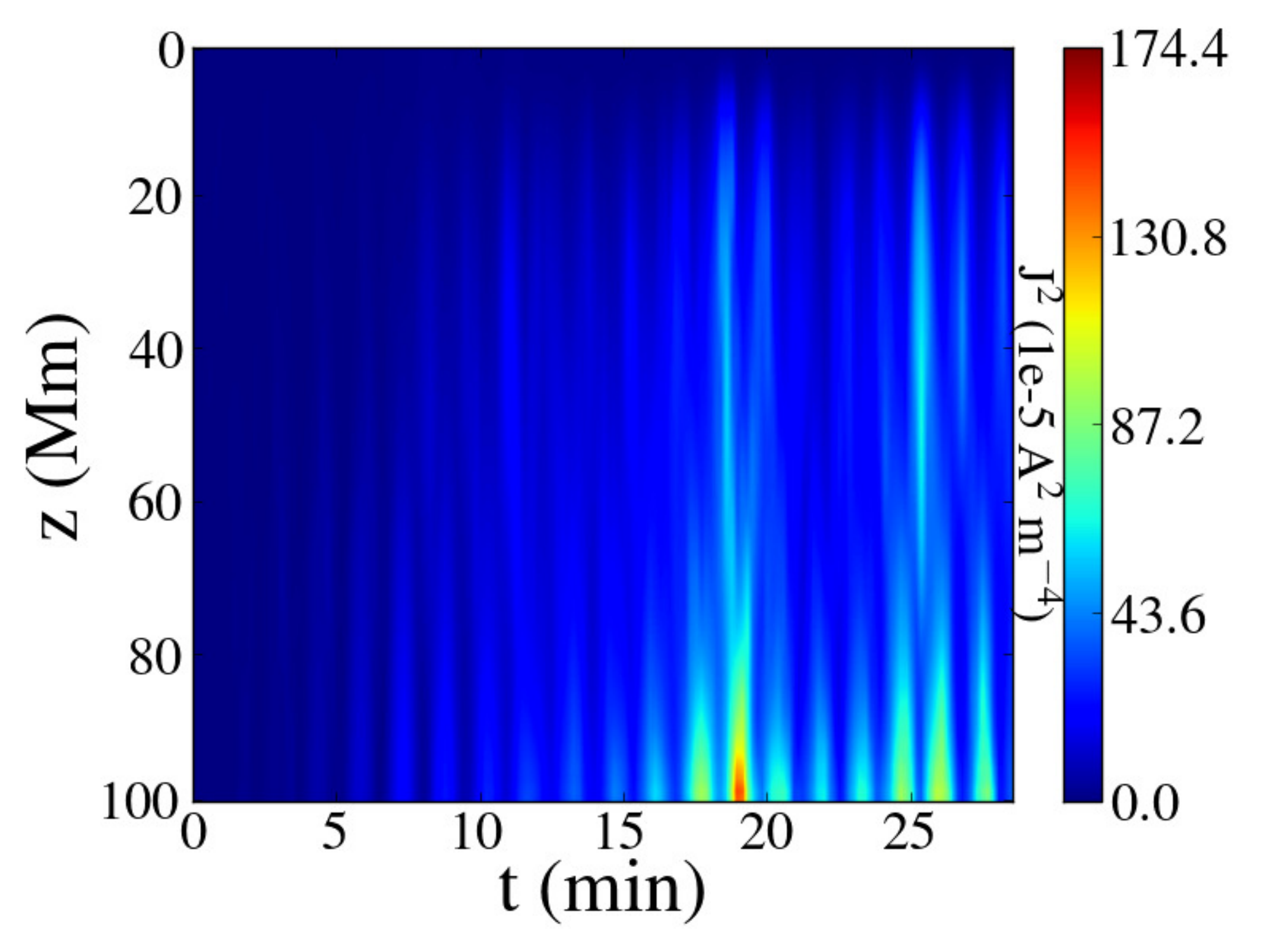}
\includegraphics[trim={0cm 0cm 0cm 0cm},clip,scale=0.16]{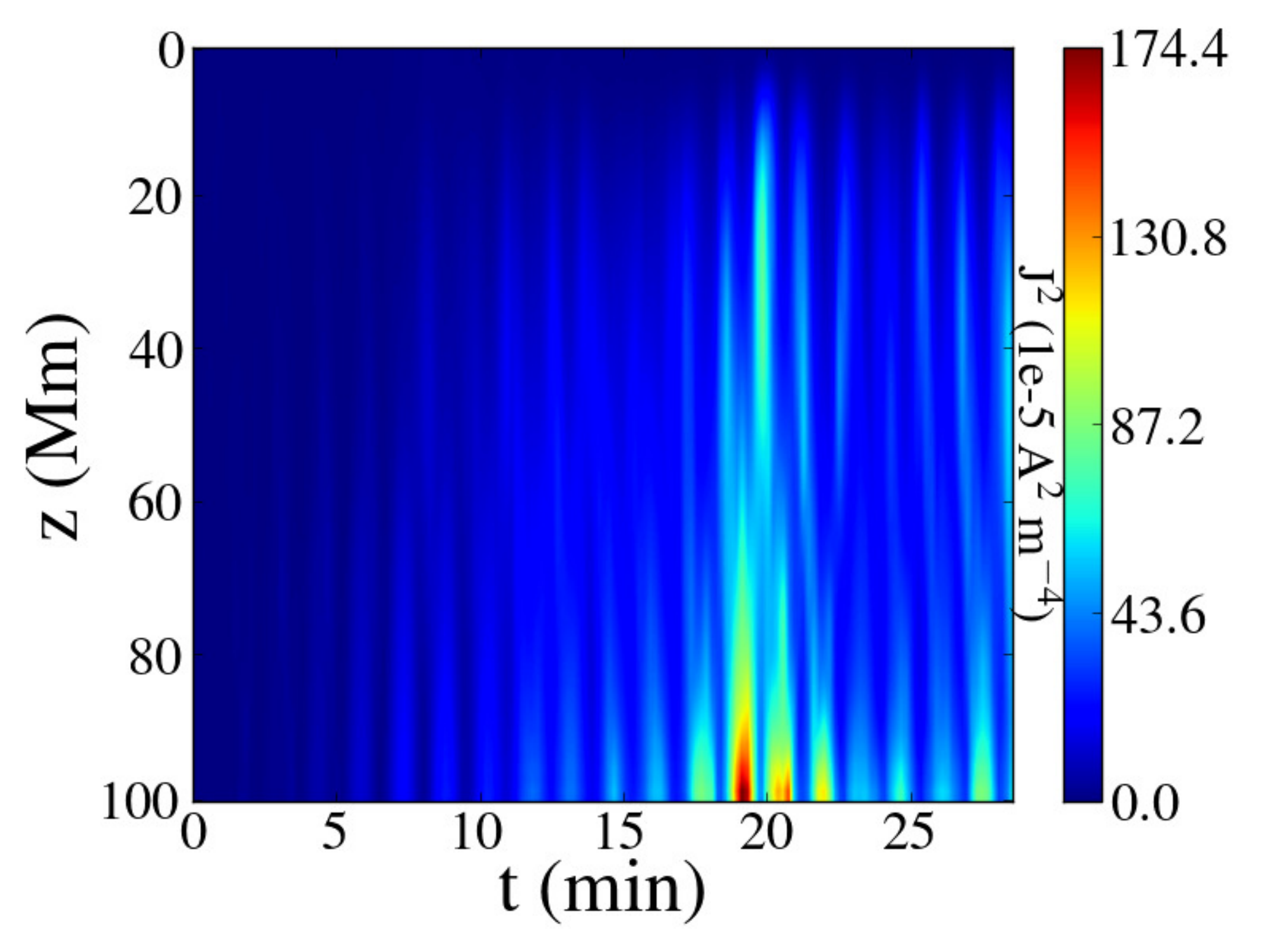}}
\caption{From top row to bottom: Average values of temperature or the whole domain, internal energy variation for the whole domain, flux tube cross section surface area for $\rho \geq 0.9 \, f(z)$, and square current densities (for $\rho \geq 0.9 \, f(z)$) along the z-axis and over time. From left to right, the cases for ideal MHD, resistive MHD, and viscous MHD (models ColdI, ColdR and ColdV) are shown. The apex is located at $z=0$.}\label{fig:456zt}
\end{figure*}

In order to find the location of energy dissipation for models ColdI, ColdR, and ColdV, we plot in Fig. \ref{fig:456zt} the temperature profiles along the $z$-axis, over time, for the aforementioned models. In the initial stages of the simulation, we observe a small temperature increase propagating from the footpoint towards the apex. These paths are attributed to slow waves initiated by the driver, which travel along the loop axis. Once the KHI manifests, the mixing between the colder loop and the hot corona drops the average temperature of our domain. This drop is more prominent at the quarter length of the flux tube, where both resistive and viscous heating are expected to have a lesser effect than at the footpoint and apex, respectively. Energy dissipation in that area is not strong enough to counter the apparent temperature drop due to the mixing, which becomes stronger at  later stages of the simulation. However, a temperature increase is observed near the footpoint and apex, as the simulations reach their final stages. This temperature increase becomes even stronger for the ColdR model, reaching its maximum values in the ColdV set-up. As we can observe, all three models show their strongest heating near the footpoint and minimum differences take place for the values of resistivity and viscosity.

Focussing on the ideal MHD case, we see that the internal energy density is increasing all along the flux tube, the highest values are found near the apex, and gradually lower values are found as we travel towards the loop footpoints. Considering the initial gradient of internal energy, we would intuitively expect an apparent drop in the average internal energy from the mixing of the different regions. This, however, is not observed. Instead, a constant increase of the internal energy density along the loop is observed over time. This increase is a combination of resistive and viscous dissipation due to numerical dissipation, as we have already seen in the UniT model. The highest values near the apex seem to contradict the results of \citet{tvd2007resist}, where it was proved that resistive heating should be the strongest for the fundamental standing mode of transverse oscillations. However, the higher observed values of temperature near the footpoint are still in agreement with that work. Looking at the flux tube surface area variation over time for model ColdI, we see that the loop is expanding. The highest values of the expansion are found near the apex where the cross-sectional surface area doubles in size as a consequence of the TWIKH rolls. Therefore, we can conclude that the observed temperature increase is not an apparent phenomenon but the result of wave heating.

\begin{figure*}
\centering
\resizebox{\hsize}{!}{\includegraphics[trim={0cm 0cm 0cm 0cm},clip,scale=0.16]{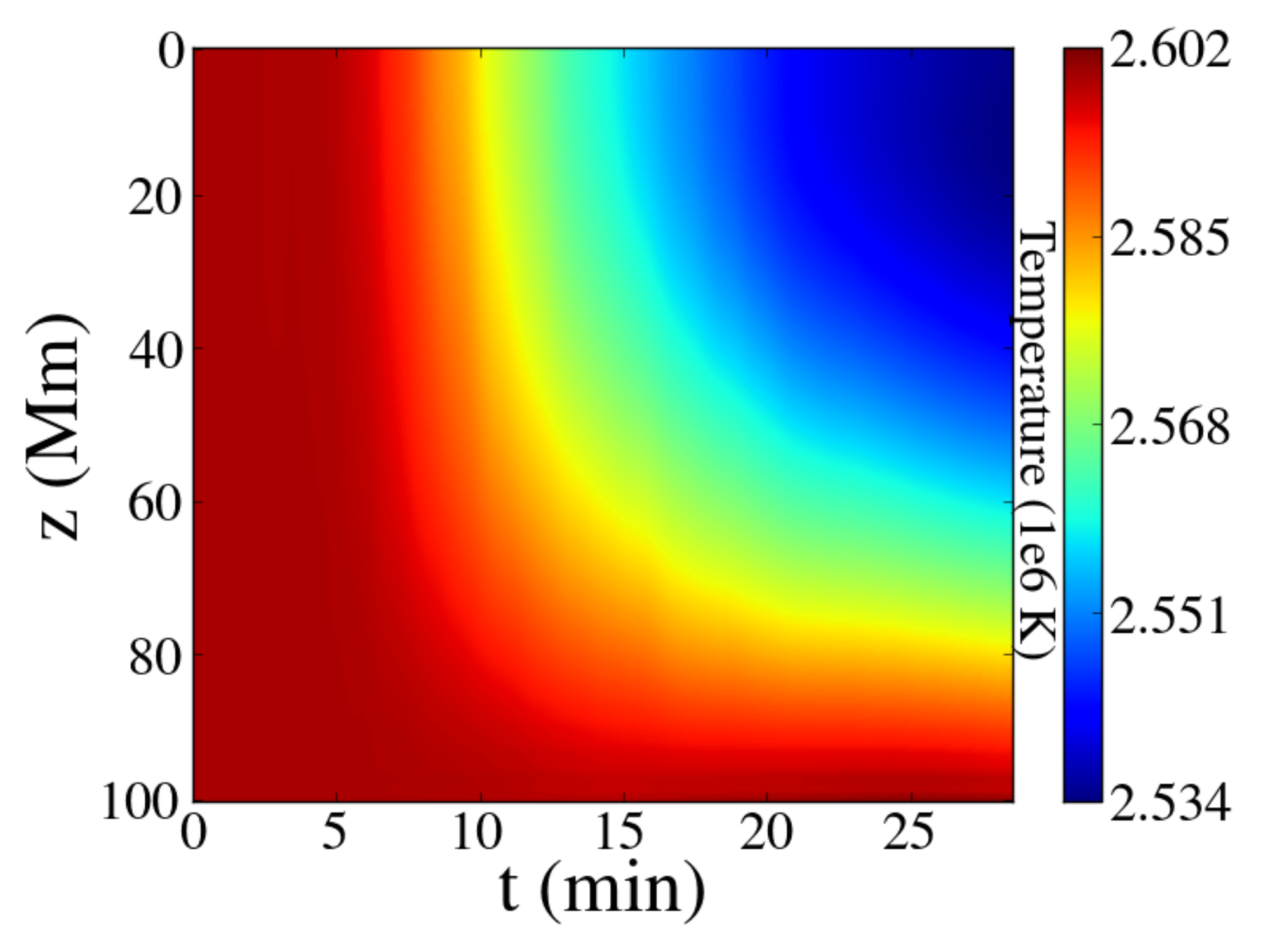}
\includegraphics[trim={0cm 0cm 0cm 0cm},clip,scale=0.16]{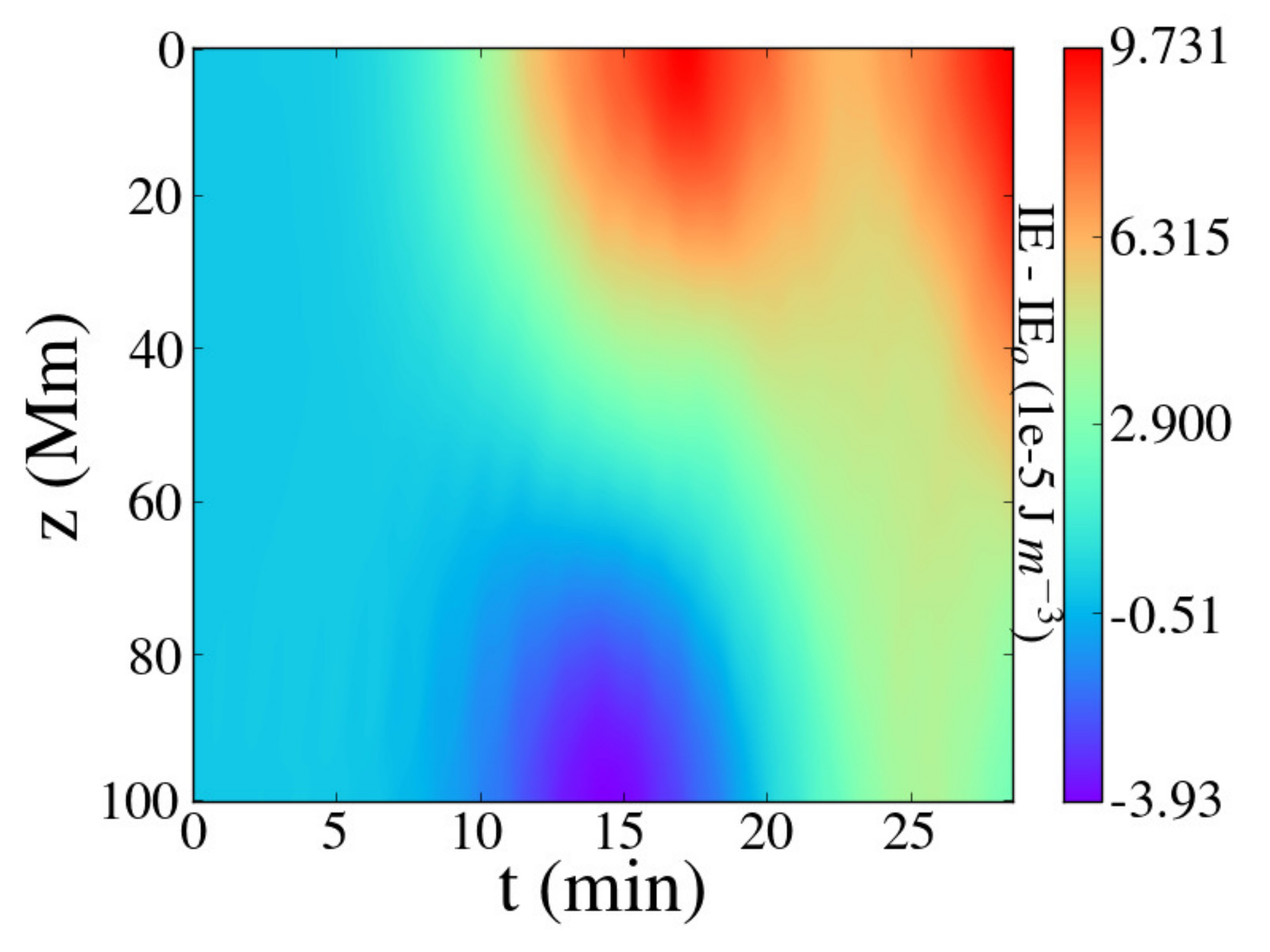}
\includegraphics[trim={0cm 0cm 0cm 0cm},clip,scale=0.16]{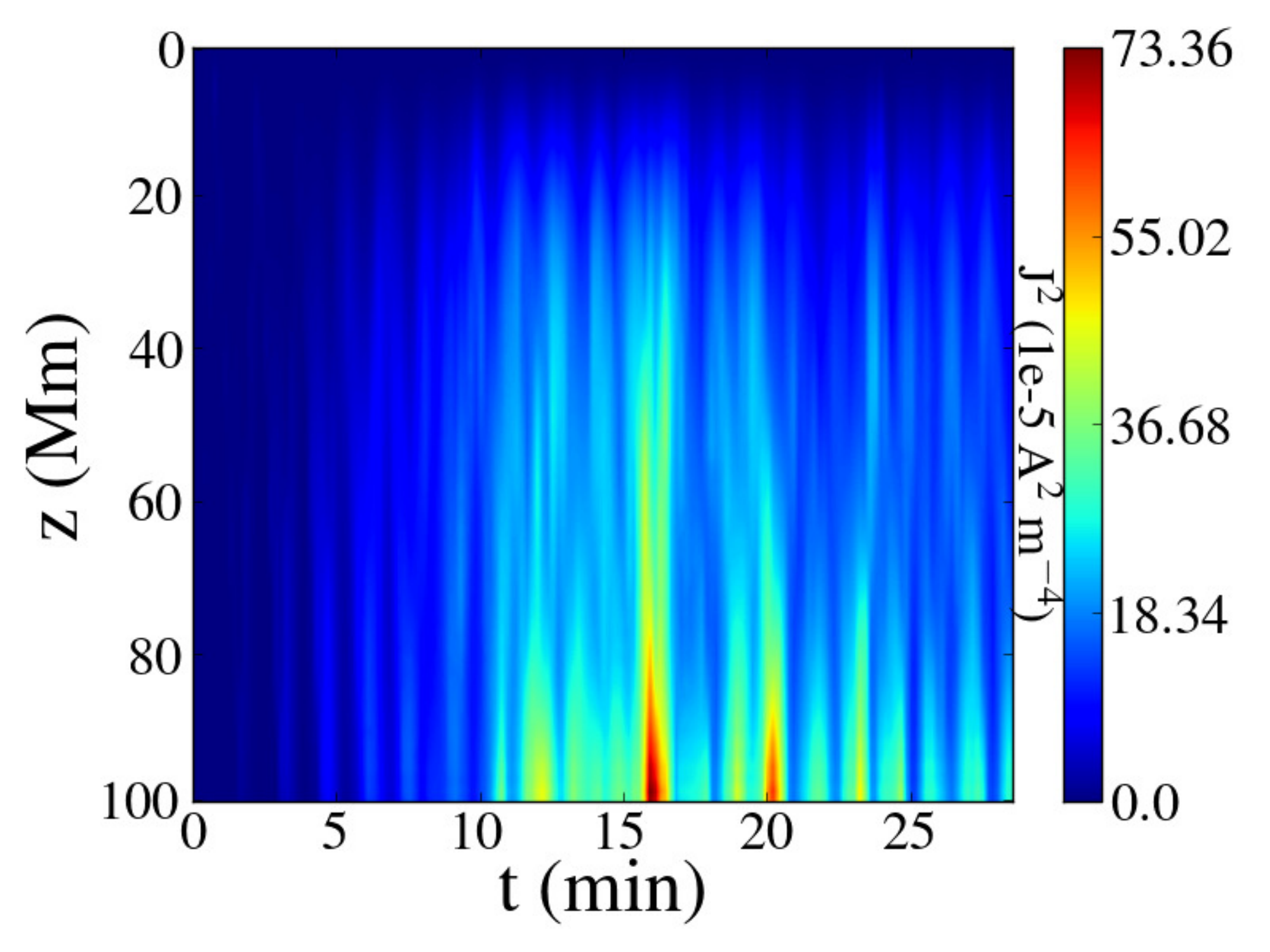}}
\caption{Top row: Average temperature (left) and internal energy (middle) of the entire $x-y$ plane and average square current density (right) of the flux tube (for $\rho \geq 0.9 \, f(z)$) per height and over time. Bottom row: Temperature (left), internal energy (middle), and density (right) at the $x-y$ plane at the apex ($z=0$). Data depict the ColdIngr model. The period of the driver is $P \simeq 172$ s.}\label{fig:ngr}
\end{figure*}

In our past study of a cold loop inside a hotter corona without gravity \citep{karampelas2017}, the mixing effects were effectively masking the results of energy dissipation in that set-up and the average temperature of our domain drops as a result of the cold loops expansion. In the present study, we reproduce the same results for the ColdIngr model, which are shown in Fig. \ref{fig:ngr}. The square current density again shows higher values near the footpoint, and an increase of the internal energy is again observed along the loop axis. The heating due to the driver generated propagating slow waves that were observed in the stratified case are not prominent in this set-up, which produces only very slight changes in the initial state of the simulations. The temperature shows a slight increase near the footpoint from ohmic dissipation due to numerical dissipation. However, the average temperature shows an apparent drop as we move higher up the loop as a consequence of TWIKH rolls developing in our domain. From our current results, we see that the introduction of gravity leads to a more complex evolution of the average temperature.

In the gravitationally stratified models of ColdI, ColdR, and ColdV, we observe a fluctuation of internal energy near the footpoint. The same fluctuation is clearer in the ColdIngr model, where we also have signs of a low frequency periodic fluctuation of the internal energy at the apex. This periodic fluctuation at the apex was also observed in \citet{magyar2016damping} for impulsive standing oscillations in coronal loops, and is associated with the ponderomotive force on loops performing standing oscillations \citep{terradasofman2004ApJ}. This perturbation is comparable to the effects of phase mixing the ColdIngr model. However, its effects are quickly negated by wave heating in the gravitationally stratified cases, while the overall dynamics seem to remain unaffected.

Looking again at models ColdR and ColdV (Fig. \ref{fig:456zt}), we see that the highest internal energy is achieved by the viscous case, and both models show stronger heating than the ideal case. The spatial and temporal profile of the internal energy is still the same as in ideal MHD, and there are very small differences between the three models for the values of resistivity and viscosity that we used in this work. These differences are the result of the dissipation parameters on the dynamical evolution of the oscillating loops. The higher temperatures at the apex for the viscous case are what we expected from our past work. Near the footpoint, we would expect the resistive case to lead to the highest temperature increase, since the square current densities (dominated by $J_z^2$) have their highest values there for all three models. The viscous case also shows higher average temperatures there because of the shrinking of the tube cross section, as observed in the $zt$ profile for the tube surface area of the ColdV  model. This shrinking of the cross section of the cold loop increases the contribution of the hot corona in the calculation of the average temperature. This is combined with the resistive heating due to numerical dissipation, resulting in the apparent effect of higher average temperature than in the ideal MHD model.

Another interesting result that we obtain for models ColdR and ColdV are the evolution of the currents. As mentioned before, the $J^2$ has the highest values near the footpoint ($z=100$ Mm) for all three models because of the strong $J_z$ currents there. Both the resistive and viscous case show on average a reduced amount of currents, and there are some temporary high values higher up the loop. The spikes in current densities and the reduced amount of ambient low current densities, unlike the ideal case, is similar in the dynamical evolution of these systems. This similarity leads to the conclusion that when we have comparable values for the Reynolds and magnetic Reynolds number in a system, increased resistivity can act as a form of turbulent viscosity and viscosity can disrupt the development of smaller scales and currents comparable to a form of anomalous resistivity. 

\subsection{Heating in very viscous cold loops ($R_e=10^2$)}
The temperature evolution of the ColdV2 model is a special case that needs to be examined separately. We have already seen in Fig. \ref{fig:TWIKH} that the very high value for shear viscosity ($R_e=10^2$) resulted in a complete suppression of the KHI in that loop. This suppression eventually leads to a different behaviour for temperature. In Fig. \ref{fig:viscous} we plot the average temperature and average internal energy per height and over time, over the entire $x-y$ plane. We observe an increase of the average temperature towards the apex, with a maximum value of $2.4\times 10^4$ K. These values are higher than the corresponding values for the ColdV model in Fig. \ref{fig:456zt}, and are the result of the very high values of the shear viscosity. We also observe a heating of around $10^3 - 10^4$ K near the footpoint. These values are lower than those for the ColdI model, and are caused by the resistive effects of numerical dissipation. The smaller values of currents inside the flux tube ($\rho \geq 0.9 \, f(x,y,z)$), as shown in Fig. \ref{fig:viscous}, can explain the lower values for temperatures.

The maximum values for temperature at the apex ($\Delta T = 0.298\times 10^6$ K) are near the boundary of the loop and in the wake behind the oscillating loop, and are caused by the viscous dissipation of energy. This can be seen in the second row of panels in Fig. \ref{fig:viscous} for the cross section at the apex. The heating inside the loop is far less ($\Delta T = 3.2\times 10^4$ K) and is the effect of energy dissipation, since we observe no mixing with the surrounding plasma.

The average internal energy per height and over time of Fig. \ref{fig:viscous} show a similar profile to the corresponding profiles for models ColdI, ColdR, and ColdV. However, the values of the average internal energy are lower than in those cases. This is caused by the lack of any TWIKH rolls for the ColdV2 model, which reduces the amount of smaller scales developing in our loop. Thus the energy dissipation is hindered, and is now confined near the loop boundary layer and the wake that is created behind the oscillating flux tube. The agreement of position between the temperature and internal energy prove that the wave heating is predominately a result of dissipation.

\subsection{Energy profiles and heating rate}
In our attempt to study the energy evolution within our system for models ColdI, ColdR, and ColdV, we need to first calculate the energy fluxes from all the boundaries, including the energy input provided by the driver, and the energy densities. We start from the MHD equation of energy 
\begin{equation}\label{eq:energymhd}
\dfrac{\partial e}{\partial t} = \nabla\cdot \left[ \left( e + p_{tot} \right)\vec{\upsilon} - \dfrac{\vec{B}\vec{B}}{\mu_0}\cdot\vec{\upsilon} \right] = -\nabla\cdot \left( \eta \vec{J}\times\vec{B}\right) - \nabla \Phi\cdot \rho \vec{\upsilon},
\end{equation}
where the total pressure and gas pressure are given by
\begin{equation}
p_{tot}=p+\dfrac{B^2}{2\mu_0},
\end{equation}
\begin{equation}
p = (\gamma-1)\left( e-\dfrac{\rho\upsilon^2}{2}-\dfrac{B^2}{2\mu_0}\right), \qquad \gamma = \dfrac{5}{3}
,\end{equation}
where $\eta$ is the electrical resistivity, $\gamma$ is the ratio of specific heats, and $\mu_0$ is the permeability of vacuum. We rework eq. \eqref{eq:energymhd} and reach the following equation, in compact form:
\begin{equation}
K(t) + M(t) + I(t) + G(t) = S_{tot} + F_{tot}
,\end{equation}
where the kinetic ($K(t)$), magnetic ($M(t)$), internal ($I(t)$), and gravitational energy density variations ($G(T)$), the total ($\mathcal{E}(t)$) energy density variation and energy density variation due to Poynting flux ($S_{tot}$) and plasma flow ($F_{tot}$) through the boundaries are calculated as in \cite{belien1999ApJ} as follows:
\begin{equation}
K(t) = \dfrac{1}{V} \int_V \dfrac{1}{2} \rho(t)\upsilon(t)^2 dV' - \dfrac{1}{V} \int_V \dfrac{1}{2} \rho(0)\upsilon(0)^2 dV',
\end{equation}
\begin{equation}
M(t) = \dfrac{1}{V} \int_V \dfrac{B(t)^2}{2\mu_0}  dV' - \dfrac{1}{V} \int_V \dfrac{B(0)^2}{2\mu_0}  dV',
\end{equation}
\begin{equation}
I(t) = \dfrac{1}{V} \int_V \dfrac{1}{\gamma -1} p(t) dV' - \dfrac{1}{V} \int_V \dfrac{1}{\gamma -1} p(0) dV', 
\end{equation}
\begin{equation}
G(t) = \dfrac{1}{V} \int_V \rho(t)\Phi(t) dV' - \dfrac{1}{V} \int_V \rho(0)\Phi(0) dV',
\end{equation}
\begin{equation}
\mathcal{E}(t) = K(t) + M(t) + I(t) + G(t),
\end{equation}
\begin{equation}\label{eq:poynting}
S_{tot}=-\dfrac{1}{V}\int_0^t \int_A \left[ \eta \vec{J}\times\vec{B} - \left( \vec{\upsilon}\times \vec{B} \right)\times\vec{B}  \right]\cdot d\vec{A'}dt', 
\end{equation}
\begin{equation}
F_{tot}=-\dfrac{1}{V}\int_0^t \int_A \left( \dfrac{\rho \upsilon^2}{2} +\rho \Phi + \dfrac{\gamma}{\gamma-1}p \right)\vec{\upsilon}\cdot d\vec{A'}dt' 
.\end{equation}
The energy input from the driver is the component of eq. \eqref{eq:poynting} from the bottom boundary. The top boundary, which is the location of the apex, has practically zero average input because of the considered symmetry there; the same amount of energy "enters" and "leaves" the domain through that boundary. From eq. \eqref{eq:poynting} we see that the dominant terms regarding the input energy are the velocities, currents, and magnetic fields. This is a strong hint that once this values are initially the same, the differences in the input energy will be caused by the different dynamical evolution of our systems.

\begin{figure*}
\centering
\resizebox{\hsize}{!}{\includegraphics[trim={0cm 0cm 0cm 0cm},clip,scale=0.16]{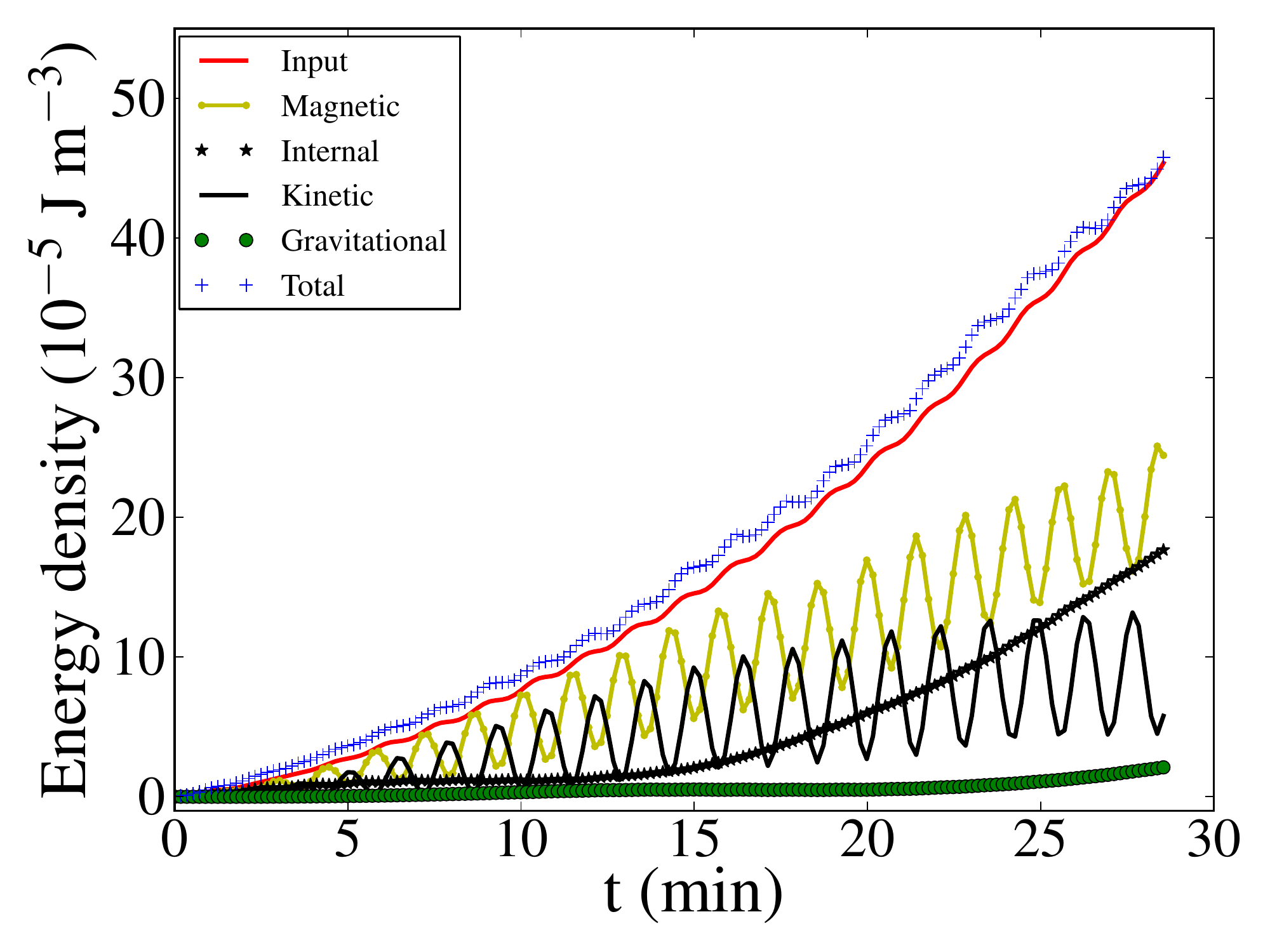}
\includegraphics[trim={0cm 0cm 0cm 0cm},clip,scale=0.16]{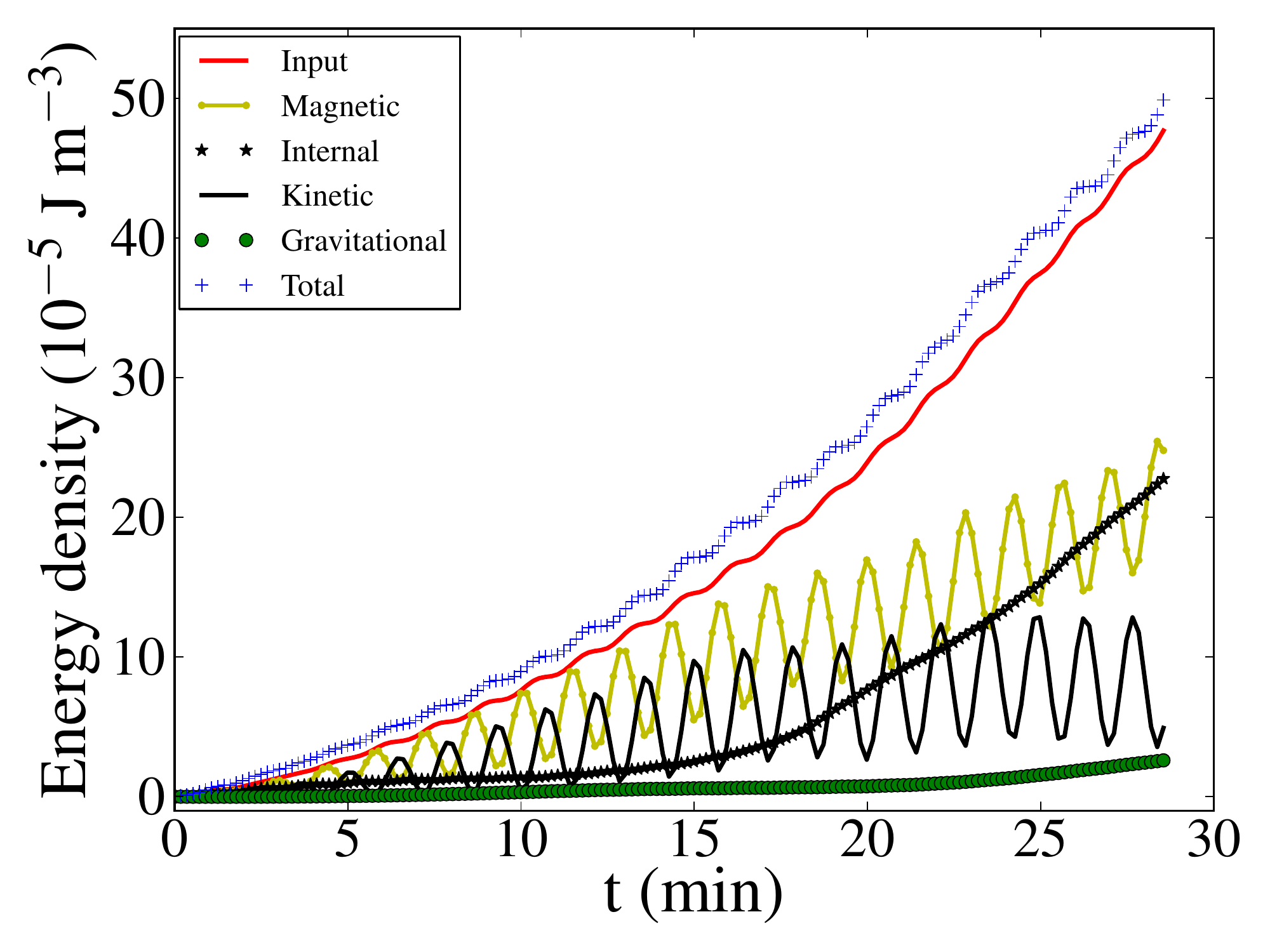}
\includegraphics[trim={0cm 0cm 0cm 0cm},clip,scale=0.16]{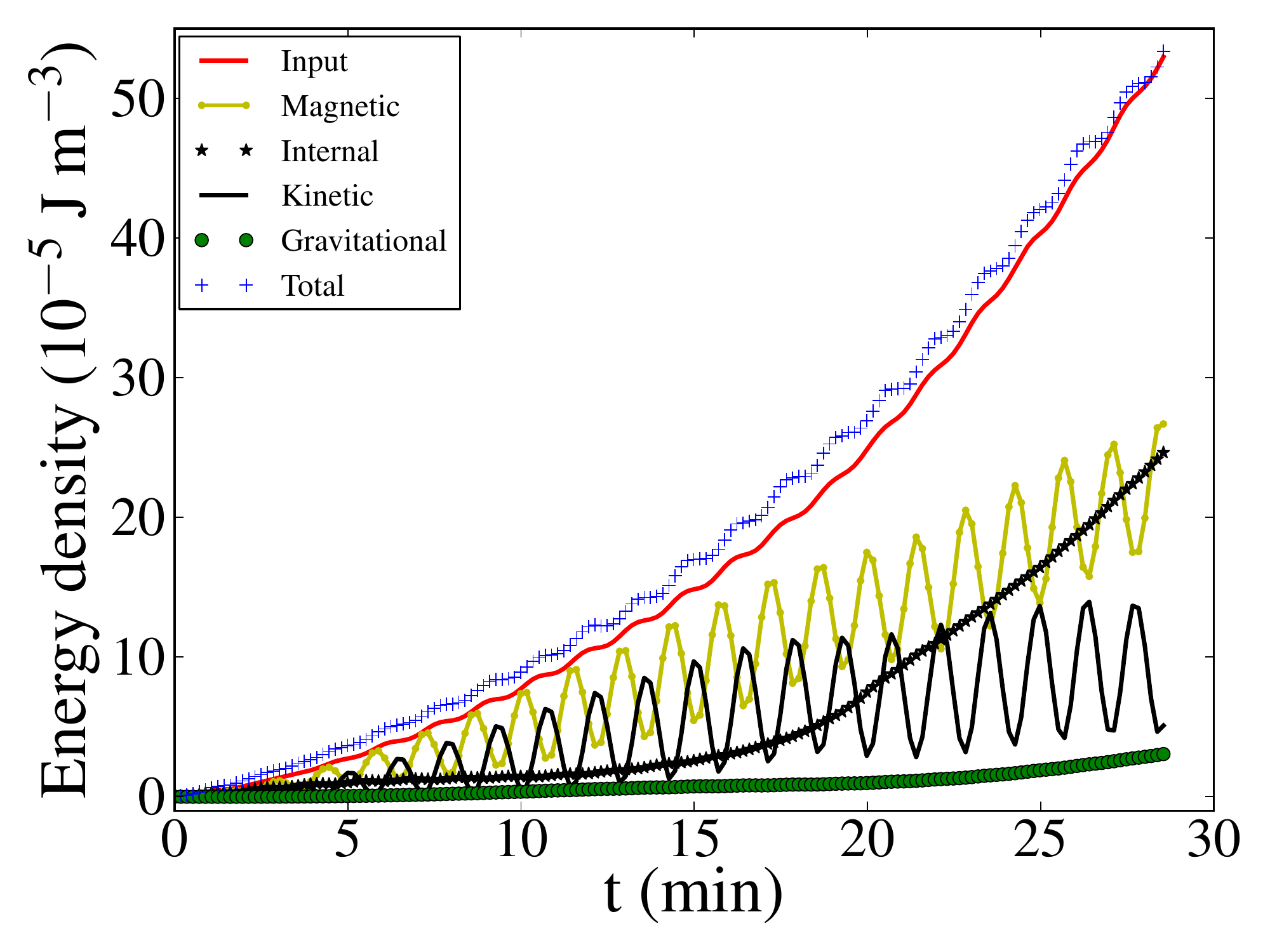}}
\resizebox{\hsize}{!}{\includegraphics[trim={0cm 0cm 0cm 0cm},clip,scale=0.16]{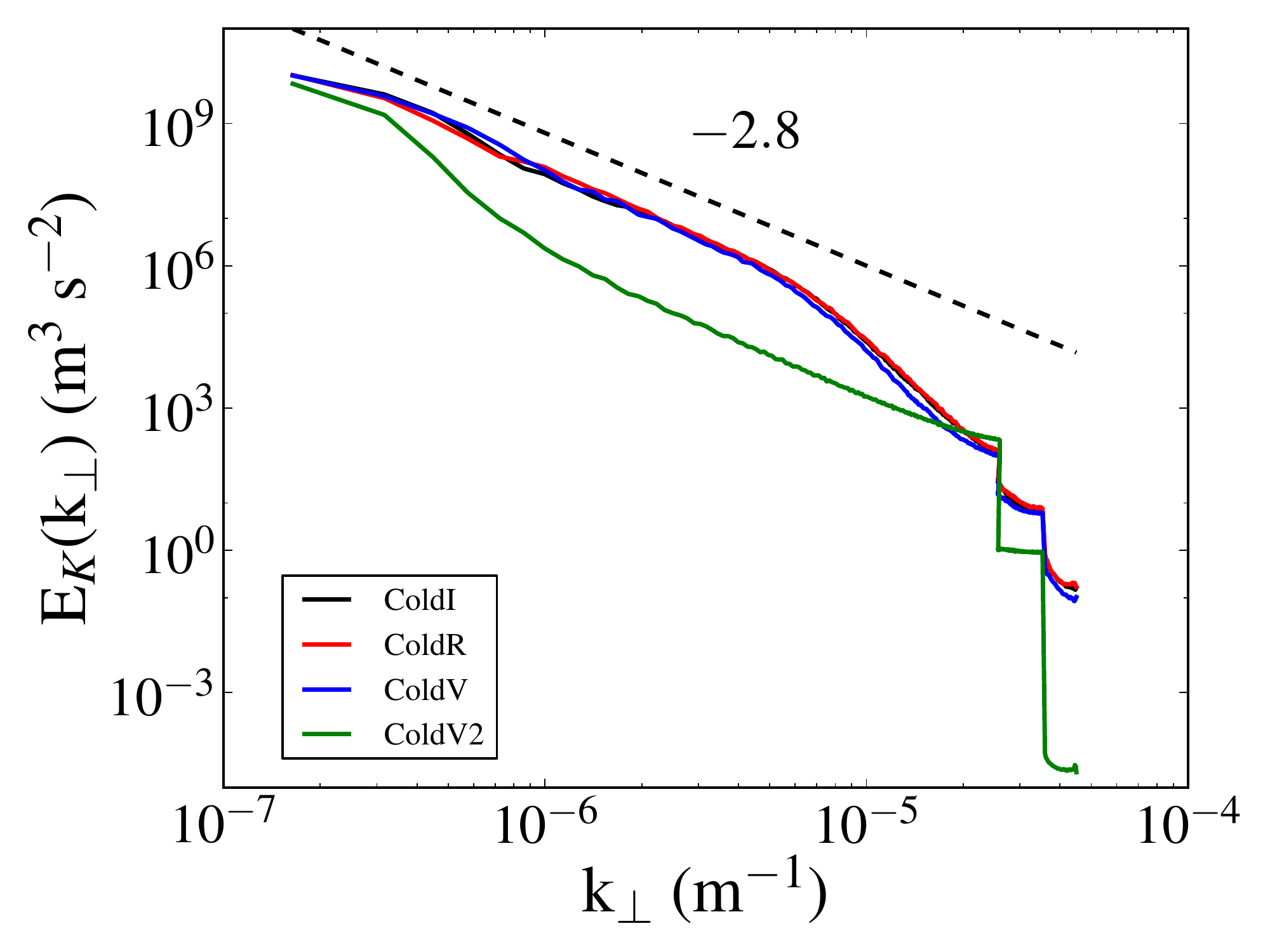}
\includegraphics[trim={0cm 0cm 0cm 0cm},clip,scale=0.16]{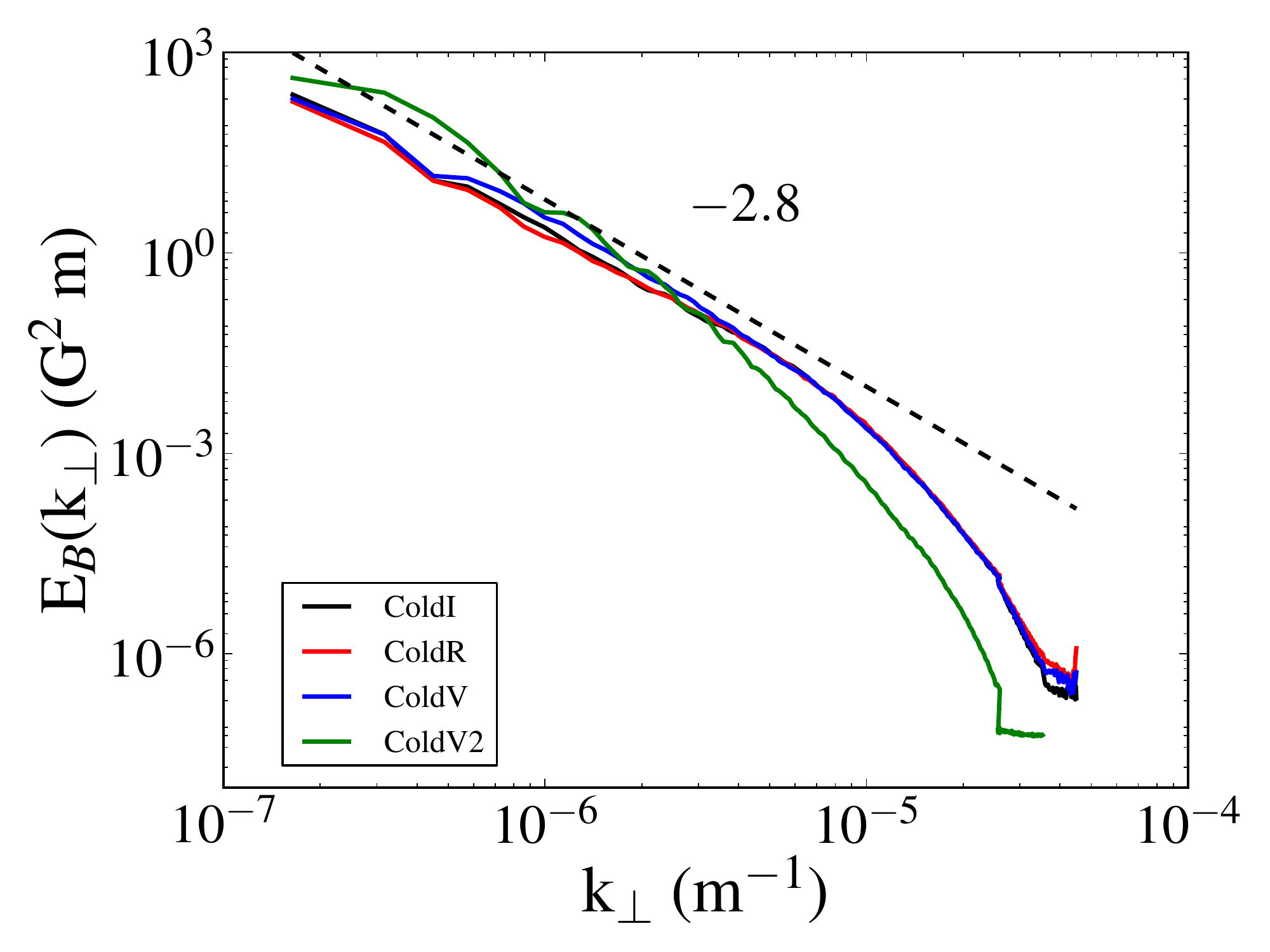}
\includegraphics[trim={0cm 0cm 0cm 0cm},clip,scale=0.16]{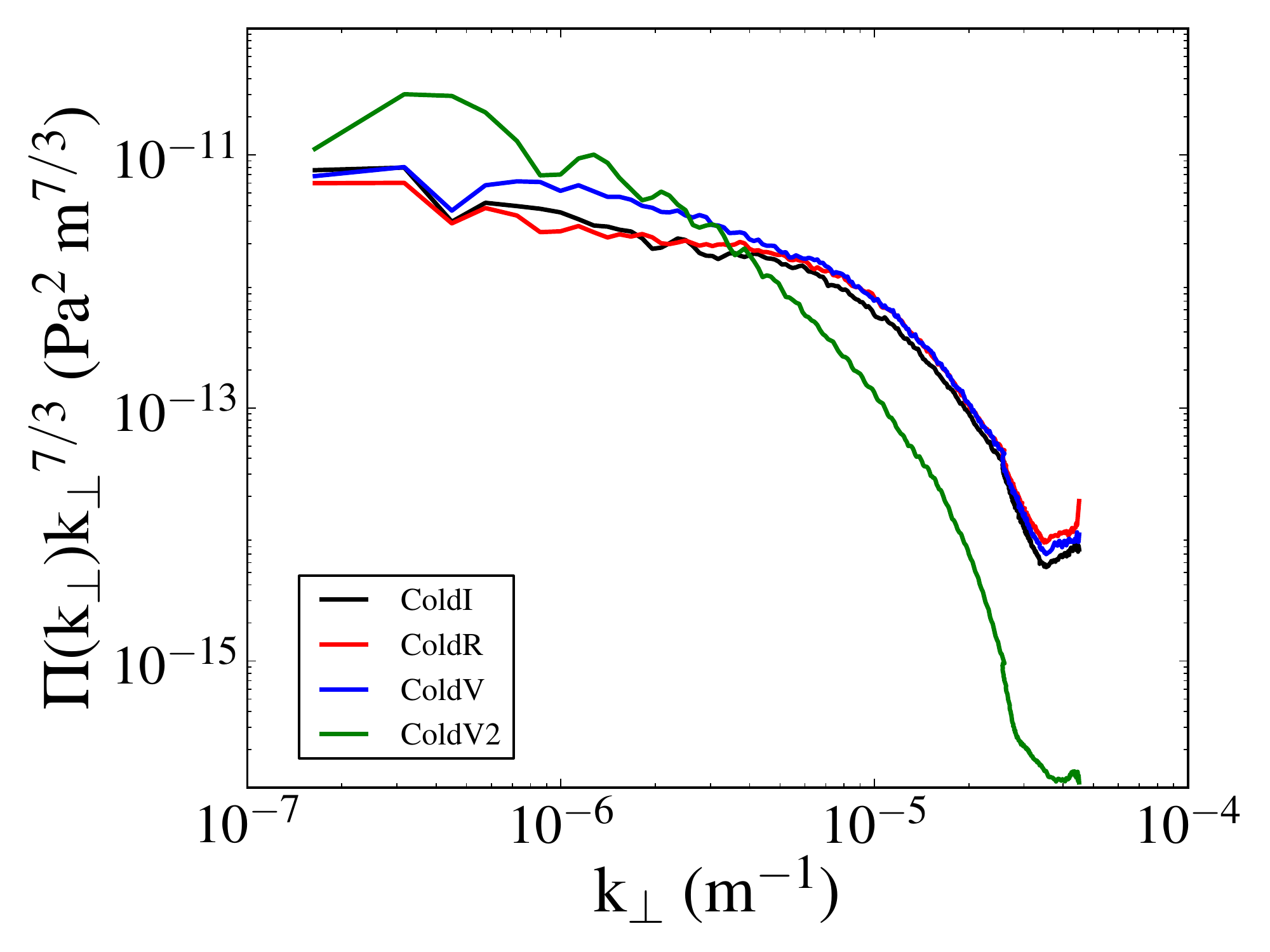}}
\caption{Top row: Time profiles for the internal, magnetic, kinetic, and gravitational energy density variations relative to the initial state, total (internal$+$magnetic$+$kinetic$+$gravitational) energy density difference, and energy density provided by the driver. All the quantities are volume averaged for the whole computational domain. From left to right: ColdI, ColdR, and ColdV models. The contributions from the Poynting flux and energy flux due to the plasma displacement through the side boundaries, have been incorporated to the magnetic and internal energy densities.
Bottom row: Time averaged $1D$ power spectra of kinetic energy, magnetic energy, and pressure at the apex, averaged over the last oscillation period, for the ColdI, ColdR, ColdV, and ColdV2 models.}\label{fig:energies}
\end{figure*}

In \citet{karampelas2017}, we plotted the energy density diagrams per time for the total, kinetic, internal, and magnetic energy density, alongside the energy input from the driver. There, we saw the rise of the internal and kinetic energy for the case of the driven oscillations. The observed drop in magnetic energy density there was attributed to the Poynting fluxes through the side boundaries, which had not been considered. In our current analysis, we calculated the energy variation due to Poynting flux ($S_{tot}$) and plasma flow ($F_{tot}$) through the side boundaries. After incorporating $F_{tot}$ into the internal energy density variation and $S_{tot}$ into the magnetic energy density variation in our domain, we plot the energy diagrams for the three models of the cold loops (models ColdI, ColdR, and ColdV) in Fig. \ref{fig:energies}. In the calculation of the energy density variations, we consider the entire simulation domain. Since the region of interest has a constant volume, the changes in the energy densities are directly translated into changes in the energies. From now on, we use the terms energy density/energy interchangeably, while discussing the results of Fig. \ref{fig:energies}. Finally, we need to stress an important factor in our analysis. By redefining the internal and magnetic energy variation so that they include the fluxes through the side boundaries, we are essentially calculating the contribution of the input energy to the magnetic and internal energy in each model. 
  
Starting from the magnetic energy (minus the Poynting fluxes from the side boundaries), we observe a similar and relatively  steady linear growth in all three models. Part of the driving energy is used for increasing the energy of the magnetic field. The highest values of the input energy are observed for the viscous set-up, and its lowest for the ideal MHD set-up. The reasons for that are the slight differences in the magnetic field (and consequently the current density) at the foot point, as a result of the dynamical evolution of its system. Another reason that causes this difference is the different value for the electrical resistivity in model ColdR.   

Studying the kinetic energy in these three models, we observe an almost linear growth for a total of six periods, followed by decelerating growth until around the eighth period. After that, the kinetic energy seems to reach a saturation and only small variations of the average values are observed until the end of the simulations. This is an interesting result when combined with the evolution of the (redefined) internal energy. The contribution to the internal energy shows initially only a small and non-steady increase. However, once the kinetic energy enters the phase of decelerating growth and eventual saturation, the internal energy exhibits a rapid growth. Given the slower growth of magnetic energy compared to the growth of input energy in the three models, once the kinetic energy starts saturating, wave heating gets stronger. The saturation of kinetic energy takes place when the loop cross section becomes turbulent and smaller scales have developed. These smaller scales reinforce energy cascade which in turn leads to more efficient dissipation through (numerical and physical) resistivity and viscosity. Finally, the higher final values of the internal energy for the ColdR and ColdV models are connected to the corresponding higher values of the input energy for these set-ups. 

In all three models, we observe a similar increase of the gravitational energy. This slight increase is caused by the redistribution of plasma along the loop, where plasma is moving from the footpoint higher up the loop because of the evolution of the scale height. A small oscillation in the profile of the gravitational energy exists owing to the ponderomotive force, but its amplitude is significantly less than the overall increase, which is thus attributed to wave heating. Finally, once we take the energy fluxes through the side boundaries into account, the total energy variation in our domain is equal to the energy input from the driver. The small differences that are observed between the two quantities can be attributed to the accuracy of the calculations and the inevitable creation of small numerical errors ($\nabla\cdot\vec{B}\neq 0$). 

In Fig. \ref{fig:energies}, we also plot the time averaged $1D$ power spectra of kinetic energy, magnetic energy, and pressure at the apex, averaged over the last period. This is justified by the fact that small-scale generation is purely perpendicular to the mean magnetic field and is at its peak during the last period of the simulation. We used a similar approach to \citet{magyar2017}, where we used the \textit{python numpy} version of the $2D$ fft routine to calculate $\vert f_{k_{x}k_{y}} \vert^2$, where $f = \vec{\upsilon}$, $\vec{b}$ and $p$ is the Fourier transform of the velocities, magnetic field, and pressure. The power spectra of velocity, magnetic field, and pressure are then calculated by integrating over a unit bandwidth as follows: 
\begin{equation}
E_K(k_\perp) = \sum_{k_xk_y} |\vec{\upsilon}_{k_xk_y}|^2,
\end{equation}
\begin{equation}
E_B(k_\perp) = \sum_{k_xk_y} |\vec{b}_{k_xk_y}|^2,
\end{equation}
\begin{equation}
\Pi(k_\perp) = \sum_{k_xk_y} |p_{k_xk_y}|^2, \qquad k_\perp = \sqrt{k_x^2+k_y^2}
.\end{equation}
During the integration, we assumed isotropy and axisymmetry of the turbulence. The slope of the inertial range seen in the power spectra for the velocities and the magnetic field are steeper ($k_\perp^{-2.8}$) than the expected spectra of $k_\perp^{-5/3}$ for strong and $k_\perp^{-2}$ for weak incompressible turbulence. This deviation is most likely caused by our assumption of isotropy and homogeneity, which is not the case in our highly structured domain, as well as the inclusion of compressibility. The isotropy is also violated by the imposition of a directional flow from the continuous driving of the oscillation. The compensated power spectrum for pressure shows a closer proximity to the expected value of $k_\perp^{-7/3}$ in the inertial range, and also reveals the effects of dissipation on the length of the inertial range. Including resistivity and viscosity reduces the length of the inertial range. 

\begin{figure*}[t]
\centering
\resizebox{\hsize}{!}{\includegraphics[trim={0cm 0cm 0cm 0cm},clip,scale=0.16]{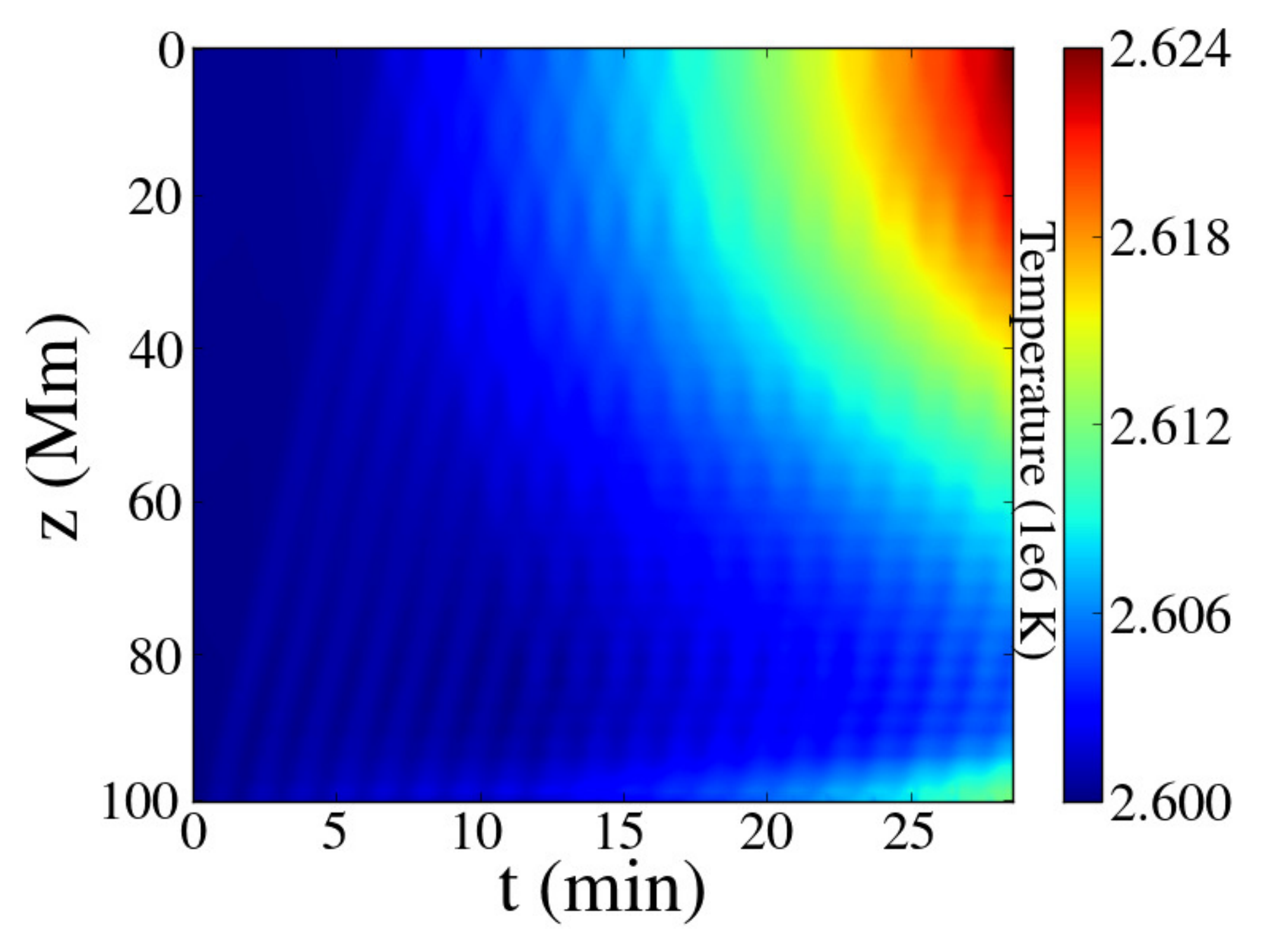}
\includegraphics[trim={0cm 0cm 0cm 0cm},clip,scale=0.16]{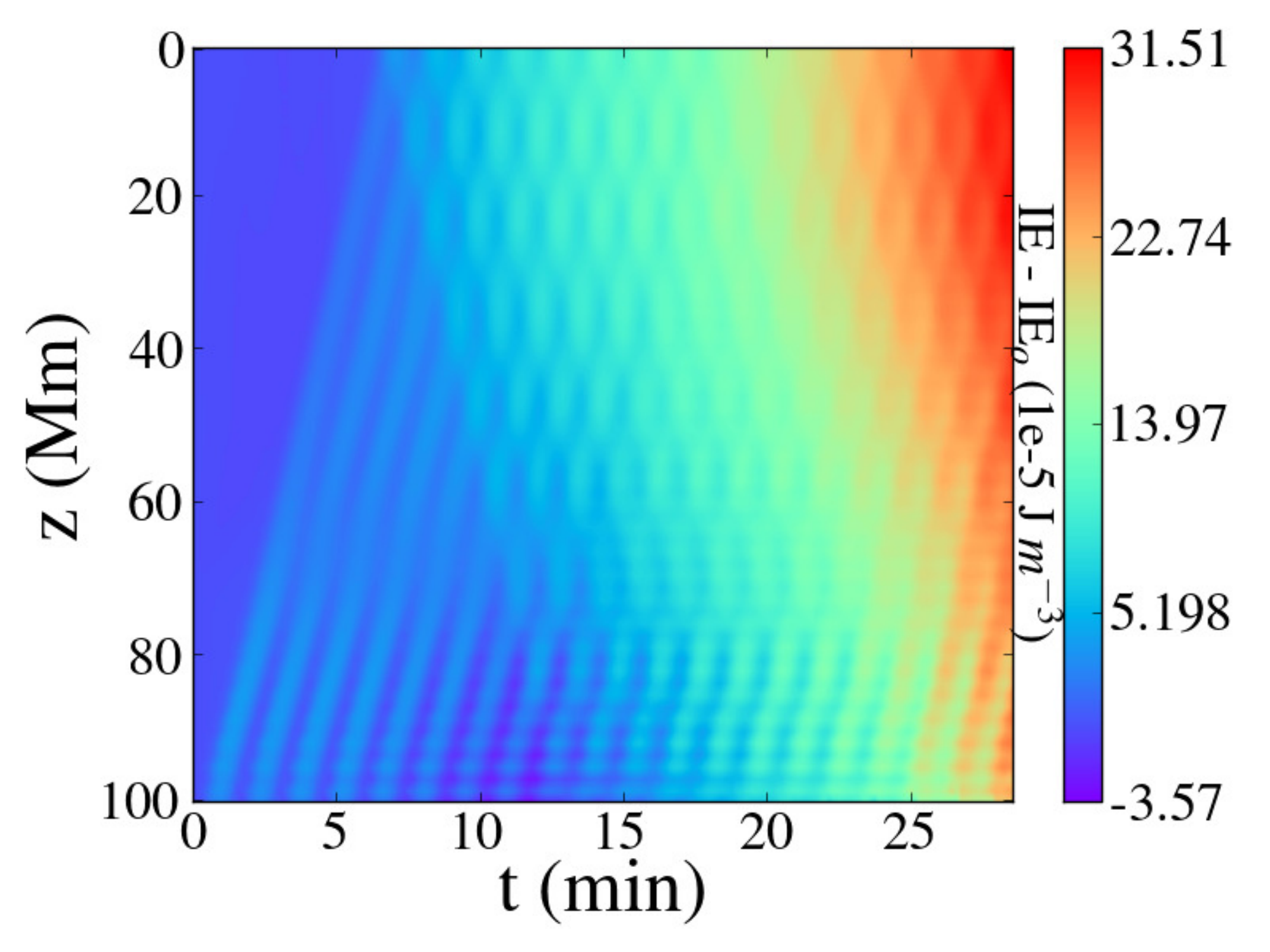}
\includegraphics[trim={0cm 0cm 0cm 0cm},clip,scale=0.16]{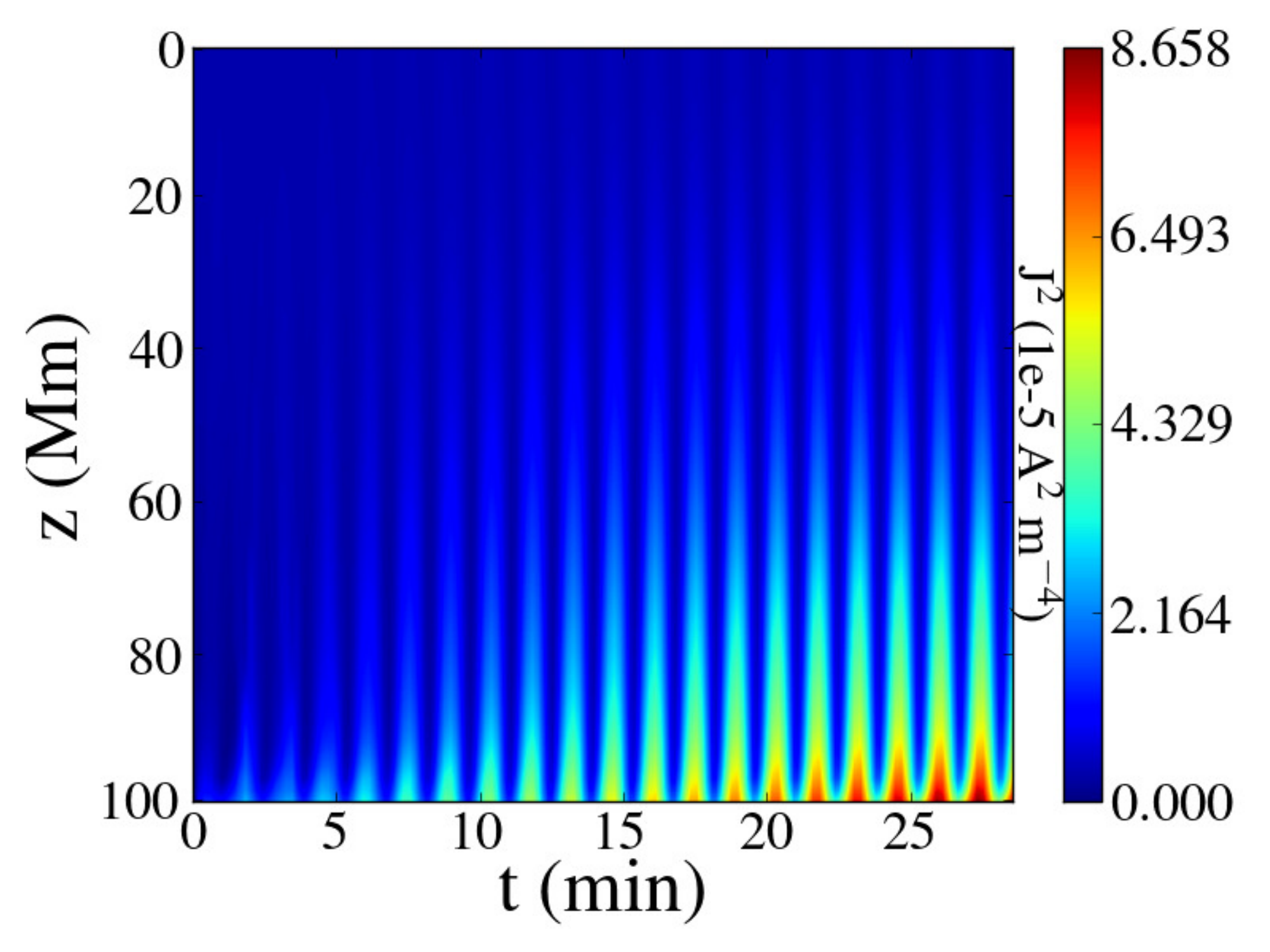}}
\resizebox{\hsize}{!}{\includegraphics[trim={0cm 0cm 0cm 0cm},clip,scale=0.16]{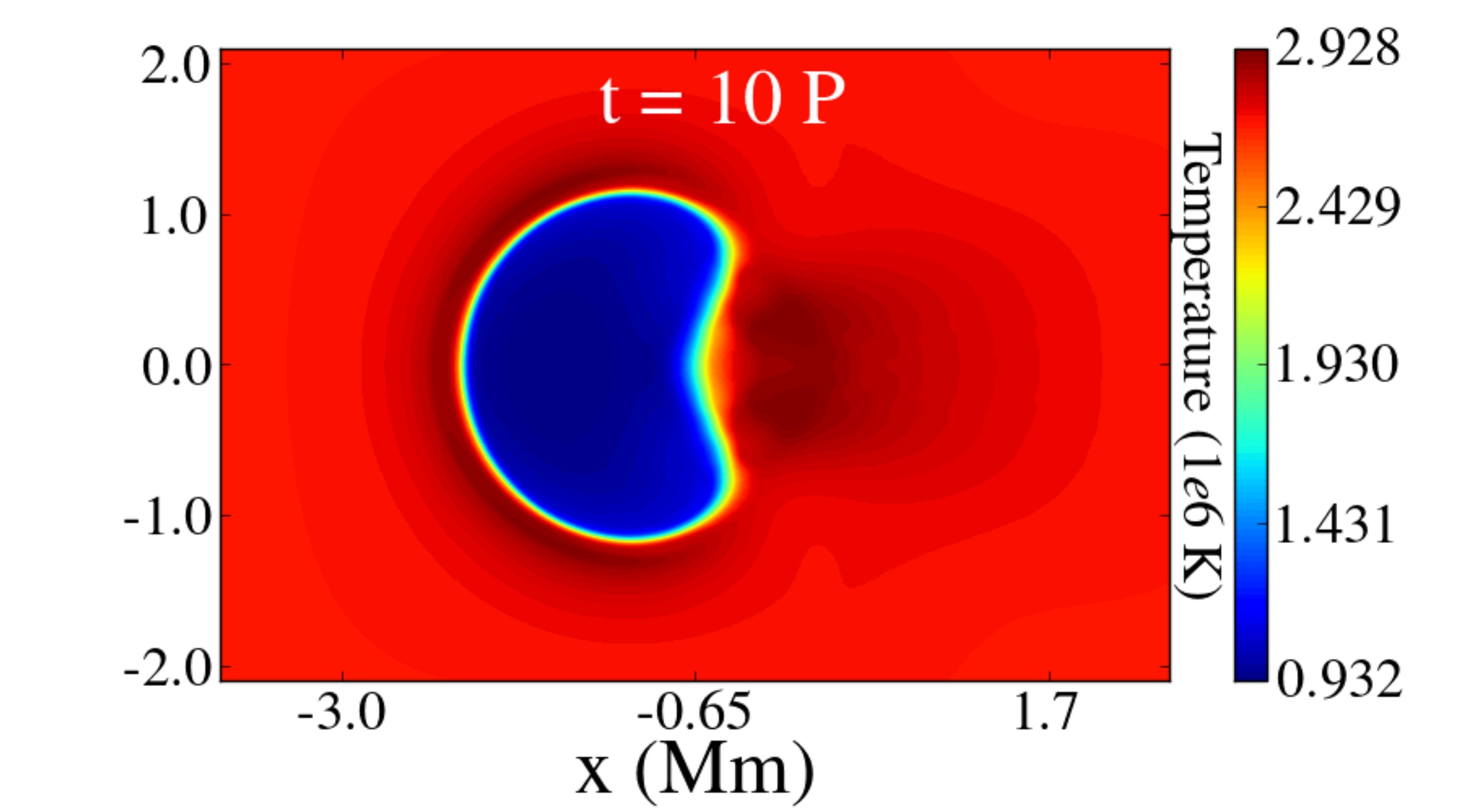}
\includegraphics[trim={0cm 0cm 0cm 0cm},clip,scale=0.16]{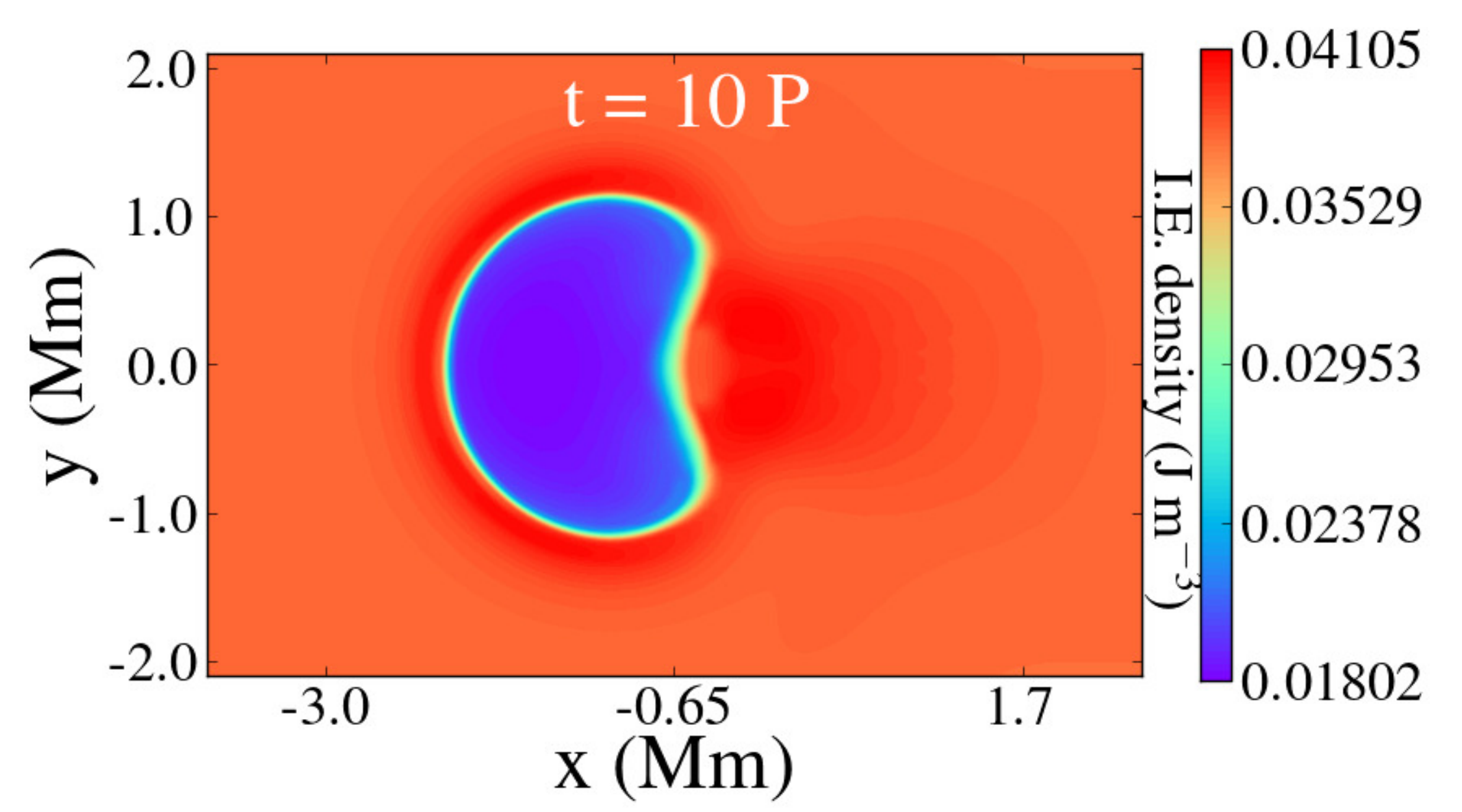}
\includegraphics[trim={0cm 0cm 0cm 0cm},clip,scale=0.16]{a34309-18_pic29}}
\caption{Top row: Average temperature (left) and internal energy (middle) of the entire $x-y$ plane and average square current density (right) of the flux tube (for $\rho \geq 0.9 \, f(z)$) per height and over time. Bottom row: Temperature (left), internal energy (middle) and density (right) at the $x-y$ plane at the apex ($z=0$). Data depict the ColdV2 model. The period of the driver is $P \simeq 172$ s.}\label{fig:viscous}
\end{figure*}

The differences between the ColdI, ColdR, and ColdV models are relatively small because of the high value of the numerical dissipation. Once we use a very high value for viscosity in the ColdV2 model, we cannot observe an inertial range anymore owing to the limitations of the current resolution. Instead, the spectrum passes rapidly to the dissipation range, even at very small wavenumbers. The amount of energy available at smaller scales is now less than in the other models, hindering dissipation. This can explain the lower values of average internal energy along the loop that we saw in Fig. \ref{fig:viscous}.

\begin{figure}
\centering
\includegraphics[trim={0cm 0cm 0cm 0cm},clip,scale=0.33]{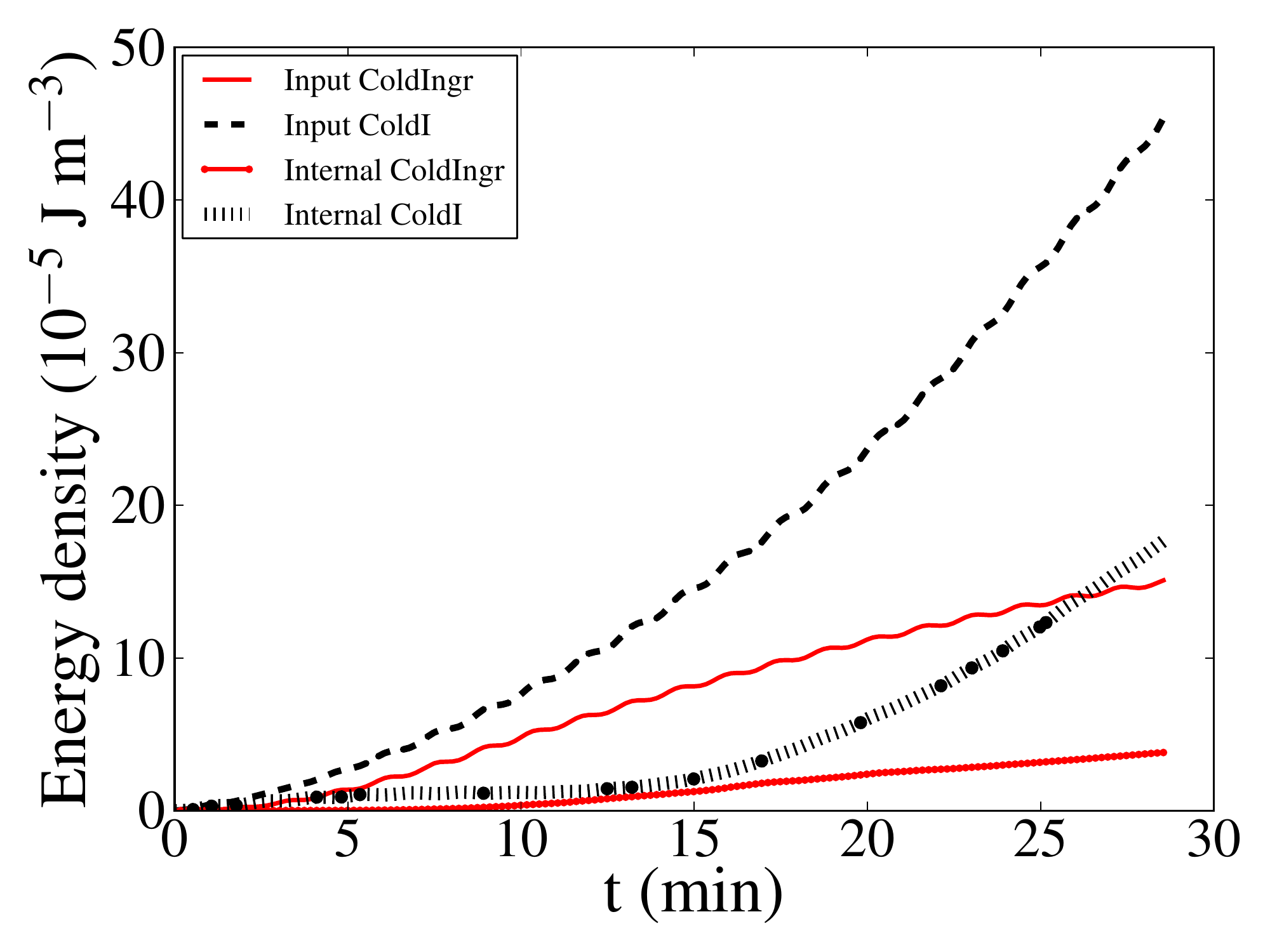}
\caption{Time profile of the input energy density from the driver and internal energy density variation relative to the initial state for the ColdI and ColdIngr models. The quantities are volume averaged for the whole computational domain; the contributions from the energy flux due to the plasma displacement through the side boundaries have been incorporated to internal energy densities.}\label{fig:energiescompare}
\end{figure}

Finally, in order to explain the differences between the gravitationally stratified and non-stratified models, we plot the time profiles of the input energy for models ColdI and ColdIngr (Fig. \ref{fig:energiescompare}). As we see the total input energy at the end of the simulation is almost three times higher in the ColdI models than it is in the ColdIngr model. From eq. \ref{eq:poynting} we know that the dominant terms in the Poynting flux are the velocities, magnetic field and, once considering the existence of physical or effective numerical resistivity, the currents. From our chosen set of parameters for the initial set-ups, both models have the same driver amplitude, driver frequency, an almost identical initial magnetic field and the same eigenfrequency for the fundamental transverse kink mode. As a result, all the differences in the final value for the total input energy are non-linearly caused by the different dynamical evolution of the oscillating loop owing to the presence (absence) of gravity. This energy input difference inevitably affects the energy evolution of the two models. A similar, but less pronounced behaviour is observed between the ColdI, ColdR, and ColdV models as well, which is caused by the differences in the dissipation parameters. 

If we consider the last $8.7$ minutes of these simulations in Fig. \ref{fig:energies} and Fig. \ref{fig:energiescompare}, when the curves of both the input and internal energy can be approximated by a linear function, we estimate the following values for the input flux ($F_{input}$ and heating rate ($H_r$) in our domain:
\begin{itemize}
\item $F_{input} = 7.5$ J m$^{-2}$ s$^{-1}$ and $H_r = 2.8$ J m$^{-2}$ s$^{-1} \approx 0.37 \, F_{input}$ for the ColdIngr model 
\item $F_{input} = 42$ J m$^{-2}$ s$^{-1}$ and $H_r = 28$ J m$^{-2}$ s$^{-1} \approx 0.67 \, F_{input}$ for the ColdI model 
\item $F_{input} = 46$ J m$^{-2}$ s$^{-1}$ and $H_r = 37$ J m$^{-2}$ s$^{-1} \approx 0.80 \, F_{input}$ for the ColdR model
\item $F_{input} = 55$ J m$^{-2}$ s$^{-1}$ and $H_r = 40$ J m$^{-2}$ s$^{-1} \approx 0.72 \, F_{input}$ for the ColdV model.
\end{itemize} 
These values are less than half from the $F_{radiative}=100$ J m$^{-2} s^{-1}$, which is the value for the radiative losses in the quiet corona \citep{withbroenoyes1977ara}. This shows that the driver in our models does not provide enough energy to sustain the density and temperature profile in the corona. One important observation is the relation between the input energy flux and the heating rate. As we see, the heating rate of the ColdIngr model is $\approx 37 \%$ of the corresponding input flux, which is significantly less than the corresponding percentages of the gravitationally stratified models. This shows that the inclusion of gravity generally leads to a more effective wave energy dissipation, for our given models.

\section{Discussion and conclusions}
In the current study, we wanted to quantify the effects of gravitational stratification and finite resistivity and
viscosity on the magnitude and location of wave heating caused by standing transverse waves in coronal loops. We performed 3D numerical simulations of single $3$D, gravitationally stratified, density-enhanced straight flux tubes in ideal, resistive, and viscous MHD. Through a parameter study, we estimated the effective values of numerical resistivity and viscosity present in our set-ups, which are many orders of magnitude larger than the expected values in the solar corona. The effects of physical dissipation, gravity, and driver strength were studied for a cold loop embedded in a hot corona and hot loops inside a colder corona. A non-stratified loop with uniform temperature was used in ideal MHD  to determine the effects of numerical dissipation on our results. The standing transverse waves in our models were produced with the use of a continuous, monoperiodic, footpoint driver; the frequency was equal to the analytically predicted value for the standing fundamental kink oscillations of uniform flux tubes \citep{edwin1983wave,andries2005A&A430.1109A}.

The effects of numerical resistivity were addressed in a model of a non-stratified loop with uniform initial temperature (UniT). In that simulation we observe an increase of the average temperature near the footpoint and near the apex (Fig. \ref{fig:isot}). This heating is caused by the numerical dissipation found in our code, which effectively acts as resistivity and viscosity in the case where ideal MHD is used. As we expect from \citet{karampelas2017} for loops undergoing a standing kink oscillation, resistivity is the main cause for heating near the footpoint, while shear viscosity is responsible for heating in the area of the apex. This simulation was used as a template to better understand the temperature evolution in our other set-ups. 

Expanding upon our previous work \citep{karampelas2018fd}, we observe the creation of spatially extended  TWIKH rolls. As a consequence of continuous driving, these TWIKH rolls expand across the loop cross section, fully deforming the initial monolithic density profile. Just like in the non-stratified case studied in that paper, the TWIKH rolls created elongated strand-like structures along the flux tube. These strands, which resemble those studied in \citet{antolin2016} for an impulsive standing kink wave, are also visible in the cases of resistive and viscous MHD (see Fig. \ref{fig:fomo}). In \citet{howson2017}, the development of KHI in impulsively oscillating flux tubes is hindered in the presence of resistivity and especially viscosity. For the driven oscillations considered in this work, the KHI was still delayed for similar values a physical dissipation. However, the TWIKH rolls eventually expand throughout the loop cross section, similar to the set-up with ideal MHD (Fig. \ref{fig:TWIKH}). 

By increasing the value of shear viscosity, we found that we need very low Reynolds numbers (we used $R_e \sim 10^2$) to suppress the development of the KHI when continuous drivers are used. The use of such a high value for shear viscosity, however, leads to an unusual temperature profile for an oscillating cold loop. The suppression of KHI prevents the mixing of plasma between the loop and the surrounding plasma. At the same time, we found a strong heating taking place near the apex, rather than the footpoint as is expected for loops transversely oscillating in the fundamental kink mode \citep{tvd2007resist}. This heating is observed both in the temperature and internal energy profiles. This shows that very high values of viscosity in the corona should not be expected, unless the corresponding heating signatures at the apex of oscillating loops are also observed.  

Studying the temperature profiles along the loop axis and over time for a non-stratified cold loop (model ColdIngr), we observed a slight temperature increase near the footpoint owing to ohmic dissipation due to numerical dissipation. However, the average temperature shows an apparent drop higher up the loop due to KHI induced mixing between plasma of different temperatures. These results are in agreement with our past studies \citep{karampelas2017}, where a weaker driver was employed. Energy dissipation takes place all along the loop axis; the strongest values are acquired near the apex. A different temperature profile was acquired when comparing with
the models of cold gravitationally stratified loops. An increase of the average temperature of our domain was observed near the footpoint and apex, despite the mixing effects (Fig. \ref{fig:456zt}). Gravity seems to affect the evolution of our systems greatly, since the corresponding set-up in \citet{karampelas2017} and the model ColdIngr predominately showed apparent cooling over our domain. The temperature increase observed in the stratified loops was located near the footpoints and apex in accordance with the results of the loop with uniform temperature. This temperature increase was also accompanied by an increase of the internal energy all along the loop length, and took its highest values near the apex. This proves that the observed temperature increase is not just an apparent phenomenon, but the result of actual wave heating. The temperature profiles are in agreement with the expected results from \citet{tvd2007resist} for these types of standing modes. However, the heating for the gravitationally stratified models is still between $28\%$ and $40\%$ of the radiative losses ($F_{radiative}=100$ J m$^{-2} s^{-1}$) for the quiet corona \citep{withbroenoyes1977ara}, so it is still not enough to sustain the observed coronal temperatures. 

By including physical dissipation ($R_e=10^4$ for the ColdV and $R_m=10^4$ for the ColdR model), the internal energy increased more along the entire loop. Both resistivity and viscosity seem to increase the internal energy near the apex (Fig. \ref{fig:456c}), and viscosity causes higher temperatures there ($\sim 5\times 10^4$ K increase). Near the footpoint, we would expect the resistive case to lead to the highest temperature increase, since the square current densities (dominated by $J_z^2$) have their highest values there for all three models. The reason why the viscous case shows higher average temperatures at the footpoint as well is the shrinking of the viscous cross section there, as observed in the $zt$ profile for the tube surface area of the ColdV  model, combined with the resistive heating due to numerical dissipation. This leads to an apparent effect of higher average temperature than in the ideal or resistive MHD model.

The similar evolution of the square current densities in Fig. \ref{fig:456zt} for models ColdR and ColdV, the similar dynamical evolution, and their differences from the ideal MHD case hints at a description of resistivity in terms of turbulent viscosity, or (shear) viscosity as in terms of anomalous resistivity. Additional simulations with very low values of magnetic Reynolds numbers ($R_e \leq 10^2$) need to be considered to determine whether such high values of resistivity exist in the solar corona.

Studying the energy profiles for models ColdI, ColdR, and ColdV, we see the average value of the magnetic energy density (minus the Poynting fluxes from the side boundaries) showing a steady, almost linear growth, similar in all models. A similar small growth was also observed for the gravitational energy density because of the redistribution of plasma along the loop. The input energy from the driver showed a faster growth, reaching different values for each set-up due to the differences in the models dynamical evolution over time. For the kinetic energy of the three models, we identified a phase of a linear growth during the first six periods, a phase of decelerating growth for two more periods, and saturation phase at the later stages of the simulation. Once the kinetic energy enters the phase of decelerating growth and eventual saturation, the internal energy (minus the fluxes due to plasma flow through the boundaries) exhibits a rapid growth.

During the kinetic energy saturation, the loop develops a turbulent profile, leading the cascade of energy into smaller scales and more efficient heating. The lack of smaller scales for the highly viscous ColdV2 model leads to a lack of efficient energy cascade, as proved by the practically non-existent inertial range in its power spectra in Fig. \ref{fig:energies}. This results in less energy at higher wavenumber, and less efficient dissipation, as was proven in Fig. \ref{fig:viscous}. This proves that a turbulent loop profile is needed for more efficient wave heating. In combination with past results \citep{magyar2016strands,magyar2017,karampelas2018fd}, we conclude that the use of monolithic density profiles for flux tubes should be done with care. 

We saw that the inclusion of gravity in our models plays an important role when it comes to the development and efficiency of wave heating. More specifically, stratified loops of the same eigenfrequency and initial magnetic field as a non-stratified loop showed a greater increase of internal energy with respect to their corresponding input energy, when compared to non-stratified loops. The lack of gravity seems to underestimate the efficiency of wave heating in straight flux tube models of coronal loops. Therefore, this should not be ignored in future studies.

To sum up our conclusions, we see that the inclusion of gravity in our models seems to play an important role when it comes to the development and efficiency of wave heating. The inclusion of physical dissipation should also be considered in any attempts to map the location of wave energy dissipation. Energy dissipation seems to be more efficient once the kinetic energy of our loops reaches a saturation phase, for a turbulent loop profile. Another important result is that resistivity and shear viscosity lead to the development of the smaller scales in a similar fashion. Driver induced TWIKH rolls develop in our set-ups unless very high dissipation is used (for example a Reynolds number of $R_e=10^2$). In case of very high viscosity, the development of smaller scales is hindered, heating inside the loop is suppressed, and temperatures are increased predominately near the apex. Future steps should include more physical mechanisms (thermal conduction and radiation) and a more realistic atmosphere than that considered here. Finally, drivers of different amplitudes and frequencies need to be considered in an attempt to determine the amount of energy required for sustained wave heating, while obtaining results that agree with the observed oscillation profiles in loops.
\\

\begin{small}
\textit{Acknowledgements.} We would like to thank the referee whose review helped us improve the manuscript. We also thank the editor for his comments. K.K. was funded by GOA-2015-014 (KU Leuven). T.V.D. was supported by GOA-2015-014 (KU Leuven). M.G. is also supported by the National Natural Science Foundation of China (41674172). This project has received funding from the European Research Council (ERC) under the European Union's Horizon 2020 research and innovation programme (grant agreement No 724326). The computational resources and services used in this work were provided by the VSC (Flemish Supercomputer Center), funded by the Research Foundation Flanders (FWO) and the Flemish Government – department EWI. The results were inspired by discussions at the ISSI-Bern and at ISSI-Beijing meetings. 
\end{small}

\bibliographystyle{aa}
\bibliography{aa34309-18}

\end{document}